\documentclass[10pt]{article}

\usepackage{amssymb} 
\usepackage{amsmath} 

\usepackage[margin=1in]{geometry}
\usepackage{graphicx,ctable,booktabs}
\usepackage{verbatim}
\usepackage{enumerate}
\usepackage{float}

\usepackage{epstopdf}
\usepackage{url}
\usepackage{xcolor}

\begin{document}
\emergencystretch 8pt

\title{A Design-Based Model of the Aortic Valve \\ for Fluid-Structure Interaction}
\author{Alexander D. Kaiser$^{1,2,3}$, Rohan Shad$^{3,4}$, William Hiesinger$^{3,4}$, Alison L. Marsden$^{1,2,3,5}$}
\date{
{\small
$^{1}$Institute for Computational and Mathematical Engineering, Stanford University; \\
$^{2}$Department of Pediatrics (Cardiology), Stanford University; 
$^{3}$Stanford Cardiovascular Institute; \\
$^{4}$Department of Cardiothoracic Surgery, Stanford University; 
$^{5}$Department of Bioengineering, Stanford University 
}
\\
\vspace{10pt}
\today}

\maketitle

\thispagestyle{empty}

\begin{abstract}

This paper presents a new method for modeling the mechanics of the aortic valve, and simulates its interaction with blood. 
As much as possible, the model construction is based on first principles, but such that the model is consistent with experimental observations. 
We require that tension in the leaflets must support a pressure, then derive a system of partial differential equations governing its mechanical equilibrium. 
The solution to these differential equations is referred to as the predicted loaded configuration; it includes the loaded leaflet geometry, fiber orientations and tensions needed to support the prescribed load. 
From this configuration, we derive a reference configuration and constitutive law. 
In fluid-structure interaction simulations with the immersed boundary method, the model seals reliably under physiological pressures, and opens freely over multiple cardiac cycles. 
Further, model closure is robust to extreme hypo- and hypertensive pressures. 
Then, exploiting the unique features of this model construction, we conduct experiments on reference configurations, constitutive laws, and gross morphology.
These experiments suggest the following conclusions:
(i) The loaded geometry, tensions and tangent moduli primarily determine model function. 
(ii) Alterations to the reference configuration have little effect if the predicted loaded configuration is identical.
(iii) The leaflets must have sufficiently nonlinear material response to function over a variety of pressures. 
(iv) Valve performance is highly sensitive to free edge length and leaflet height.
These conclusions suggest appropriate gross morphology and material properties for the design of prosthetic aortic valves. 
In future studies, our aortic valve modeling framework can be used with patient-specific models of vascular or cardiac flow.

\end{abstract}

\section{Introduction}

The aortic valve is one of four valves in the human heart. 
It lies between the left ventricle, the main pumping chamber of the left heart, and the aorta, the central artery through which oxygenated blood leaves the heart. 
The valve serves to prevent backflow of blood during diastole, the filling phase of the left ventricle, in which blood enters the chamber through the mitral valve, and opens in systole as the heart beats. 
In a typical anatomy, the valve is composed of three thin, flexible leaflets. 
A highly oriented system of collagen fibers provides the primary mechanical stiffness of the leaflets, and leads to highly anisotropic material properties. 
The leaflets are anchored to a non-planar ring called the aortic annulus. 
During forward flow, the leaflets bend to create an approximately circular orifice at the inlet of the aorta. 
During closure, the free edges of the leaflets coapt, and the orifice is roughly trisected by the lines of contact.

In this paper, we present a new method for modeling the mechanics of the aortic valve and simulate its interaction with blood. 
We use nearly first-principles techniques, yet tune the results to empirical knowledge from experiments on the gross morphology, kinematics and material properties of valve leaflets. 
This framework allows us the flexibility to adapt the model to a range of patient specific anatomies, without reliance on data that is not typically available in a clinical setting. 
We assume that the valve leaflets must support a pressure, and then derive the valve geometry and material properties from the resulting differential equations. 
The solution of these differential equations is referred to as the \emph{predicted loaded configuration}; this includes the loaded geometry, its fiber orientations and the tensions required to support such a load. 
The formulation in this work allows the leaflets to be tuned for a given gross morphology, to fit on a known annular geometry, and directly provides information about the material properties in both the fiber and cross-fiber directions. 
Since the geometry and material properties are derived and tuned, rather than measured and assigned, we refer to this as a \emph{design-based} approach to elasticity.

A central challenge in fluid-structure interaction (FSI) studies of heart valves is to simulate multiple cardiac cycles under physiological pressures, and achieve behavior that qualitatively matches that of a real valve. 
The model should open freely, allowing a jet of forward flow during the ejection phase, then close under back pressure, sealing without leak or regurgitation, and finally open again and repeat the cycle. 
To simulate closure is especially challenging, as leaks must be prevented only by the balance of elastic forces in the leaflets and fluid forces \cite{kheradvar2015emergingIV}. 

The first goal of this paper is to demonstrate that our methods achieve the above central challenge. 
We perform FSI simulations with the immersed boundary (IB) method and show that the model seals under physiological pressures and opens freely over multiple beats. 
Further, its closure is robust to pressures much lower and higher than physiological pressures. 
The second goal is to study the effect of differing reference configurations, constitutive laws and gross morphology for aortic valve tissue. 
Using the design-based model generation scheme, the reference configuration and constitutive law can be modified while maintaining an identical predicted loaded geometry and tension. 
We exploit this to alter the rest lengths associated with given loaded lengths, and study a number of constitutive laws -- each of which are equivalent in the predicted loaded configuration -- to determine the pressure ranges for which each are effective. 
These numerical experiments suggest functional explanations for observed native valve material properties and geometry. 
Further, observed ranges for good model valve function in turn suggest optimal ranges for aortic valve prosthetics.

Our techniques are directly adapted from prior modeling methods for the mitral valve \cite{thesis,kaiser2019modeling}, for which there are similar goals but a very different anatomy, as well as prior models of the aortic valve \cite{PeskinH319}.  
In their original study, Peskin and McQueen used the simplification that a single fiber family bears all of the load; they treated the cross-fiber direction as not bearing any load and did not include its potential deformation. 
The entire loaded geometry of the leaflets was determined by the free edge, so the leaflet geometry could not be forced to conform to a general annular geometry and the gross morphology of the leaflets could not be readily tuned. 
Subsequent work using this predicted loaded configuration was effective in FSI simulations at physiological pressures \cite{Griffith_aortic,flamini2016immersed}. 
More recently, in vitro experiments revealed that the radial, or cross-fiber direction undergoes large strains \cite{yap2009dynamic}, and material tests suggests that the radial direction exerts significant stress under such strains \cite{may2009hyperelastic}. 
Thus, we seek a formulation with more flexibility in the emergent geometry and that includes radial tension and strain.

In the current approach, the fiber structure emerges from the solution of a differential equation, and therefore can be considered a new ``rule based'' method for assigning fiber structure.  
In one such technique, a modified Laplace equation was solved to interpolate the fiber orientation from known points at the boundary of the leaflets \cite{hasan2017image}. 
The modified Laplace equation was selected because its solutions are smooth and readily computable with standard numerical techniques. 
In contrast, the differential equation that we solve is motivated directly by the valve's function, and its solution provides simultaneous information about the valves geometry, fiber orientation, and material properties.

There are a variety of other methods for FSI and aortic valve modeling. 
Griffith and collaborators use the IB method with a finite element formulation on the structure and achieve flow rates comparable to clinical measurements \cite{hasan2017image}. 
Their geometry is based on a simple analytic shape from \cite{Driessen}, which is manually modified to match scan data. 
Hsu et al. use the immersogeometric framework, which uses non-conforming meshes as in the IB method, and manually designed geometry with computer aided design (CAD) software \cite{hsu2015dynamic}. 
Fictitious domain methods are also non-conforming, but require contact forces \cite{astorino2009fluid,shadden2010computational}. 
Mao et al. use smooth particle hydrodynamics and geometry measured from a CT scan \cite{sun_left_heart}. 
Marom et al. use a two way coupling approach that requires contact forces and use a geometry based on a simple analytic shape, tuned to gross morphology from echocardiography \cite{marom2012fluid}. 
Arbitrary Lagrangian-Eulerian (ALE) schemes, which maintain non-overlapping domains and a sharp interface between them, appear to have limited success in simulating heart valves. 
One study with ALE required the addition of a fictitious surface to close the valve orifice once near-contact was detected, thus largely prescribing closure rather than having it emerge from the system's dynamics \cite{spuhler20183d}.
Bavo et al. compared ALE methods with IB methods, and did not succeed in simulating an entire cardiac cycle with ALE because of remeshing problems, and they required complex and nonphysical handling of contact \cite{bavo2016fluid}. 
While some studies focus solely on solid mechanics, studies comparing solids-only simulations with FSI simulations suggest that simulating FSI is essential to study the full dynamics of the valve, revealing qualitative and quantitative differences \cite{LAU20101057,marom2012fluid,bavo2016fluid,hsu2015dynamic}.
Questions related to heart valve prosthetics that require FSI to simulate are reviewed in \cite{yoganathan2005flow}. 
See Le et al. \cite{lee2020fluid} for additional review and discussion. 
Of results reported in these studies, methods using non-conforming, IB approaches appear to work best, with multiple studies reporting realistic behavior through at least one cardiac cycle. 
Most model geometries are constructed from a simple analytic shape or a scan, however, analytic shapes are not anatomical, and scan data at sufficient resolution may not be available to construct a good patient-specific valve geometry. 
In contrast, our method allows us to create arbitrarily sized model valves with generic material properties, without relying on individual measurements of valve anatomy or material properties that are typically not available clinically. 
In future work, an appropriately sized model valve could be constructed to fit in a patient-specific vascular or heart geometry.

The remainder of this paper is organized as follows. 
In Section \ref{anatomy} we review literature on aortic valve anatomy and physiology. 
In Section \ref{Methods}, we derive a system of partial differential equations, the solution of which gives the predicted loaded configuration. 
Using this configuration, we derive a constitutive law for the valve.
Then, we outline methods for FSI including the model test chamber and boundary conditions.
In Section \ref{Results}, we present the model geometry, the emergent heterogeneous constitutive law, and results of FSI simulations. 
We then change the reference configuration, the constitutive law, and the model geometry and report results on each. 
We discuss these results in Section \ref{Discussion}, study limitations in Section \ref{Limitations}, and finally conclusions in Section \ref{Conclusions}. 

Source code for the project is freely available at \url{github.com/alexkaiser/heart_valves}.

\section{Relevant aspects of aortic valve anatomy and physiology}
\label{anatomy}

\begin{figure}[th!]
\centering
\includegraphics[width=\columnwidth]{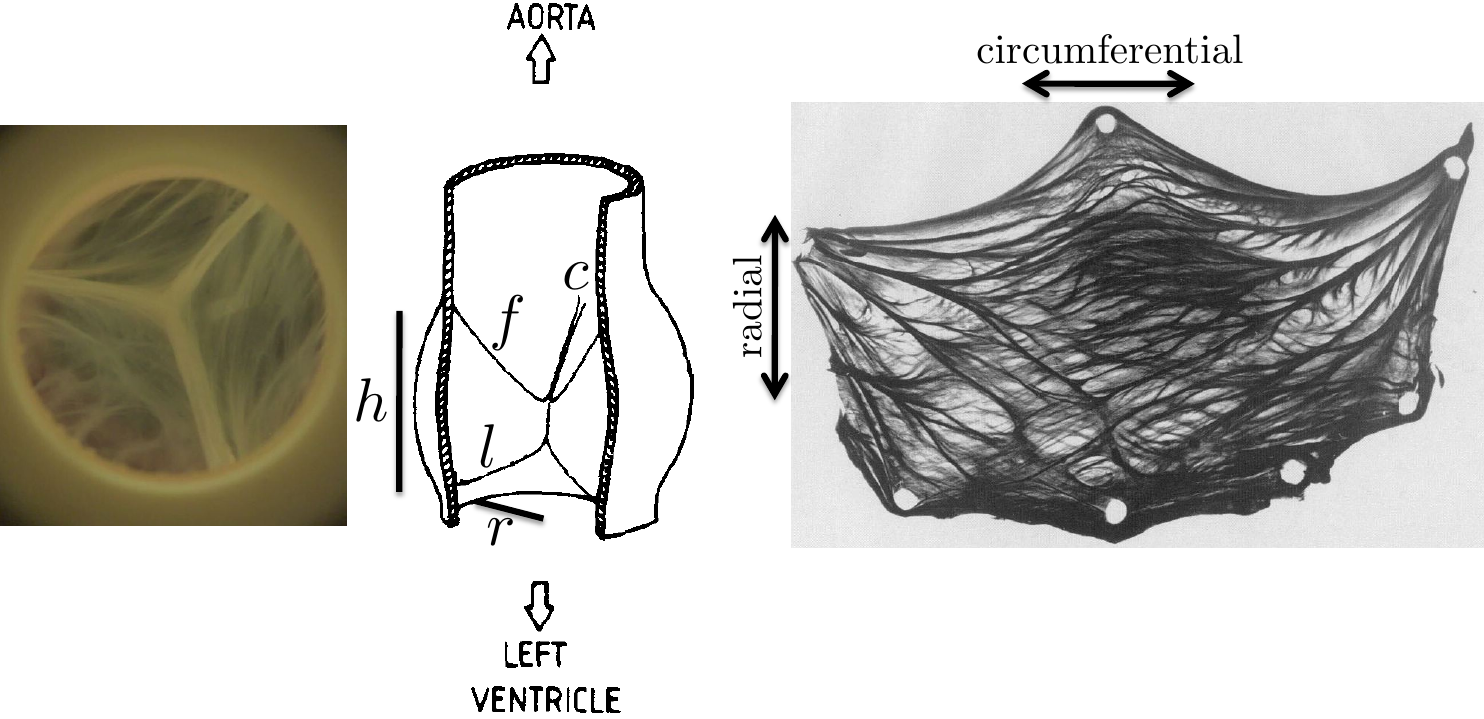}   
\caption{Anatomy of the aortic valve.
Left: Closed aortic valve from above.
Center: Schematic in a dissected view from the side, illustration reprinted from \cite{SAUREN1983327} with permission from Elsevier. 
Labeled are the free edge (f), the leaflet height (l), the annular radius (r), the annular height (h) and a commissure point where leaflets meet (c).
Right: Excised aortic valve leaflet stained for collagen, reprinted from \cite{sauren_thesis}. 
Tissue-scale collagen fiber bundles are visible with a heavily circumferential orientation. 
} 
\label{anatomy_diagrams}
\end{figure}

We wish to construct a model that is consistent with real valves in four categories: the gross morphology of the valve, its fiber structure, its material properties, and its loaded strain. 
Figure \ref{anatomy_diagrams} reviews the anatomy of the aortic valve.
The expected properties that we use as targets are: 
\begin{description}

\item[{\bf Gross morphology:}] 
For a given radius $r$, the loaded leaflet height is approximately 1.4$r$, the loaded free edge length of each leaflet is 2.48$r$, and the height of the annulus is $1.4$r \cite{swanson1974dimensions}. 
The leaflet thickness is approximately 0.044 cm \cite{sahasakul1988age}. 

\item[{\bf Fiber structure:}] 
Fibers run circumferentially, from commissure to commissure \cite{SAUREN198097}. 
Material response is highly nonlinear, with more stiffening at lower strains in the circumferential direction than in the radial direction \cite{may2009hyperelastic}. 
Tangent stiffness in the circumferential and radial directions are of order $10^{8}$ and $10^{7}$ dynes/cm$^{2}$ respectively, at sufficiently high strains such that collagen fibers are fully recruited \cite{pham2017quantification}. 

\item[{\bf Strain:}] 
Fully-loaded strains in the circumferential and radial directions are approximately 0.15 and 0.54, respectively  \cite{yap2009dynamic}. 

\end{description}

\noindent 
A short review of experimental literature associated with these four categories follows.  

{\bf Gross morphology:} 
Swanson and Clark provide a representative description of the gross morphology of human aortic valve leaflets in a loaded configuration  \cite{swanson1974dimensions}. 
They claim that the radius at which the leaflet attaches is constant through the height of the annulus. 
For a radius $r$, the height to the commissure is 1.42$r$, the free edge length of each leaflet is 2.48$r$, the leaflet height in the radial direction is 1.4$r$.
Sch{\"a}fers et al. report similar mean leaflet heights of 2.0 cm with a mean radius of 1.38 cm \cite{schafers2013cusp}. 
A study on human cadaver hearts reported a mean leaflet thickness of 0.044 cm \cite{sahasakul1988age}. 

{\bf Fiber structure:}
Aortic leaflet material properties are highly nonlinear and anisotropic, attributed to oriented, wavy collagen fibers that are recruited, or straightened, under load and then provide additional stiffness \cite{SAUREN1983327}. 
These fibers run from commissure to commissure, or circumferentially, and fiber bundles can be seen with even moderate magnification (Figure \ref{anatomy_diagrams}). 
These bundles are hierarchical, meaning that fiber bundles visible reflect alignment of smaller scale collagen fibers, fibrils and molecules \cite{gautieri2011hierarchical,doi:10.1146/annurev.biochem.77.032207.120833,ABHILASH20141829}. 
The radial direction is referred to as the cross-fiber direction. 

{\bf Material properties:}
Material testing of aortic leaflet tissue reveals that the circumferential (fiber) direction is stiffer, and has a more nonlinear response at smaller strains compared to the radial direction. 
May-Newman et al. measured nonlinear material properties in both the fiber and radial directions  \cite{may2009hyperelastic}.
The curves appear approximately exponential with a monotonically increasing slope; there is no clear linear region after a certain strain. 
Pham et al. report a tangent modulus of $9.9 \cdot 10^{7}$ dynes/cm$^{2}$ circumferentially, and $2.3 \cdot 10^{7}$ dynes/cm$^{2}$ radially for an approximately 4:1 ratio of circumferential to radial tangent modulus \cite{pham2017quantification}. 
Clark found an affine region of material response begins at strains of 0.13 circumferentially and 0.24 strain radially, and the fully recruited tangent modulus to be $5.8 \cdot 10^{7}$ dynes/cm$^{2}$ circumferentially, and $1.7 \cdot 10^{7}$ dynes/cm$^{2}$ radially, for an approximately 4:1 ratio of circumferential to radial stiffness \cite{clark1973stress}. 
Sauren et al. reported an approximately 20:1 ratio of circumferential to radial stiffness at maximum \cite{SAUREN1983327}. 
Literature on the shear response appears to be limited, but one modeling study found that shear response had little influence on the closing kinematics of the valve \cite{hammer2011mass}.

{\bf Kinematics and expected strains:}
To consistently define strain, let $E$ denote engineering strain $E = (L-R)/R$ where $L$ denotes the current length and $R$ denotes reference length.
Yap et al. tested porcine leaflets in vitro, and found that the principle axes of strain aligned with the radial and circumferential directions of the leaflets, and found a near-constant circumferential strain of 0.15 and a radial strain of 0.54 during diastole \cite{yap2009dynamic}.
Additionally, there may be strain at all times in the cardiac cycle or even at rest due to prestrain, which may substantially change the values of strains that are reported in studies \cite{rausch2013effect}. 
One study found prestrain through the cardiac cycle relative to an excised state fixed in glutaraldehyde \cite{aggarwal2016vivo}. 
Gluteraldehyde fixation, however, can markedly change the stress-strain response of the leaflet tissue \cite{billiar2000biaxial,billiar2000biaxial2}.
However, we show in Section \ref{varied_strain} that the model functions well under a variety of strains assigned to the predicted loaded configuration.

\section{Methods}
\label{Methods}

\subsection{Construction of the model aortic valve}
\label{problem_formulation}

The following assumptions form an idealized summary of the anatomy and function of the loaded aortic valve, and we will use them directly to construct the model valve. 

\begin{enumerate}
  
\item The valve is composed of three leaflets, each of which is anchored to the aortic annulus. 

\item Fibers run from commissure to commissure on each leaflet. 
Curves in the cross-fiber direction run from the annulus to the free edge of each leaflet. 

\item Each of the leaflets can exert tension in the circumferential, or fiber, direction and the radial, or cross-fiber, direction. Shear tension is assumed to be identically zero. 

\item Tension in the leaflets supports a uniform pressure load, creating a static mechanical equilibrium in which all forces balance. 

\end{enumerate}

We expect analysis of the closed configuration in equilibrium to be a good predictor of the loaded configuration when interacting with fluid. 
This is because the inertio-elastic timescale, an estimate of the duration of time it takes the leaflets to deform when pressurized, is much shorter than the expected duration of valve closure. 
Previously \cite{kaiser2019modeling}, we estimated this timescale to be 
$ r \sqrt{ \rho/\eta } $
where $r = 1.25 $ cm denotes the radius of the annulus, $\rho$ = 1 g/cm$^{3}$ is the density of the leaflets and $\eta = 10^{7}$ dynes/cm$^{2}$, which is the lowest order of magnitude for the fully-recruited tangent modulus that we found in the literature.
This gives an inertio-elastic timescale of $4.0 \cdot 10^{-4}$ s. 
Since the valve is closed for approximately 0.5 s per cardiac cycle, analysis based on the closed configuration should be a good predictor of the dynamics in general. 

We represent the leaflet surfaces as an unknown parametric surface in three dimensional space, 
\begin{align}  
\mathbf X(u,v) : \Omega \subset \mathbb{R}^{2} \to \mathbb{R}^{3} . 
\end{align}
The surface $\mathbf X$ has units of length, or cm, and the parameters $u,v$ are taken to be dimensionless. 
Curves on which $u$ varies and $v$ is constant run circumferentially and conform to and thus represent the fibers. 
Curves on which $v$ varies and $u$ is constant run radially, in the cross-fiber directions. 
The unit tangents to these two families of curves determine the local directions at which the leaflet exerts force in the fiber and cross-fiber directions, and are defined as 
\begin{align}
\frac{\mathbf X_{u}}{ | \mathbf X_{u} |} \quad \text{ and } \quad \frac{\mathbf X_{v}}{ | \mathbf X_{v} |}, 
\end{align}
respectively. 
These directions are not required to be orthogonal.

\begin{figure}[t]
\centering
\includegraphics[width=.6\columnwidth]{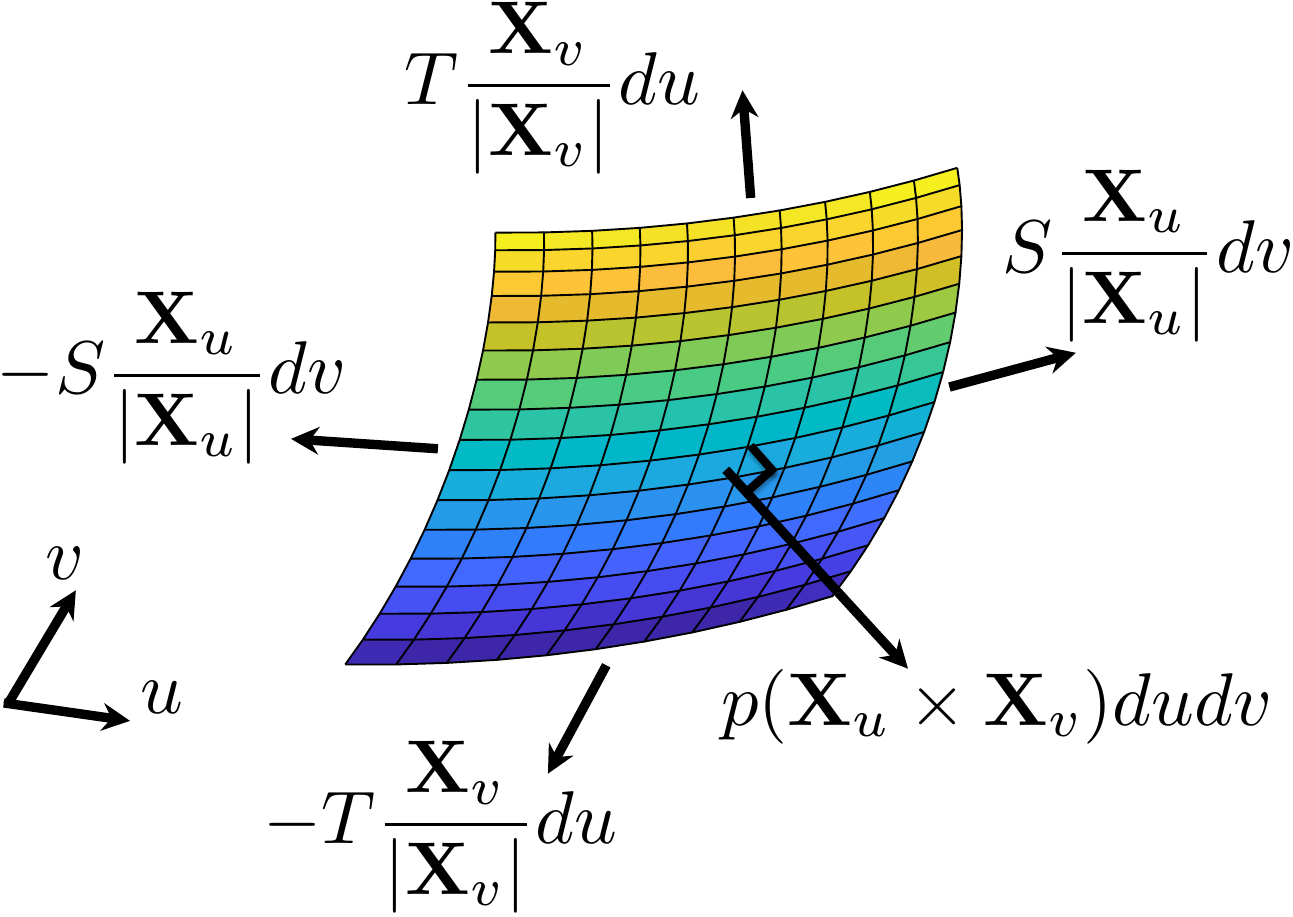}
\caption{Free body diagram of forces on the leaflets. 
Each of the tension forces is evaluated on one edge of the patch; the pressure force acts on the entire patch. 
Tensions on opposite sides of the patch are not equal and do not cancel, but we omit arguments for visual clarity. 
}
\label{free_body_diagram_leaflet.pdf}
\end{figure}

Now, consider the mechanical equilibrium on an arbitrary patch of leaflet specified by $[u_{1}, v_{1}] \times [u_{2}, v_{2}]$. 
The pressure $p$ acts normal to the leaflet across the entire patch. 
Let $S$ denote circumferential tension and $T$ denote radial tension. 
Circumferential tension acts on the curves specified by $v = v_{1}$ and $v = v_{2}$, and radial tension acts on the curves specified by $u = u_{1}$ and $u = u_{2}$. 
Figure \ref{free_body_diagram_leaflet.pdf} shows a free body diagram of these forces. 
Summing these forces gives the integral form of the mechanical equilibrium, or 
\begin{align} 
&0 = \int_{v_{1}}^{v_{2}}   \int_{u_{1}}^{u_{2}}    p  \left(  \mathbf X_{u}(u,v) \times \mathbf X_{v}(u,v) \right)    du dv  \\ 
&\hspace{-3pt}+ \int_{v_{1}}^{v_{2}}  \left(  S(u_{2}, v) \frac{ \mathbf X_{u} (u_{2}, v) }{|  \mathbf X_{u} (u_{2}, v) |} - S(u_{1},v)   \frac{ \mathbf X_{u} (u_{1}, v) }{|  \mathbf X_{u} (u_{1}, v) |}   \right)    dv \nonumber  \\ 
&\hspace{-3pt}+ \int_{u_{1}}^{u_{2}} \left(  T(u, v_{2})  \frac{\mathbf X_{v} (u, v_{2})}{|\mathbf X_{v} (u, v_{2})|} - T(u,v_{1})  \frac{\mathbf X_{v} (u, v_{1}) }{|\mathbf X_{v} (u, v_{1})|} \right)  du . \nonumber
\end{align}
Tension forces apply on the boundary of the patch, and so appear under single integrals, whereas pressure force appears in an area integral. 
Next, apply the fundamental theorem of calculus to convert each of the single integrals to double integrals. 
Then, swap the order of integration formally as needed to obtain
\begin{align} 
0 = &\int_{v_{1}}^{v_{2}}  \int_{u_{1}}^{u_{2}} \bigg(   p  (  \mathbf X_{u} \times \mathbf X_{v} )    +  \frac{\partial}{\partial u}  \left( S \frac{ \mathbf X_{u} }{ |\mathbf X_{u}| } \right) +  \frac{\partial}{\partial v}  \left( T \frac{ \mathbf X_{v} }{|\mathbf X_{v}|} \right)    \bigg)  \;  du dv . \nonumber
\end{align}
Since the patch is arbitrary, the integrand must be identically zero and the integrals can be dropped. 
This gives a partial differential equation for equilibrium of the leaflets 
\begin{align} 
0 = p  (  \mathbf X_{u} \times \mathbf X_{v} )  +   \frac{\partial}{\partial u}  \left( S \frac{ \mathbf X_{u} }{ |\mathbf X_{u}| } \right)  +  \frac{\partial}{\partial v}  \left( T \frac{ \mathbf X_{v} }{|\mathbf X_{v}|} \right).    
\label{eq_eqns}
\end{align}
Note that changes in the tension, that is, curvature in the fiber and cross-fiber directions, as well as local variation in the magnitude of tension, directly balance the pressure force. 

Equation \eqref{eq_eqns} has three components and five unknowns, the leaflet configuration $\mathbf X$ and the tensions $S$ and $T$, so additional information is required to close this equation. 
We wish to work directly with the loaded configuration, since the loaded configuration determines whether the valve will seal. 
Further, we do not immediately possess a realistic reference configuration, nor do we wish to use a simple analytic shape, since it is not anatomical. 
Thus, we specify a tension law that does not rely on a reference configuration. 
The simplest such law would be to prescribe constant tension, but this is not effective, as there are no parameters to tune to alter the gross morphology of the leaflets. 
Further, the radial tension $T$ is likely to be nonconstant, because at the free edge it must be supported by curvature in circumferential tension in a single fiber at the free edge only, whereas radial tension near the annulus can be supported by curvature in a number of fibers in the leaflet. 
Additionally, there is nothing to prevent fibers from nearly colliding during nonlinear iterations to solve the discretized equations equations of equilibrium. 
Adjacent points become close and evaluation of finite differences becomes ill-conditioned, and the nonlinear solvers fail to converge. 
(Discussion of the discretization and numerical methods follows.) 

To close equation \eqref{eq_eqns}, we prescribe the maximum tension but allow the value of tension to vary below that prescribed value. 
We therefore define $S$ and $T$ as 
\begin{align}
S(u,v) &= \alpha \left( 1 - \frac{1}{1 + |\mathbf X_{u}|^{2} / a^{2} } \right), \quad
T(u,v) = \beta \left( 1 - \frac{1}{1 + |\mathbf X_{v}|^{2} / b^{2} } \right). \label{dec_tension}  
\end{align}
Here, $\alpha$ denotes the maximum tension in the fiber direction, $\beta$ denotes the maximum tension in the cross-fiber direction. 
The parameters $a,b$ are tunable free parameters with units of length that can be adjusted to control the fiber spacing and gross morphology of the loaded configuration of the valve.  
Note that these parameters are not required to be constants, and indeed using a nonconstant value of $a$ was necessary to achieve anatomical gross morphology of the leaflets. 

Substituting Equation \eqref{dec_tension} into Equation \eqref{eq_eqns} gives the final form of the equilibrium equations  
\begin{align} 
0 &= p  (  \mathbf X_{u} \times \mathbf X_{v} )  
+  \frac{\partial}{\partial u}  \left( \alpha \left( 1  -  \frac{1}{1 +  |\mathbf X_{u}|^{2} / a^{2}  } \right) \frac{ \mathbf X_{u} }{ |\mathbf X_{u}| } \right)  
+  \frac{\partial}{\partial v}  \left(  \beta \left( 1  -  \frac{1}{1 +  |\mathbf X_{v}|^{2} / b^{2} }     \right) \frac{ \mathbf X_{v} }{|\mathbf X_{v}|} \right) . \label{eq_eqn_dec_tension}   
\end{align}

Equation \eqref{eq_eqn_dec_tension} is discretized using a centered finite difference scheme. 
Let $\mathbf X^{j,k}$ denote an arbitrary point internal to the leaflets, and we refer to the connection between vertices in the discretized model as links or edges. 
The nonlinear system of equations associated with this point is given by 
\begin{align}
&0 =  p  \left(  \frac{(\mathbf X^{j+1,k} - \mathbf X^{j-1,k} )}{2\Delta u}  \times \frac{ ( \mathbf X^{j,k+1} - \mathbf X^{j,k-1} ) }{2\Delta v} \right)  \label{equilbrium_eqn_discrete}  \\ 
            &+  
             \frac{\alpha}{\Delta u} 
	     \left(  1 - 
	     \dfrac{1}{1 + \dfrac{ | \mathbf X^{j+1,k}  -  \mathbf X^{j,k} |^{2}}{  a^{2} (\Delta u)^{2} } }  \right)            
             \frac{ \mathbf X^{j+1,k}  -  \mathbf X^{j,k}  }{ | \mathbf X^{j+1,k}  -  \mathbf X^{j,k} |  }  \nonumber  \\  
            &  -    
             \frac{\alpha}{\Delta u} 
	     \left( 1 - 
	     \dfrac{1}{ 1 + \dfrac{ | \mathbf X^{j,k}  -  \mathbf X^{j-1,k} |^{2}}{  a^{2} (\Delta u)^{2}  } } \right) 
              \frac{ \mathbf X^{j,k}  -  \mathbf X^{j-1,k}  }{ | \mathbf X^{j,k}  -  \mathbf X^{j-1,k} |  }     \nonumber   \\   
             &  + 
             \frac{\beta}{\Delta v} 
	     \left( 1 -
	     \dfrac{1}{ 1 + \dfrac{ | \mathbf X^{j,k+1}  -  \mathbf X^{j,k} |^{2}}{  b^{2} (\Delta v)^{2}  } } \right) 
             \frac{ \mathbf X^{j,k+1}  -  \mathbf X^{j,k}  }{ | \mathbf X^{j,k+1}  -  \mathbf X^{j,k} |  } \nonumber  \\ 
             &  -               
              \frac{\beta}{\Delta v} 
	     \left( 1 -
	     \dfrac{1}{ 1 + \dfrac{ | \mathbf X^{j,k}  -  \mathbf X^{j,k-1} |^{2}}{  b^{2} (\Delta v)^{2}  } } \right)
             \frac{ \mathbf X^{j,k}  -  \mathbf X^{j,k-1}  }{ | \mathbf X^{j,k}  -  \mathbf X^{j,k-1} |  }    .  \nonumber 
\end{align}

The nonlinear system is solved using Newton's method with line search. 
To compute the Jacobian of equation \eqref{equilbrium_eqn_discrete}, each term was differentiated analytically, then the entire sparse matrix is constructed using the analytic forms at each step of the Newton's iteration.

\begin{figure}[th!]
\centering
\includegraphics[width=.7\columnwidth]{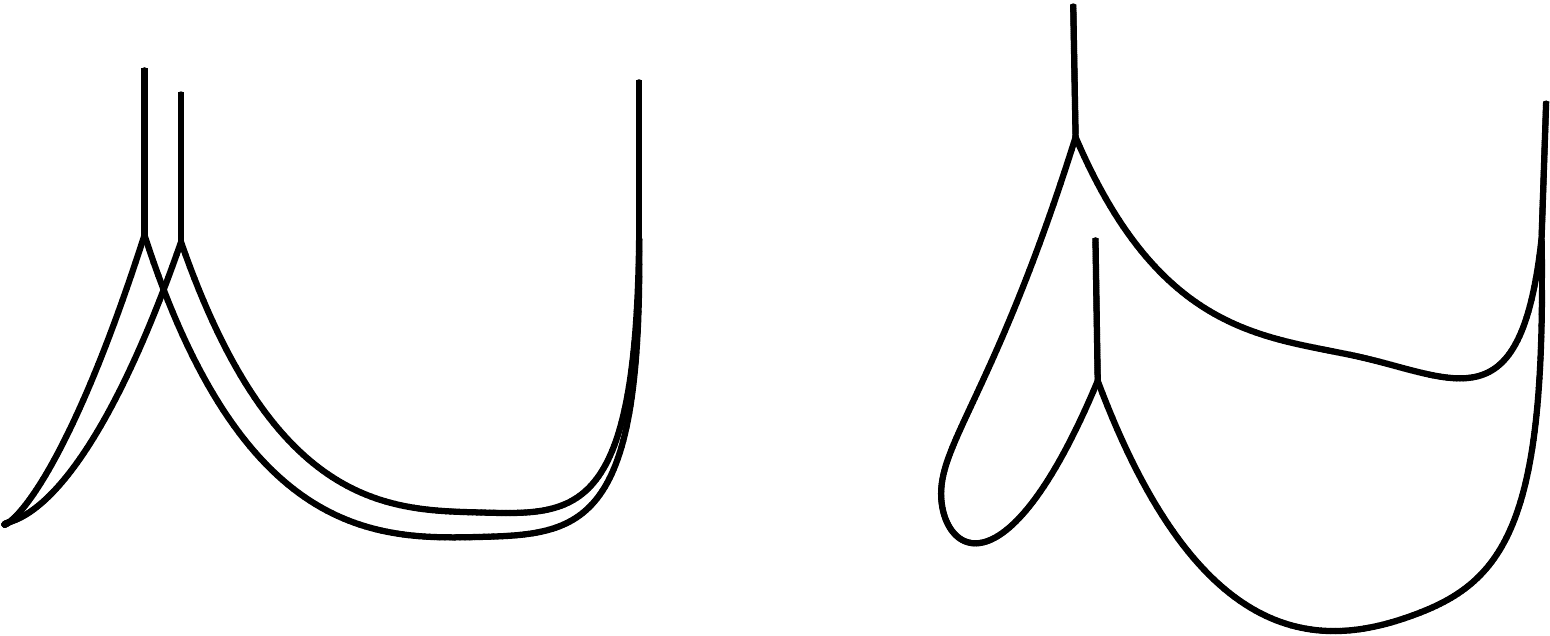} 
\caption{ Two views of the three-dimensional aortic annulus. 
This curve is prescribed as a Dirichlet boundary condition at the bottom of the leaflets and the commissures when solving the equations of equilibrium \eqref{equilbrium_eqn_discrete}.  
The position of the free edge emerges from the solution. 
There zero radial tension (homogeneous Neumann) boundary conditions, are prescribed. 
} 
\label{aortic_annulus}
\end{figure}

The equations are solved on the domain $\Omega = [0,1] \times [0,1/6]$. 
The curve $v = 0$ represents the annulus; its location is prescribed as a Dirichlet boundary condition. 
The radius $r = 1.25$ cm \cite{boraita2016reference}, and the annular height is $1.4 r$ \cite{swanson1974dimensions}. 
The curve $v = 1/6$ represents the free edge; it is treated with a homogeneous Neumann boundary condition, meaning that zero tension is applied in the radial direction, where the valve tissue ends. 
The curves $u = 0, 1/3, 2/3$ represent the commissures; their position is prescribed as a Dirichlet boundary condition. 
Circumferential fibers attach at the commissures from a height of $0.9$r to $1.4$r.   
The curve $u = 1$ is identified with the curve $u = 0$ by periodicity. 
The geometry of the aortic annulus is shown in Figure \ref{aortic_annulus}. 

We seek to match the loaded geometry provided in \cite{swanson1974dimensions}. 
The loaded leaflet height at the center of the leaflet is then targeted to be $1.4r = 1.75$ cm. 
They report the free edge length as $2.48r = 3.1$ cm, which is larger than the diameter of the annulus, accounting for the non-planar line of coaptation. 
We target the loaded free edge length to be $2.48r + 0.4 = 3.5$ cm. 
The factor of $0.4$ cm is added to create a small amount of extra free edge length that ensures firm coaptation.
Since we do not include a notion of contact here, there may be slightly less strain in FSI simulations locally at the free edge, as coaptation prevents the free edge from straining further and removes pressure locally. 
These values correspond to a reference height of 1.13 cm and a free edge length of 3.04 cm, following definitions of the reference configuration from the loaded configuration (See Section \ref{constitutive}).

The pressure is set to the constant value $p = 60$ mmHg, slightly below the nominal diastolic pressure difference of 80 mmHg across the fully-loaded valve. 
Each of the tunable parameters are selected by trial and error, to match achieve the desired gross morphology in the predicted loaded configuration. 
We select $\alpha, \beta$ and $b$ to be constants, and we tune $a$ to vary in a linear manner from the annulus to the free edge. 
The nonuniform value of $a$ was necessary to avoid bulging of the predicted loaded configuration near the annulus and maintain adequate length of the free edge simultaneously. 
Values are shown in Table \ref{coefficients_table}. 

\begin{table*}[ht]
\centering 
\begin{tabular}{  c  |  c |  c  |  c |  c | c |  }
Type	& Max tension   &  Max tension & Dec tension & Dec tension     \\ 
	&   name  &   dynes &  name  &  cm   \\
\hline 
Circumferential & $\alpha$ &   $2.68 \cdot 10^{7}$ & $a$  & 36.00-91.60 (annulus-free edge) \\ 
\hline
Radial & $\beta$ & $9.22 \cdot 10^{5}$ & $b$ &   24.00  \\ 
\hline
\end{tabular}
\caption{Values of coefficients in model. 
The value of $a$ varies linearly from the annulus to the free edge. 
}
\label{coefficients_table}
\end{table*}

Note that this phrasing does not include any notion of contact, and we allow the leaflets to interpenetrate on solving equation \eqref{equilbrium_eqn_discrete}. 
All three leaflets are solved for simultaneously. (It would be possible to solve for each individually, but previous work on the mitral valve solved for all leaflets simultaneously, and we kept this phrasing for simplicity and future flexibility \cite{kaiser2019modeling}.) 
We allow interpenetration because the overlapping creates a bit of extra length on the free edges that allows the leaflets to obtain good coaptation and seal, or more informally mash together, when running FSI simulations (in which the leaflets are not allowed to interpenetrate). 
Solving the equations is simpler without this additional force, and when simulated with fluid, the IB method prevents further interpenetration without specifying additional contact forces \cite{doi:10.1137/070699780}. 
Further, not allowing contact sets the material properties such that the leaflet \emph{could} bear a pressure load. 
In in vitro experiments, we occasionally see one leaflet begin to close and appear to support a pressure on its own and form a bowl-like shape immediately prior to any contact \cite{10.1093/ejcts/ezw317}, so rather than specifically account for different strains or loading conditions locally in the coaptation region, we load the entire leaflet uniformly.

\subsection{Setting the constitutive law}
\label{constitutive}

The solutions to equations \eqref{eq_eqn_dec_tension}, as shown in Section \ref{loaded_valve}, specify the loaded geometry of the valve, and the tensions required to support such a load. 
For each link in the discretized model, this then gives a single point on the tension/strain curve. 
In this section, we apply this information to set a reference configuration and constitutive law for the valve. 
We do this in such a way that using the new constitutive law, the equilibrium equations \eqref{eq_eqns} are still satisfied, but with the constitutive law defining tensions, rather than the formulas in equation \eqref{dec_tension}. 

First, we prescribe uniform strain to the loaded configuration. 
Let $E_{c}$ denote the circumferential strain, and $E_{r}$ denote the radial strain, $L$ the length of any link, and $R$ its associated rest length. 
Yap et al. found that the aortic valve achieves a nearly constant strain of 
\begin{align}
E_{c} = 0.15, \quad E_{r} = 0.54 . 
\end{align}
in the belly region when fully loaded \cite{yap2009dynamic}. 
We prescribe these values in each direction, then solve 
\begin{align}
E = \frac{L - R}{R}
\end{align}
for the rest length $R$ in each link in the model. 

Next, we assign each link a constitutive law. 
We base the shape (but not the local stiffness) of the constitutive law on the experimental results of \cite{may2009hyperelastic}.
Their strip biaxial tests on specimen P35 in their Figure 4 appear to be approximately exponential, and are taken to be representative curves. 
Thus, we assume that the tension/strain relationship is exponential through the origin for positive strains, and zero for compressive strains, or 
\begin{align}
\tau(E) = 
\begin{cases} 
\kappa (e^{ \lambda E} - 1)   & : E \geq 0  \label{exponential_law} \\ 
0                     & : E < 0 ,
\end{cases}
\end{align}
The exponential rate of the curves associated with data in the circumferential and radial directions were estimated using a nonlinear least squares fit using Matlab \cite{MATLAB} and take values $\lambda_{c} = 57.46$ and $\lambda_{r} = 22.40$ respectively.

The stiffnesses coefficients $\kappa$ scale the constitutive law. 
The value of $\kappa$ is set individually for each link in the model to achieve the value of tension in the solution of the equilibrium equations \eqref{equilbrium_eqn_discrete} at the prescribed strains $E_{c}$ and $E_{r}$. 
Let $S^{j + 1/2,k}$ denote the circumferential tension between $\mathbf X_{j,k}$ and $\mathbf X_{j+1,k}$. 
We solve 
\begin{align} 
S^{j + 1/2,k} = \kappa_{c}^{j + 1/2,k} (e^{ \lambda_{c} E_{c}} - 1) 
\end{align} 
for the circumferential membrane stiffness coefficients $\kappa_{c}^{j + 1/2,k}$. 
Similarly, if $T^{j,k+1/2}$ denotes the radial tension between $\mathbf X_{j,k}$ and $\mathbf X_{j,k+1}$, we solve 
\begin{align}
T^{j,k + 1/2} = \kappa_{r}^{j,k + 1/2} (e^{ \lambda_{r} E_{r}} - 1) 
\end{align}
for the radial membrane stiffness coefficient $\kappa_{r}^{j,k + 1/2}$ associated with the current link. 

Using this constitutive law, we solve an additional equilibrium problem where the pressure is set to zero, $p = 0$ mmHg. 
A Dirichlet boundary condition is prescribed at the free edge to ensure that the solution is fully open and without self-intersections. 
We then expand the leaflet membrane in the normal direction to the open configuration to obtain a model with anatomical thickness. 
Three total layers are placed $0.022$ cm apart to create a thickness of $0.044$ cm, as measured in \cite{sahasakul1988age}. 
This is to mitigate the ``grid aligned artifact'' that may appear in IB simulations when a discontinuous pressure is supported by an infinitely thin membrane  \cite{kaiser2019modeling,thesis}. 
Three layers are selected to approximately match the desired structural mesh width of 0.023 cm and because three layers produced the greatest mitigation of the artifact on a test problem with an analytical solution.
The stiffness in the circumferential and radial direction of each layer is set to one third of the membrane stiffness calculated above. 
Linear springs in the normal direction keep the layers adjacent.
The spring stiffness is tuned by trial and error to achieve minimal movement between layers in FSI simulations through three cardiac cycles without influencing time step restrictions.

This completes the constitutive law for the leaflets, and this configuration is then used as an initial condition for FSI simulations. 
This constitutive law is fiber-based, in that all forces are determined based on a network of linear and nonlinear springs, and the network is structured into curves that represent biological fiber bundles in the circumferential direction and material response in the radial direction. 
This is similar to a mass-spring model, except that the structure is assumed to be neutrally buoyant and mass is thus handled as fluid density by the IB method without additional mass at structure nodes. 
In a comparative study of the aortic valve, mass-spring models had similar deformations to those of hyperelastic finite element based discretization, and were approximately ten times faster \cite{hammer2011mass}.

\subsection{Summary of construction}

We summarize the major steps of the model construction as follows: 

\begin{enumerate}

\item Solve the equations of equilibrium \eqref{eq_eqn_dec_tension} with the tension law \eqref{dec_tension} to compute the predicted loaded configuration. This configuration includes the length of each link in the discretized model and the tension it exerts. 

\item Compute a reference configuration and constitutive law. Prescribe strains of $E_{c} = 0.15$, $E_{r} = 0.54$ in the circumferential and radial direction, respectively. Derive the reference length for each link in the model from the prescribed strain and length in the predicted loaded configuration. Scale the membrane stiffnesses for each link in the model to achieve the tension specified by the predicted loaded configuration at the prescribed strain. 

\item Solve the generic equations of equilibrium \eqref{eq_eqns} using the new constitutive law for tensions and zero pressure. This gives an open configuration (with some prestrain) without self-intersections. 

\item Thicken the model to anatomical thickness by placing two additional membrane layers in the normal direction to the current membrane. 
Assign one third of the original membrane stiffness to each layer. 
Tie the three layers together with linear springs. 

\item Use this constitutive law and configuration as initial conditions for FSI simulations.

\end{enumerate}

\subsection{Fluid-structure interaction}
\label{fsi}

The IB method is a framework for the modeling and simulation of FSI \cite{ib_acta_numerica}. 
The method uses two different frames of reference, a lab-based or Eulerian reference frame for the fluid, and a material or Lagrangian frame for the structure. 
Let $\mathbf x$ denote a physical location in the fluid domain, and $t$ denote time. 
Let the field $\mathbf u(\mathbf x, t)$ represent fluid velocity and $p(\mathbf x, t)$ represent pressure. 
These fields are defined with respect to the Eulerian frame, and accept arguments that are fixed with respect to spatial location. 
The field $\mathbf f(\mathbf x, t)$ denotes a body force exerted by the structure onto the fluid. 
This is one distinctive feature of the IB method: the fluid interacts with the structure through such a body force. 
The parameters $\rho$ and $\mu$ represent density and dynamic viscosity, respectively. 
Let $\mathbf s$ denote material points on the structure. 
(We previously used $u,v$ to denote material points on the leaflets, but change to $\mathbf s$ here to avoid confusion with fluid velocity.) 
Let $\mathbf X(\mathbf s, t)$ denote the position of the structure associated with material point $\mathbf s$ at time $t$. 
Let $\mathbf F(\mathbf s,t) d\mathbf s$ denote the force exerted by the structure onto the fluid associated with patch $d \mathbf s$. 
These two frames are coupled via convolutions with the Dirac delta function in a manner shown below.

The governing equations of the IB method are 
{\allowdisplaybreaks
\begin{align}
\rho \left( \frac{ \partial \mathbf u (\mathbf x, t)}{\partial t}  + \mathbf u (\mathbf x, t) \cdot \nabla \mathbf u (\mathbf x, t) \right) &= - \nabla p (\mathbf x, t) + \mu \Delta \mathbf u (\mathbf x, t) + \mathbf f (\mathbf x, t)  \label{momentum} \\
\nabla \cdot \mathbf u(\mathbf x, t)  &= 0   		\label{mass}		\\
\mathbf F( \, \cdot \, , t) &=   \mathcal F (\mathbf X( \, \cdot \, ,t))  \label{nonlinear_force}  \\   
\frac{ \partial \mathbf X(\mathbf s,t)}{\partial t}&=  \mathbf u(\mathbf X(\mathbf s,t), t) 	 \label{interpolate} 	 \\
		&= \int \mathbf u(\mathbf x, t)   \delta (  \mathbf x  - \mathbf X(\mathbf s,t) )  \;  d  \mathbf x   \nonumber  \\ 
\mathbf f(\mathbf x, t)   &= \int  \mathbf F(\mathbf s,t)  \delta(  \mathbf x - \mathbf X(\mathbf s,t)  )   \; d\mathbf s   .       \label{spreading} 
\end{align}  
}
Equations \eqref{momentum}, \eqref{mass} are the Navier Stokes equations describing the dynamics of a viscous, incompressible fluid.  
Equation \eqref{momentum} represents momentum conservation, and equation \eqref{mass} represents volume conservation or incompressibility. 
Equation \eqref{nonlinear_force} represents a mapping from the configuration of the structure to the force that the structure exerts on the fluid. 
This includes the nonlinear constitutive law of the leaflets, and any other forces prescribed, including those that keep the leaflet mounted in place on its edges. 
The omitted argument indicates that the mapping $\mathcal F$ takes the entire configuration of the structure $\mathbf X$ as an argument, and produces the entire function $\mathbf F$ as output. 
Equations \eqref{interpolate} and \eqref{spreading} are interaction equations that couple the two frames. 
Equation \eqref{interpolate} is the equation of velocity interpolation, and says that structure moves with velocity equal to the local fluid velocity. 
Equation \eqref{spreading} is the equation of force spreading that computes the Eulerian frame force from the Lagrangian frame force. 
When discretized, the delta function in equations \eqref{interpolate} and \eqref{spreading} is replaced with a regularized delta function; we use the 5-point delta function derived in \cite{IB5_arxiv}. 
The equations were solved with the software library IBAMR using a staggered grid discretization \cite{IBAMR,griffith2010parallel}. 
Simulations were run on Stanford University's Sherlock cluster, with 48 Intel Xeon Gold 5118 cores with a 2.30GHz clock speed.

\begin{figure}[ht!]
\centering
\includegraphics[width=0.4\columnwidth]{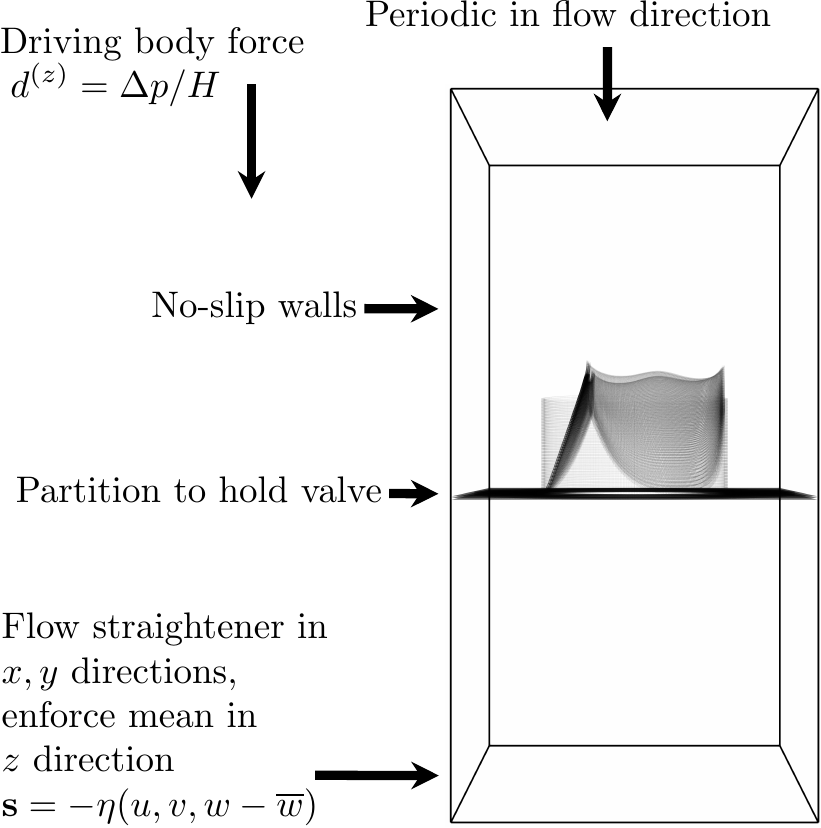} 
\caption{Simulation setup.
The model valve is mounted in a cylindrical scaffold, and attached to a flat partition. 
Sides are treated with no-slip boundary conditions. 
The domain is periodic in the flow direction. 
Ventricular pressure is prescribed, aortic pressure is determined by a lumped parameter network, and the pressure difference is prescribed as a body force. 
An additional force that acts as a flow straightener in the $x,y$ directions and approximately enforces the mean in the $z$ direction is applied on a thin slab at the bottom of the domain. 
 } 
\label{schematic}
\end{figure}

The fluid domain is a taken to be a box of dimensions $[-L,L] \times [-L, L] \times [-2L + 1, 2L + 1]$ where $L = 2.25$ cm. 
A schematic of the flow chamber is shown in Figure \ref{schematic}. 
Test chambers with similar geometry are seen in valve testing devices \cite{ViVitro} and standards \cite{ISO}, and have been shown to be sufficient for valve testing in in vitro experiments \cite{sathananthan2018overexpansion,sathananthan2019implications,tefft2019cardiac,peruzzo2019vitro,clark1978durability}. 
The domain is divided into $96 \times 96 \times 192$ points for a spatial resolution $\Delta x = 0.046875$ cm in the fluid. 
The time step is taken to be $\Delta t = 5 \cdot 10^{-6}$ s, which was the largest value that we found to be stable. 
The density is $\rho = 1$ g/cm$^{3}$, and the viscosity is $\mu = 0.04$ Poise. 
No slip walls are placed on the sides of the domain. 
The domain is taken to be periodic in the flow direction, and to hide periodic effects, we add a mathematical flow straightener to the bottom $0.2$ cm of the domain. 
This is a body force of the form $\mathbf d = -\eta (u, v, w - \bar w)$, where $\bar w$ denotes the mean of the z-component of velocity. 
This creates a friction-like force that approximately enforces zero flow in the $x$- and $y$-directions and enforces the mean in the $z$-direction. 
Overlap and interpenetration of the leaflets is prevented by the IB method without any specifically added contact forces \cite{doi:10.1137/070699780}. 

A pressure difference across the valve drives the simulations. 
This is prescribed as a uniform body force in the $z$ direction of the form $\mathbf d = (0,0,\Delta p/H)$, where $H$ is the height of the domain and $\Delta p = p_{ao} - p_{lv}$ is the difference between the aortic and left ventricular pressures.   
This is equivalent to prescribing the pressure difference directly at the top and bottom of the domain via a change of variables (See Appendix \ref{appendix_pressure}.) 
Using a periodic domain avoids the ``open-boundary instability'' associated with flow reversal at boundaries \cite{CNM:CNM2918}.
Approaches to correct this instability on a non-periodic domain are not always reliable when the sign of flow oscillates rapidly during valve closure, whereas with a periodic domain, this effect is removed entirely. 
The value of the aortic pressure is determined by the dynamics of a Windkessel or rcr (resistor capacitor resistor) lumped parameter network. 
The aortic pressure is governed by the ordinary differential equations 
\begin{align}
\frac{1}{R_{p}} \left(P_{ao} - P_{wk} \right) &=  Q_{ao} \\ 
C \frac{dP_{wk}}{dt} + \frac{1}{R_{d}} P_{wk} &= Q_{ao}, \label{rcr_ode}
\end{align}
where $P_{ao}$ is the prescribed aortic pressure, $P_{wk}$ is the pressure in the Windkessel, $Q_{ao}$ is flow through the aortic valve, $R_{p}$ is proximal resistance, $R_{d}$ is the distal resistance and $C$ is the capacitance \cite{kim2009coupling,Griffith_aortic}. 
The flow target, $Q_{mean}$ is set to 6.1 L/min, slightly higher than the nominal mean flow of 5.6 L/min, to account for some extra resistance from the flow straightener. 
We target a pressure of 120 mmHg in systole for approximately 40\% of the cardiac cycle and 80 mmHg in diastole for approximately 60\% of the cardiac cycle.  
The mean pressure is set to be be a weighted average of the systolic and diastolic pressures, giving $P_{mean} = 96$ mmHg. 
The total resistance is then computed as $R_{total} = P_{mean} / Q_{mean}$. 
The ratio of the proximal to distal resistors is taken to be $0.064$ \cite{laskey1990estimation}. 
This gives proximal resistance $R_{p} = 76.81$ s dynes cm$^{-5}$ and distal resistance $R_{d} = 1182.10$ s dynes cm$^{-5}$. 
To tune the capacitance, we select an early diastolic pressure of $P_{1} = 100$ mmHg, since the pressure drops from systolic pressure in closure, an end diastolic pressure of $P_{2} = 80$ mmHg, and the predicted time as $t$ as the time in diastole. 
Assuming perfect closure during diastole so $Q_{ao} = 0$, equation \eqref{rcr_ode} has an exact exponential solution. 
We substitute these values into the exact solution to obtain 
\begin{align}
C = \frac{-t}{ R_{d} \log(P_{1}/P_{2})} = 0.0018  \text{ cm}^{5} \text{dynes}^{-1}. \label{capacitance_formula}
\end{align}
The ventricular pressure is prescribed following the experimental measurements shown in \cite{yellin_book}. 
The heart rate is set to a nominal value of 75 beats per minute, or 0.8 s.
All simulations are run for three cardiac cycles, which is sufficient to test multiple cycles, then stopped due to high overall computational time. 
Convergence and periodicity are discussed in Appendix \ref{convergence}.

The structure mesh is targeted to approximately twice as fine as the fluid resolution, or $0.023$ cm.
The annulus has three-dimensional length $11.28$ cm. 
Since $N$ is required to be a power of two on each leaflet, this corresponds to $N$ = 384 points around the annulus. 
Note that the precise length of links in the leaflets at any time are determined as solutions to equations \eqref{momentum}-\eqref{spreading} and change according to the dynamics of system.
A flat partition is added to the plane $z = 0$ outside of the annulus. 
The leaflets are mounted to a cylindrical scaffold of height $\pi r/ 3 = 1.31$ cm that serves to cover holes between the partition and the annulus. 
The cylinder and partition are approximately rigid, and held to a fixed position using \emph{target points}. 
For a point $\mathbf X$ and its desired position $\mathbf X_{target}$, this is a force $\mathbf f = -k (\mathbf X - \mathbf X_{target})$, representing a linear spring of zero rest length.

\section{Results}
\label{Results}

\subsection{The loaded model valve}
\label{loaded_valve}

\begin{figure}[t!]
\centering
\setlength{\tabcolsep}{2pt}
\begin{tabular}[t]{c c c}
\includegraphics[width=.34\columnwidth]{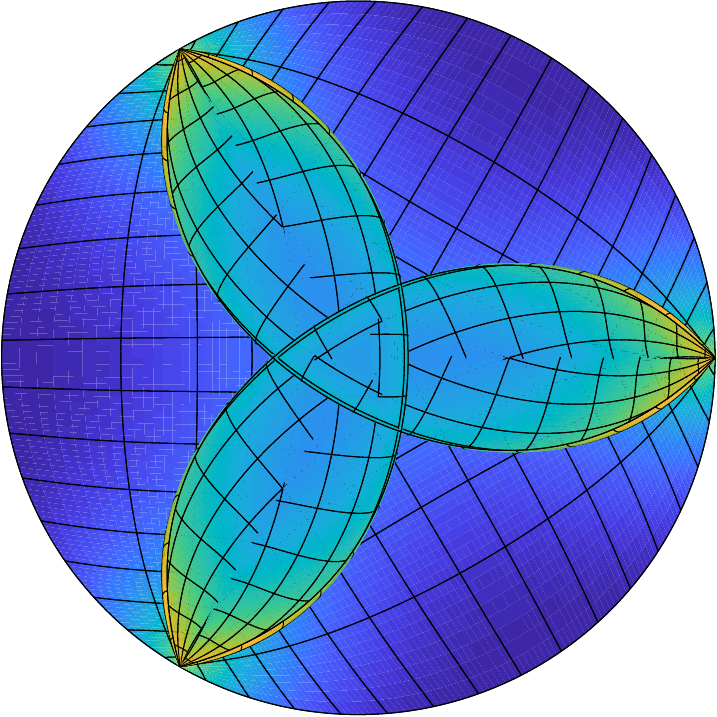} & 
\includegraphics[width=.37\columnwidth]{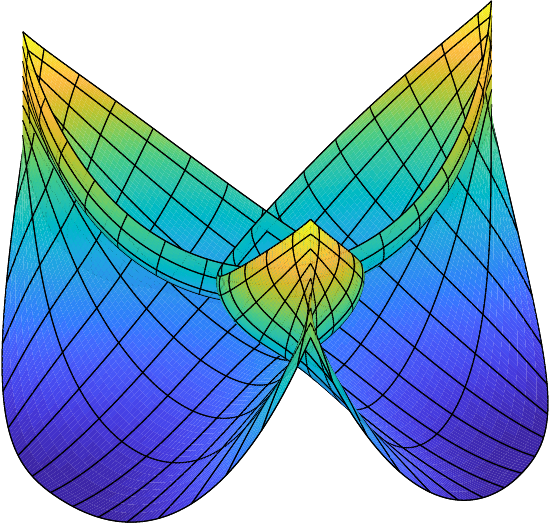} & 
\includegraphics[width=.25\columnwidth]{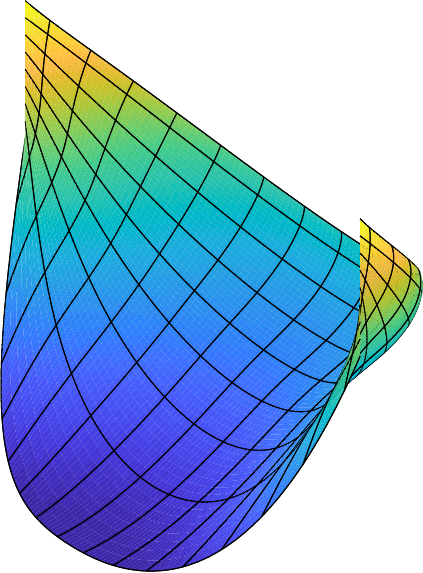}
\end{tabular}
\caption{Predicted loaded configuration of the aortic valve. 
From left to right, the valve is viewed from above, at an angle and showing one leaflet only. 
Every eighth contour in the fiber and cross-fiber directions in the computational mesh is shown in black for visual clarity. 
At this model-construction stage, we allow the leaflets to interpenetrate, as depicted on the left and center panel, so each leaflet bears pressure on its entirety. 
This configuration is used to generate the reference configuration and constitutive law for the valve. 
An open configuration without interpenetration will be prescribed as an initial condition to the FSI simulations. 
} 
\label{aortic_closed}
\end{figure}

\begin{figure*}[t!]
\centering
\setlength{\tabcolsep}{5pt}
\begin{tabular}[t]{c c c c}
circumferential tension & 
dynes & 
radial tension & 
dynes  \\ 
& 
$\cdot 10^{3}$ & 
&
$\cdot 10^{3}$ \\
\includegraphics[width=.4\textwidth]{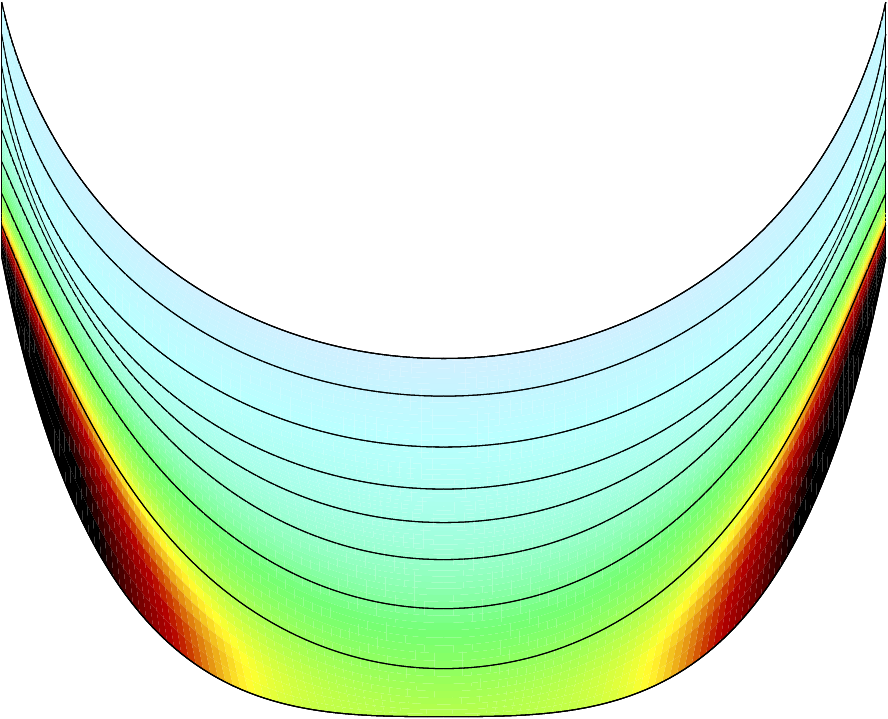} & 
\raisebox{40pt}{\includegraphics[height=120pt]{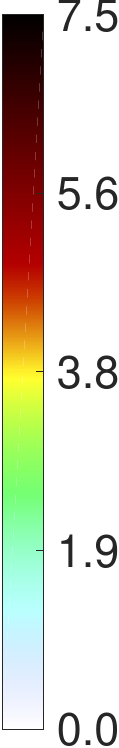} } &
\includegraphics[width=.4\textwidth]{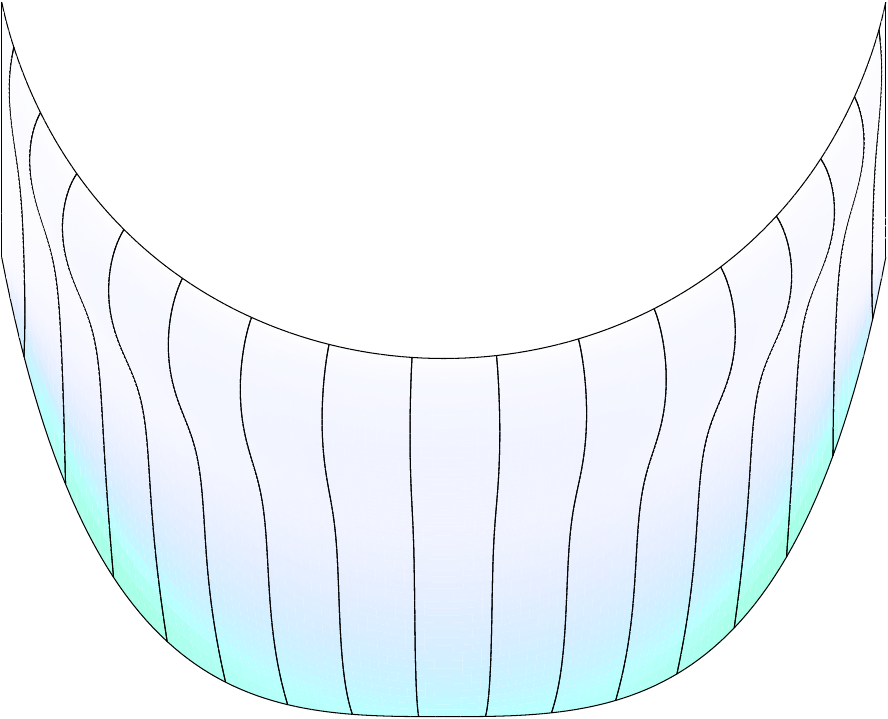} &
\raisebox{40pt}{\includegraphics[height=120pt]{colorbar_only_aortic_tension.pdf} } 
\end{tabular}
\caption{
Emergent tension in one aortic valve leaflet in the predicted loaded configuration showing the circumferential, fiber direction (left) and radial, cross-fiber direction (right). 
The tension is much larger in the circumferential direction than the radial. 
} 
\label{aortic_valve_tension}
\end{figure*}

The predicted loaded configuration of the valve is shown in Figure \ref{aortic_closed}. 
This configuration arises as the solution to the static equations of equilibrium \eqref{equilbrium_eqn_discrete}, and predicts the geometry and tension required to support the prescribed pressure load. 
Since we allow the leaflets to interpenetrate and do not include contact forces at this stage, the leaflets cross near the free edge.
The right panel shows only one leaflet to clarify the other two views. 
The emergent tension in the leaflets is shown in Figure \ref{aortic_valve_tension}. 
The values of tension that emerge are heterogeneous and much lower in the radial direction than the circumferential direction. 
The circumferential stiffness coefficients $\kappa_{c}$ vary from 0.15 to 2.35 dynes, with a mean of 0.37 dynes. 
These values correspond to minimum, maximum and mean tensions of $8.15 \cdot 10^{2}$, $1.30 \cdot 10^{4}$ and $2.02 \cdot 10^{3}$ dynes respectively. 
The radial stiffness coefficients $\kappa_{r}$ vary from $7.48 \cdot 10^{-5}$ to $8.71 \cdot 10^{-3}$ dynes, with a mean of $1.57\cdot 10^{-3}$ dynes, corresponding to minimum, maximum and mean tensions of $1.34 \cdot 10^{1}$, $1.56 \cdot 10^{3}$ and $2.81 \cdot 10^{2}$ dynes respectively. 
More meaningful than the local tensions is the predicted tangent modulus that results (See Section \ref{constitutive_results}.)

\begin{figure*}[t!]
\centering
\setlength{\tabcolsep}{0pt}
\begin{tabular}[t]{c c c c}
circumferential tangent modulus & 
$ \cdot 10^{8} \dfrac{\text{dynes}}{\text{cm}^{2}} $ & 
radial tangent modulus & 
$ \cdot 10^{8} \dfrac{\text{dynes}}{\text{cm}^{2}} $  \\ 
\includegraphics[width=.39\textwidth]{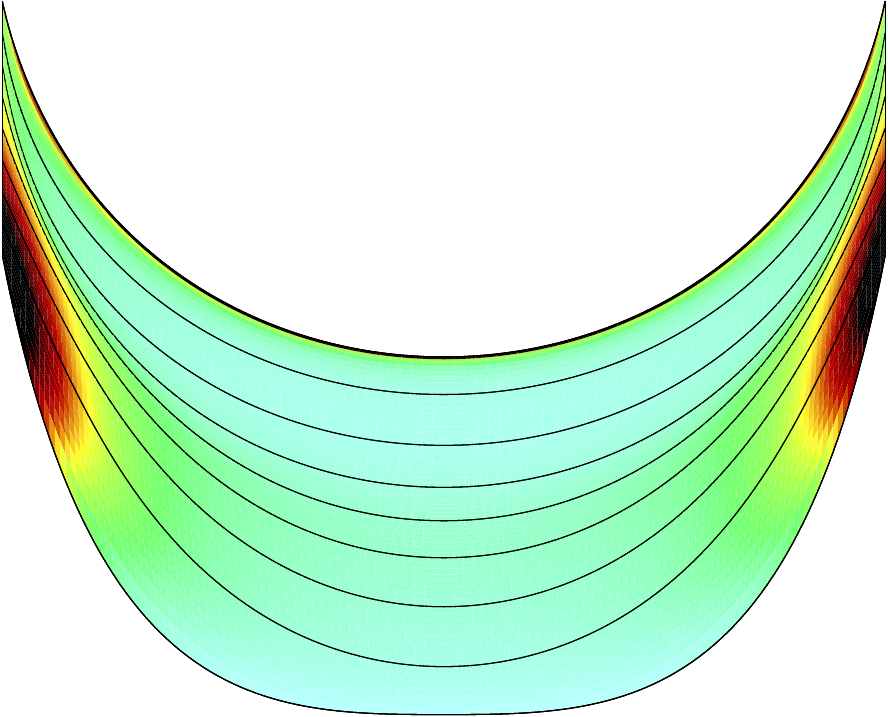} & 
\raisebox{30pt}{\includegraphics[height=120pt]{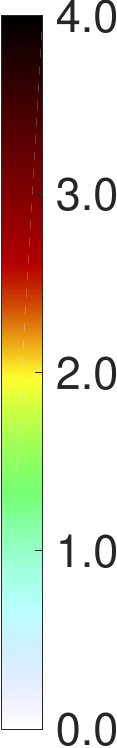} } &
\includegraphics[width=.39\textwidth]{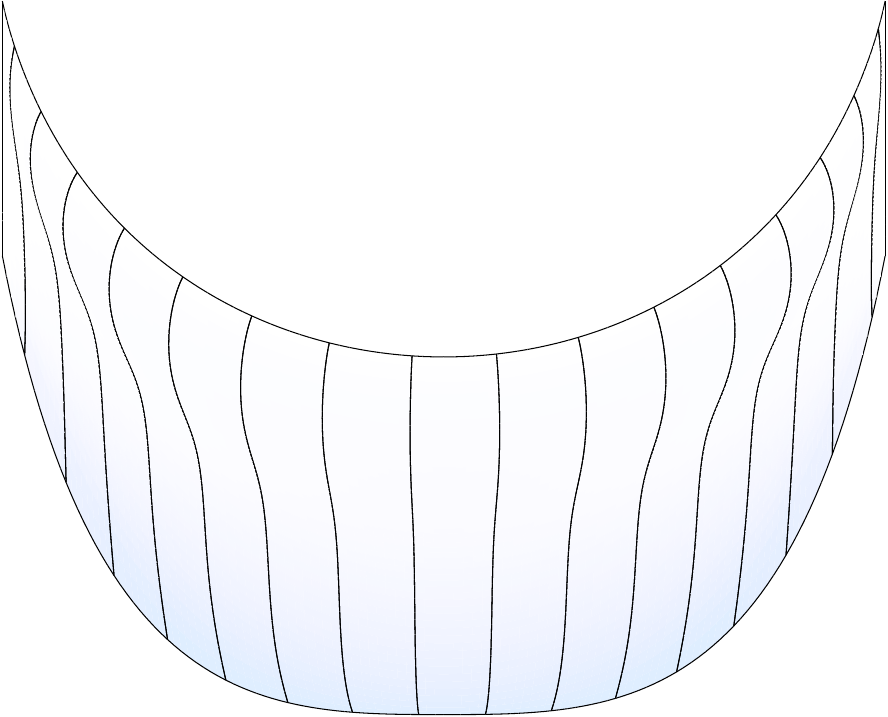} &
\raisebox{30pt}{\includegraphics[height=120pt]{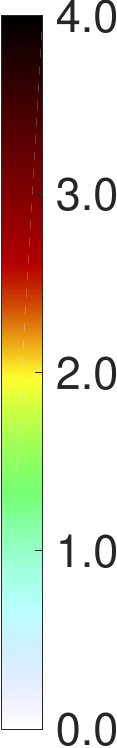} } \\
ratio circumferential/radial tangent modulus & 
& 
radial tangent modulus (rescaled) & 
$ \cdot 10^{7} \dfrac{\text{dynes}}{\text{cm}^{2}} $  \\ 
\includegraphics[width=.39\textwidth]{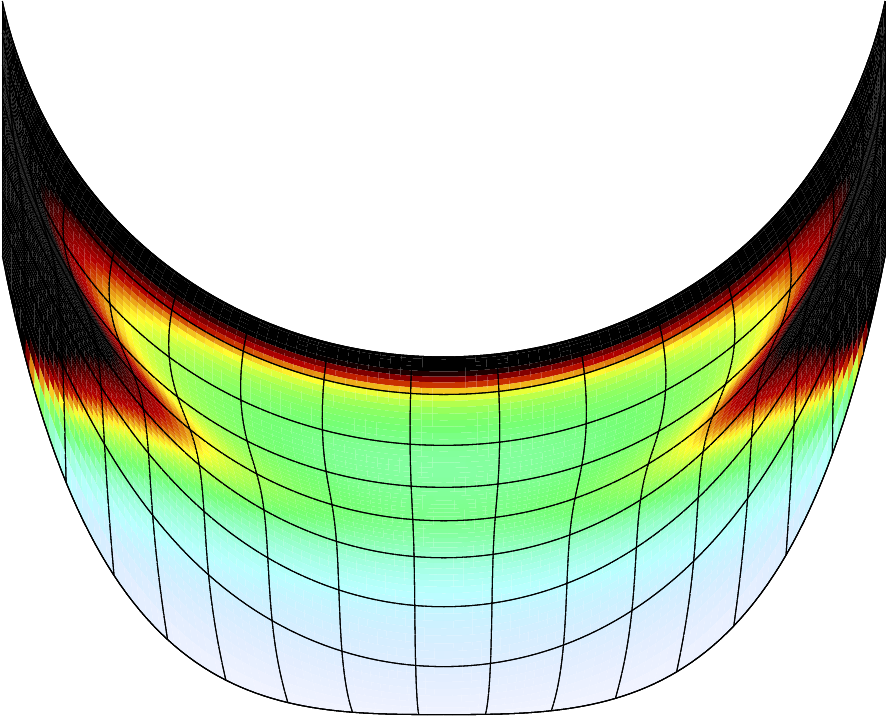} & 
\raisebox{30pt}{\includegraphics[height=120pt]{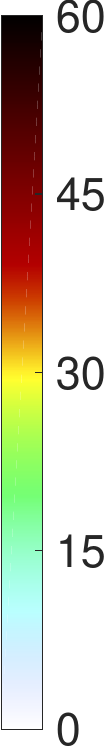} } &
\includegraphics[width=.39\textwidth]{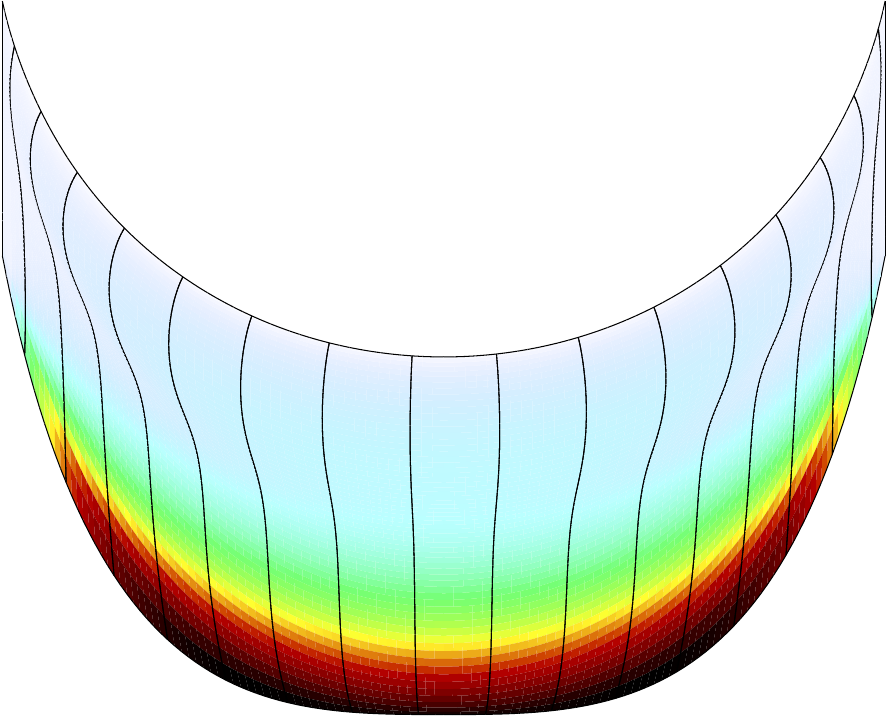} &
\raisebox{30pt}{\includegraphics[height=120pt]{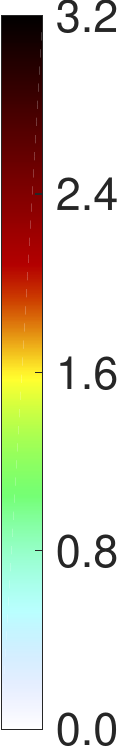} } 
\end{tabular}
\caption{
Emergent tangent modulus and ratio of tangent moduli in one aortic valve leaflet. 
The top row shows the emergent tangent modulus in the circumferential direction (left) and radial direction (right). 
The bottom row shows the ratio of circumferential over the radial tangent moduli (left) and the radial tangent modulus, but plotted on a smaller scale (right). 
} 
\label{aortic_valve_tangent_mod}
\end{figure*}

\begin{figure*}[t!]
\setlength{\tabcolsep}{0.5pt}
\centering
\begin{tabular}{ccccccccc}
\raisebox{110pt}{\includegraphics[width=.04\textwidth]{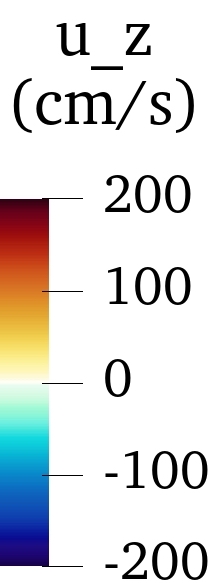}} 
&
\includegraphics[width=.117\textwidth]{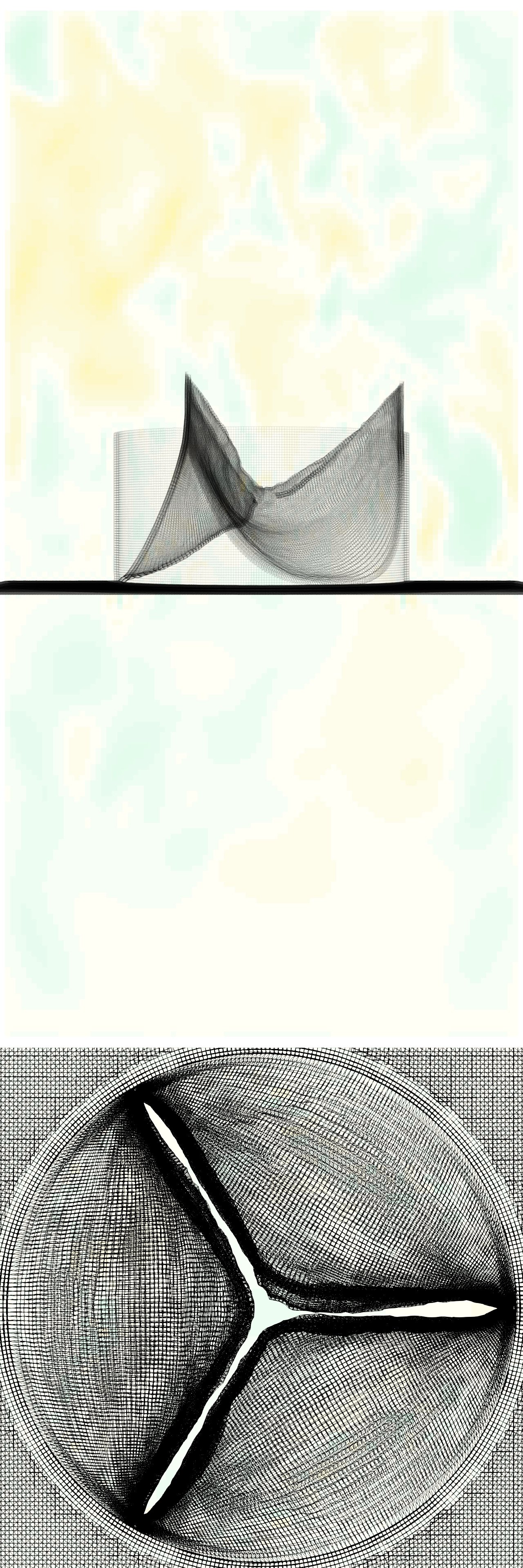} & 
\includegraphics[width=.117\textwidth]{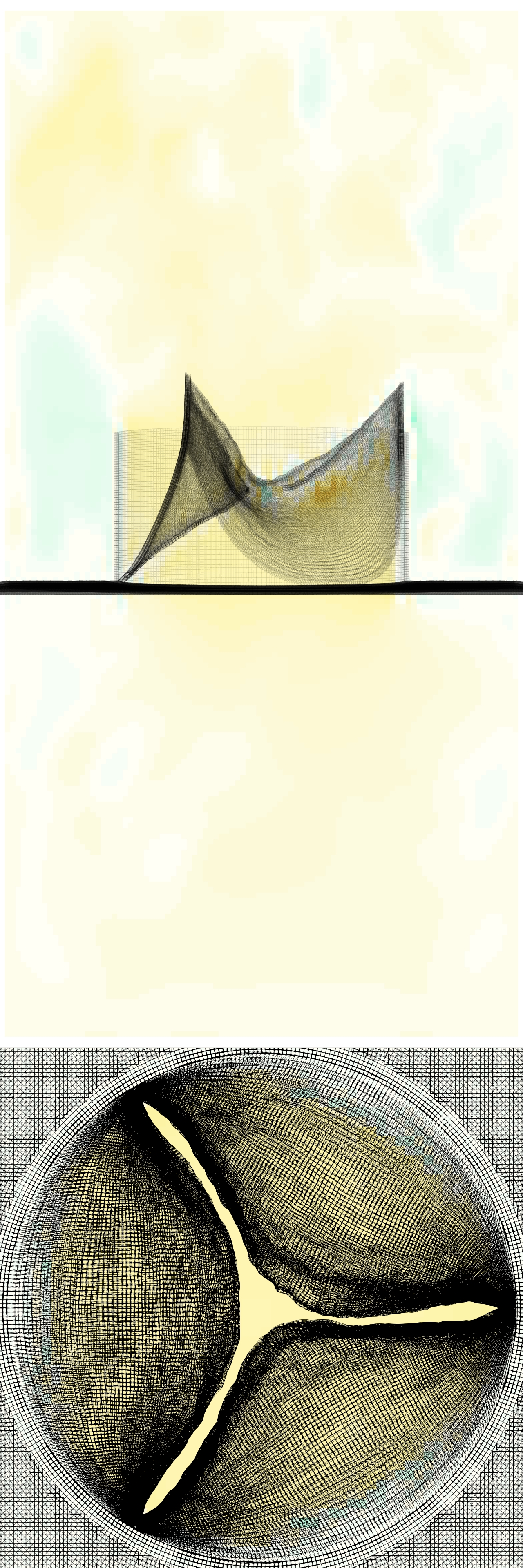} & 
\includegraphics[width=.117\textwidth]{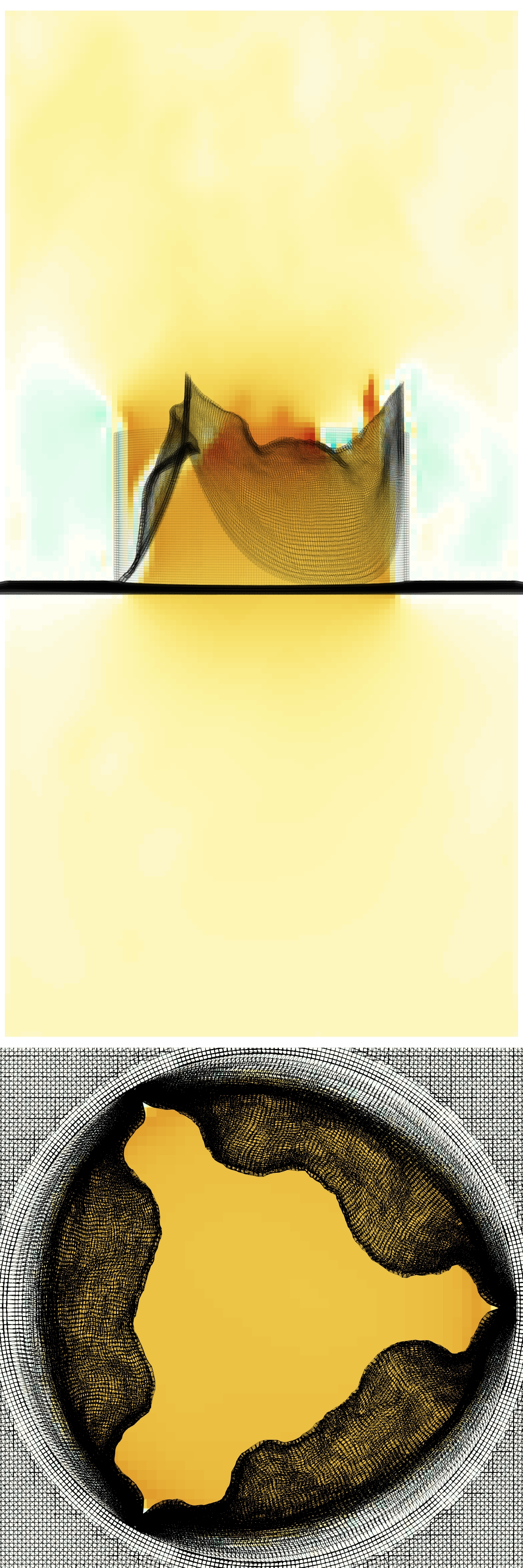} &
\includegraphics[width=.117\textwidth]{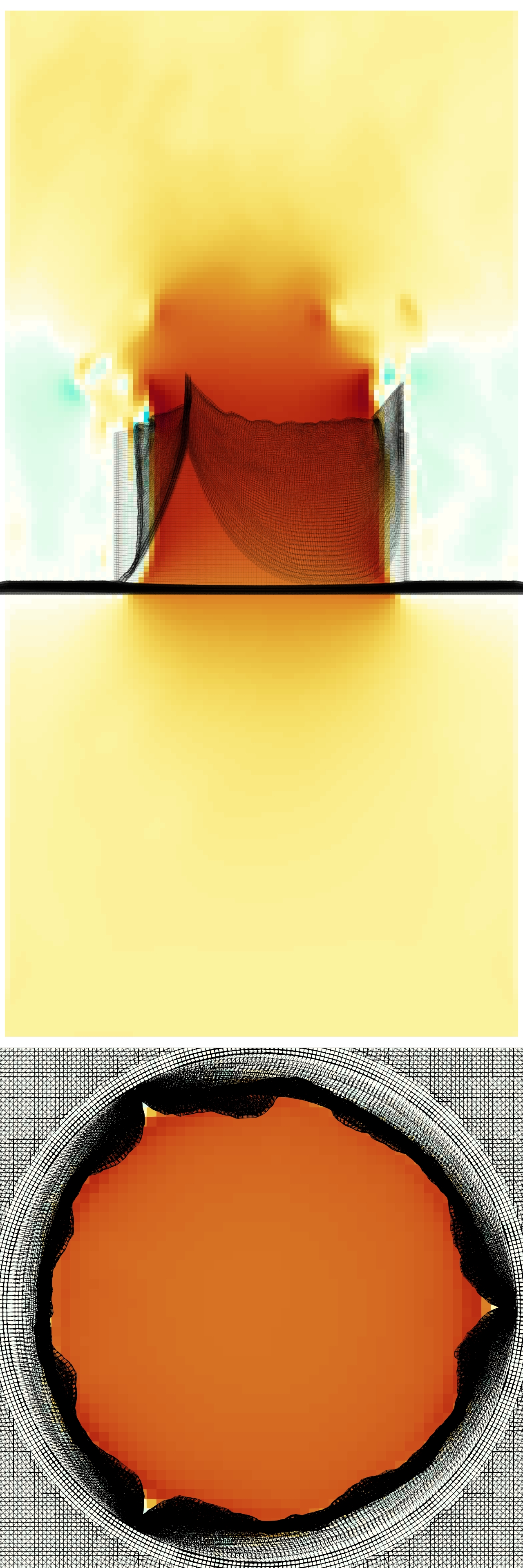} &
\includegraphics[width=.117\textwidth]{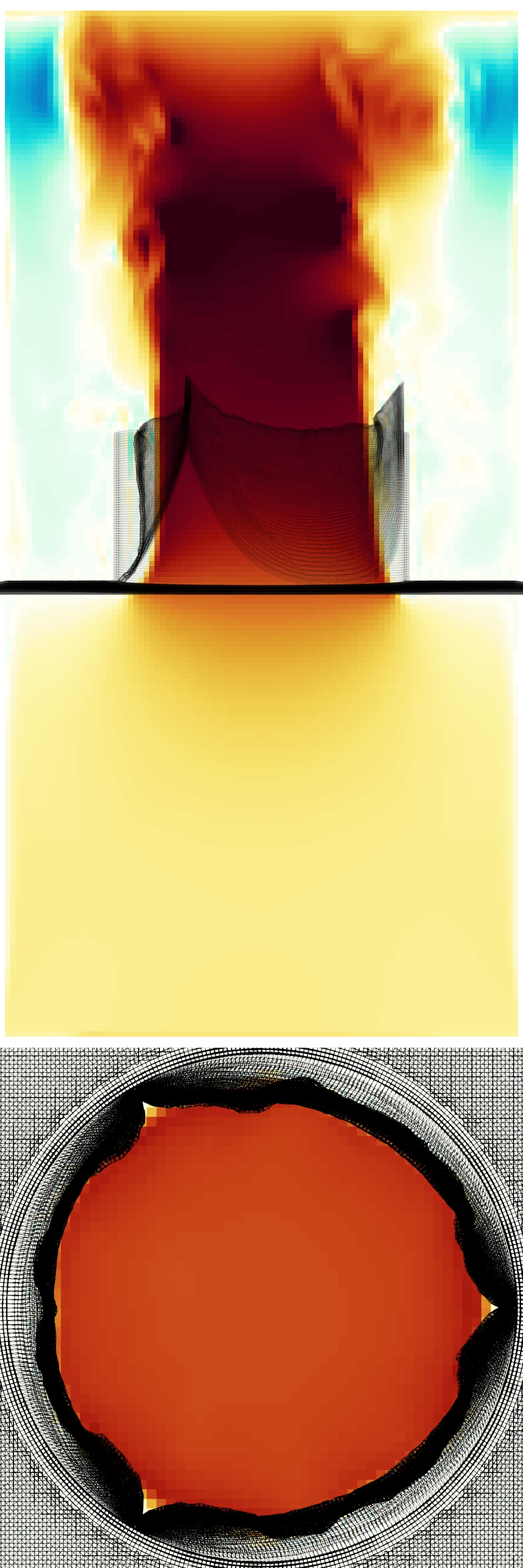} &
\includegraphics[width=.117\textwidth]{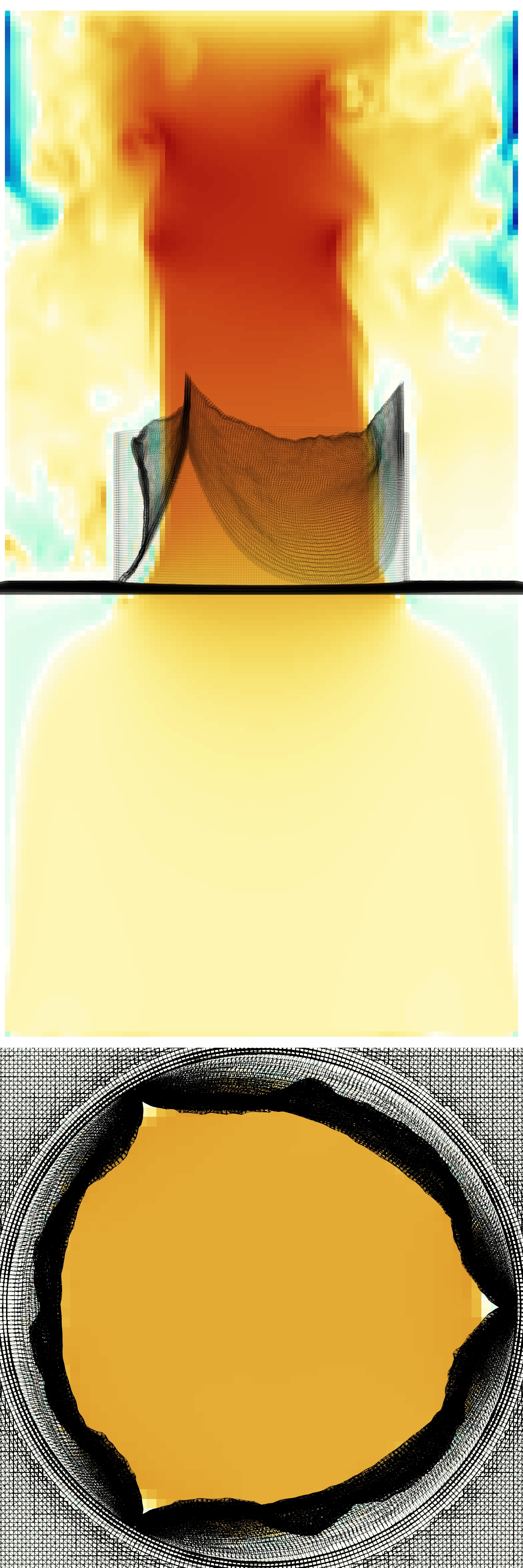} &
\includegraphics[width=.117\textwidth]{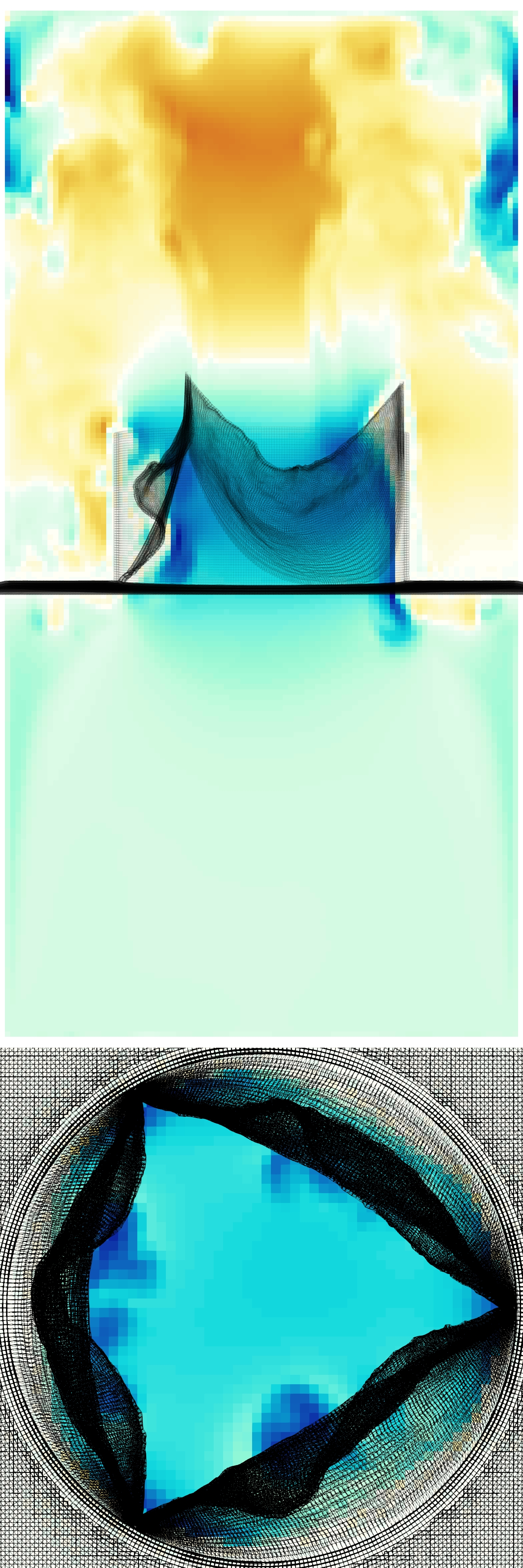} &
\includegraphics[width=.117\textwidth]{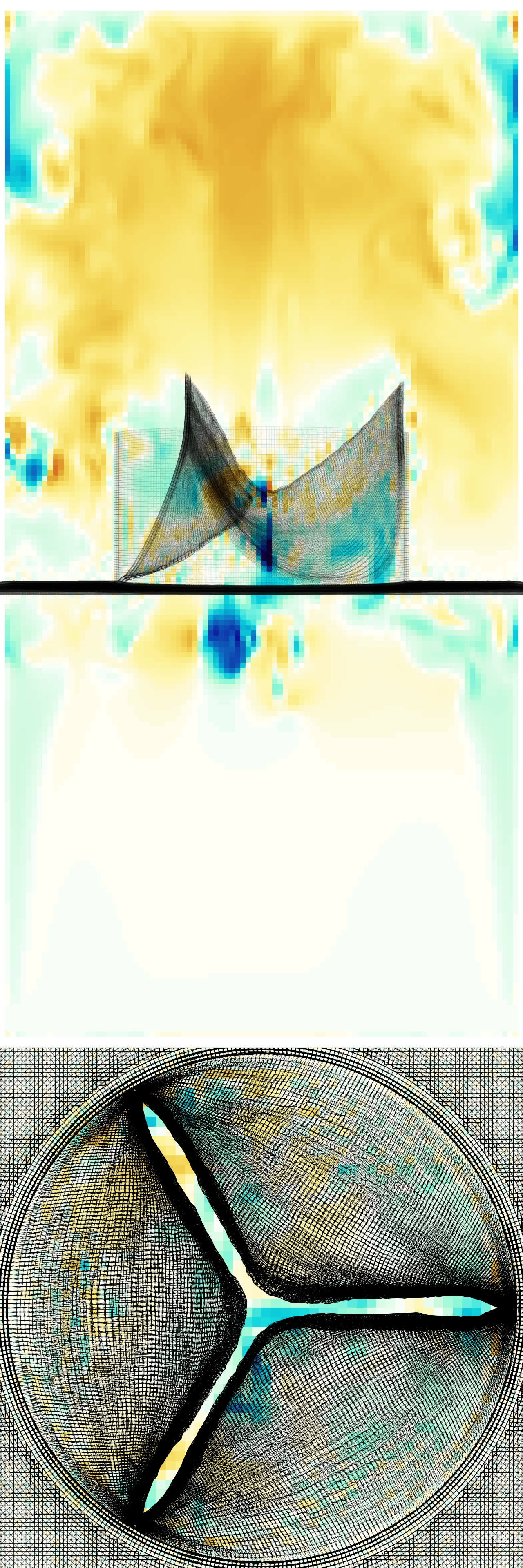}   
\end{tabular}
\caption{Slice view of the $z$ component of velocity through the cardiac cycle.
From left to right, the frames show the fully closed valve, the initial unloading prior to opening, the valve in middle of opening, the just open valve, the fully open valve at peak forward flow, the open valve as forward flow begins to slow, the valve initiating closure, and the just-closed, still-vibrating valve. 
} 
\label{basic_pressures_through_cycle}
\end{figure*}

\subsection{The constitutive law}
\label{constitutive_results}

From the steps described in Section \ref{constitutive}, a constitutive law was constructed. 
We estimate the fully-loaded tangent modulus that emerges from the process as follows. 
Equation \eqref{exponential_law} (including the local stiffness coefficient $\kappa$) is differentiated with respect to $E$ and evaluated at $E_{c}$ or $E_{r}$, depending on the direction of the link. 
This gives the tangent stiffness in units of force, which is divided by an area element to convert to the tangent modulus. 
The length of the area element is computed as half the distance to each adjacent point in the opposing direction to the link, and the thickness is $0.044$ cm \cite{sahasakul1988age}.
The mean circumferential tangent modulus is $1.43 \cdot 10^{8}$ dynes/cm$^{2}$, the mean radial tangent modulus is $ 5.75 \cdot 10^{6}$ dynes/cm$^{2}$, and the ratio of the means is approximately 25.

The emergent tangent moduli of this constitutive law at the prescribed strains of $E_{r}$ and $E_{c}$, as well as the local ratio of the moduli, are shown in Figure \ref{aortic_valve_tangent_mod}. 
Both the radial and circumferential directions have a heterogeneous modulus. 
The circumferential, fiber direction in the central, belly region of the leaflet has a local tangent modulus of order $10^{8}$ dynes/cm$^{2}$. 
Near the commissures and the annulus, the circumferential tangent modulus has a locally higher modulus. 
One experimental study found vast variety in tissue, including tendon like structures near the commissures, which suggests that there may be heterogeneous thickening or higher elastic modulus there \cite{billiar2000biaxial}. 
The radial direction, is an order of magnitude less stiff overall. 
When viewed on the same scale as the circumferential tension, the tangent modulus appears low overall, and differences in the load radial tangent modulus are barely visible. 
Viewing on a smaller scale reveals that the radial direction is most stiff near the annulus and much less stiff near the free edge.
This is consistent with experiments that show less stiffness in the radial direction near the free edge \cite{billiar2000biaxial}. 
Near the annulus, there are many circumferential fibers in the direction of the free edge to support tension in the radial direction via curvature. 
Near the free edge, the radial-direction contours end, their tension must be balanced by curvature of fewer circumferential fibers, and a much lower local tangent modulus results.

\subsection{The model valve in fluid}
\label{results_fsi}

Using the simulation setup described in Section \ref{fsi}, we simulate the model valve with the IB method. 
Figure \ref{basic_pressures_through_cycle} shows the configuration of the valve and slice views of the velocity field through the cardiac cycle in the third beat. 
The top panels show the valve from the side, and a slice view of the $z$-component of velocity in the $x = 0$ plane. 
The bottom panels at each time point show the valve from above, and a slice view of the $z$-component of velocity at the $z = 0$ plane. 
We discuss the panels from left to right. 
The first frame shows the fully closed valve. 
Its configuration has been nearly static for the previous approximately 0.4s. 
The flow has relaxed from previous cycles and is relatively still. 
Since the IB method regularizes the forces due to the structure, from the perspective of the numerical method, this configuration is closed despite small visible gaps. 
Then, the valve begins to unload. 
The free edges are still close together, but the belly of the leaflets has started to rise, and at the center a small opening has begun to form. 
Next, the valve in the middle of opening shows transient ripples on the free edge. 
Forward flow has begun, but mostly fluid that is moving with the leaflets, rather than flow moving through the open configuration. 
Then, the valve has fully opened, achieving near-maximum open area, and the forward jet is starting to develop. 
At peak systolic flow, the valve is fully open and a strong jet flows forward. 
Next, the jet continues but at a lower magnitude, while the valve remains nearly as open and forward flow has begun to slow. 
In the penultimate frame, closure begins.  
The jet breaks off and the leaflets move towards the center of the orifice.  
A slight, transient, local flow reversal appears between the leaflets. 
This is not true leakage, but rather represents fluid that had yet to ``clear'' the leaflets during the closure. 
Finally, the valve has just closed and is still vibrating.
A slight ``puff'' of reverse flow is all that remains of the closing transient below the valve.

\begin{figure}[t!]
\centering
\includegraphics[width=0.5\columnwidth]{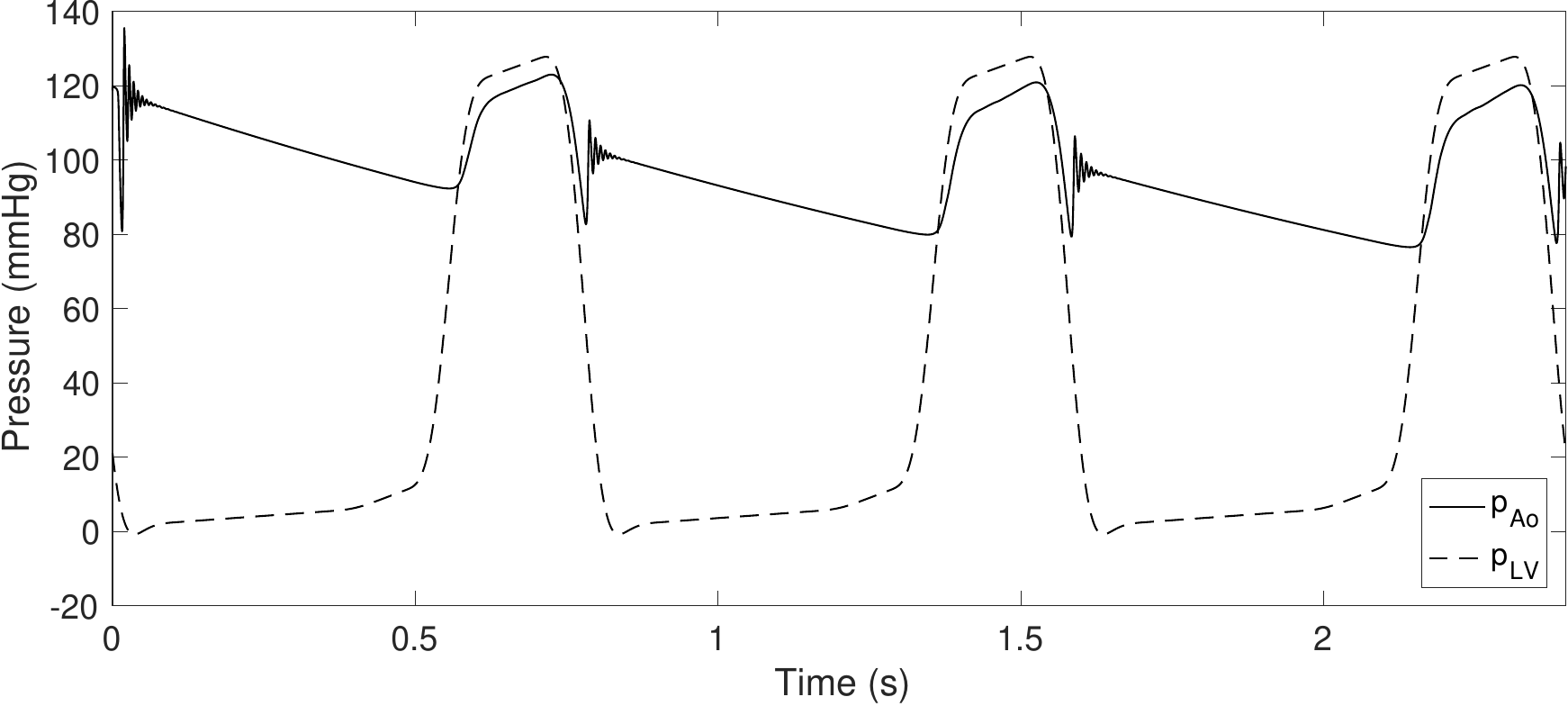} 
\includegraphics[width=0.5\columnwidth]{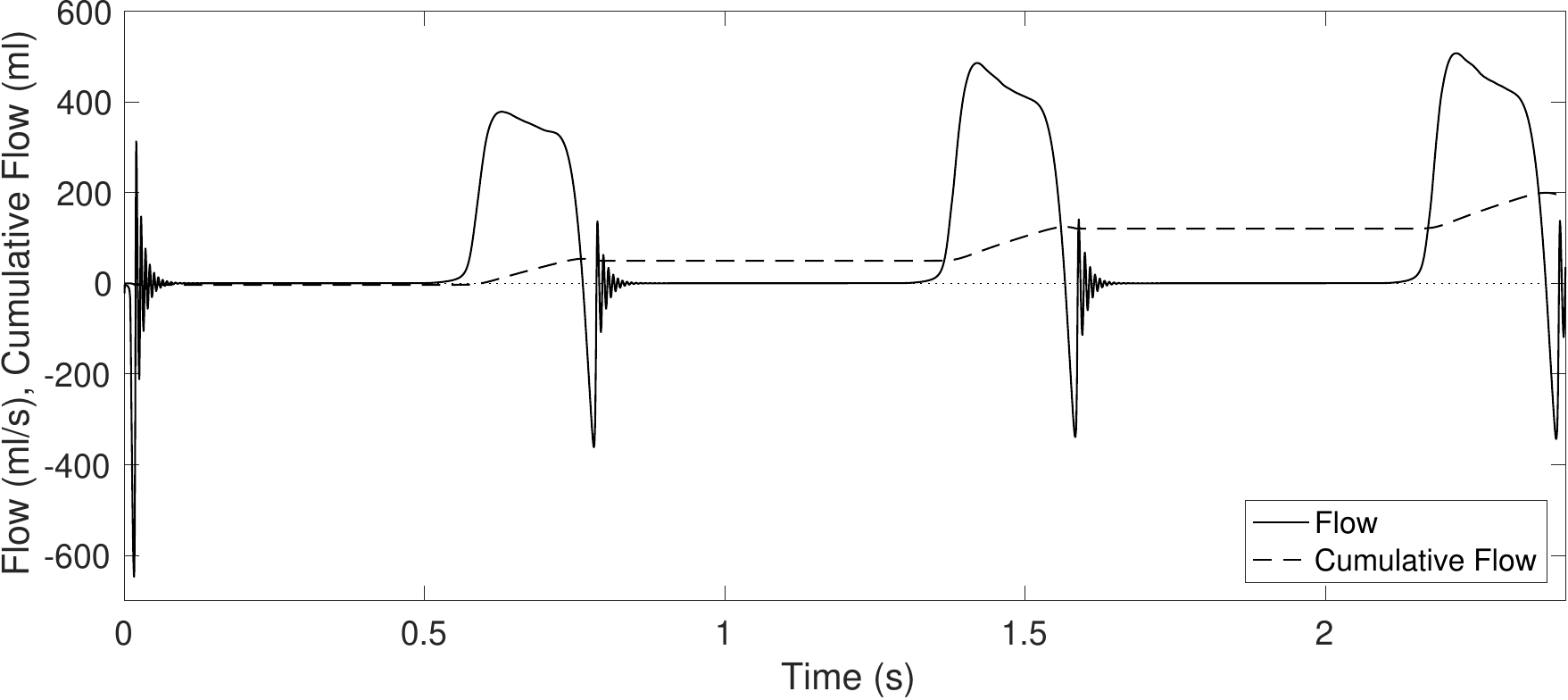} 
\caption{Driving pressures at physiological values and emergent flows.
} 
\label{flow_pressure_basic}
\end{figure}

Pressures in the physiological range and the flow rates that result are shown in Figure \ref{flow_pressure_basic}. 
The simulation begins with diastole, and the aortic valve immediately closes. 
After the initial vibration concludes, flow is approximately zero for about half a second. 
The ventricular pressure rises during systole, and with a slight lag, the aortic pressure follows. 
A forward pressure difference is established and the forward flow rate through the valve rapidly rises. 
The pressure difference declines gradually through systole, and rapid ejection continues. 
Note that the pressure difference during forward flow includes the pressure difference across the flow straightener; the pressure difference across the valve itself is lower. 

At end systole, the ventricular pressure drops below the aortic pressure and continues to fall and the valve begins to close. 
A prominent dicrotic notch appears in the aortic pressure, which emerges from the combined dynamics of the fluid, valve and lumped parameter network. 
Immediately after, a violent vibration begins then is quickly damped out. 
The aortic pressure, the dynamics of which are governed by equation \eqref{rcr_ode}, shows an oscillation at the same time. 
This vibration in pressure and flow causes the S$_{2}$ heart sound or colloquially the ``dub'' of ``lub-dub.''
Next, the flow is approximately zero; the valve is tightly sealed. 
This repeats over the next two cardiac cycles. 
Note that the back pressure is over an order of magnitude greater than the forward driving pressures. 
This asymmetry creates demanding conditions for the valve, as it must support a pressure, then open freely under a forward pressure that is much smaller. 

See also movies {\bf M1,M2}, which show this simulation visualized with slice views in slow motion and real time, respectively, and {\bf M3}, which shows pathlines in slow motion.

\begin{figure}[t!]
\centering
\includegraphics[width=0.5\columnwidth]{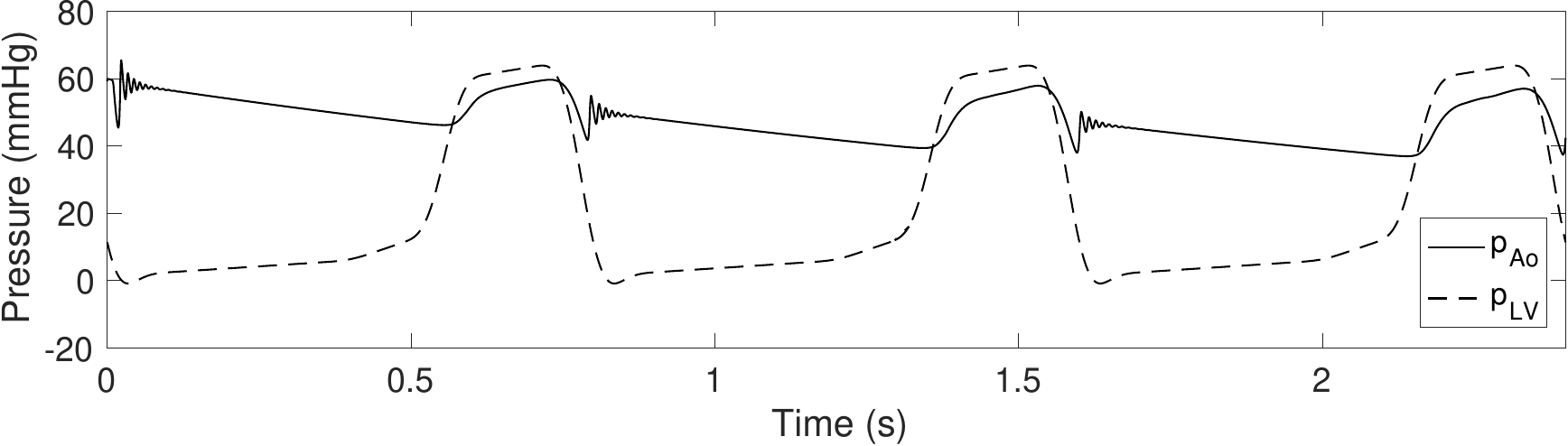} 
\includegraphics[width=0.5\columnwidth]{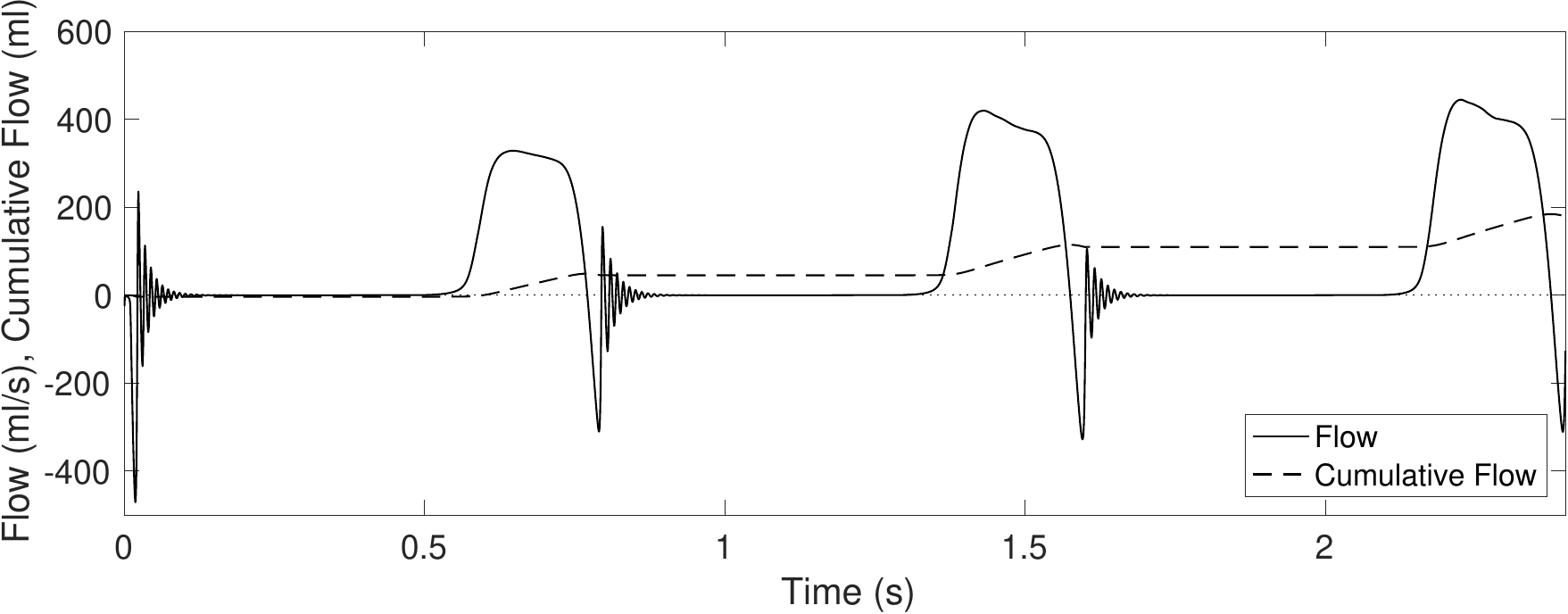} 
\caption{Hypotensive driving pressures and emergent flows. 
Despite much lower loading pressures, the valve seals competently over multiple beats. 
} 
\label{flow_pressure_low}
\end{figure}

Next, we ``stress test'' the model to ensure it functions over higher and lower driving pressures. 
First, we simulate with lower, or hypotensive, systolic aortic and ventricular pressure pressure. 
We target a pressure of 60/40 mmHg, or half of the original targets. 
In the lumped parameter network, the total resistance is turned down by half, while the ratio of proximal to distal resistance is kept constant. 
The capacitance is set using equation \eqref{capacitance_formula} with $P_{1} = 40$ mmHg and $P_{2} = 50$ mmHg. 
This gives $R_{p} = 38.41$ s dynes cm$^{-5}$, $R_{d} = 591.05$  s dynes cm$^{-5}$ and $C = 0.0036$ cm$^{5}$  dynes$^{-1}$.
Ventricular pressure remains constant though diastole, then is halved in systole, as it only needs to rise above the lower aortic systolic pressure that occurs with half resistance. 
Flows and pressures are shown in Figure \ref{flow_pressure_low}. 
Despite lower pressures, the valve is fully competent and seals without leak over three cardiac cycles. 

\begin{figure}[t!]
\centering
\includegraphics[width=0.5\columnwidth]{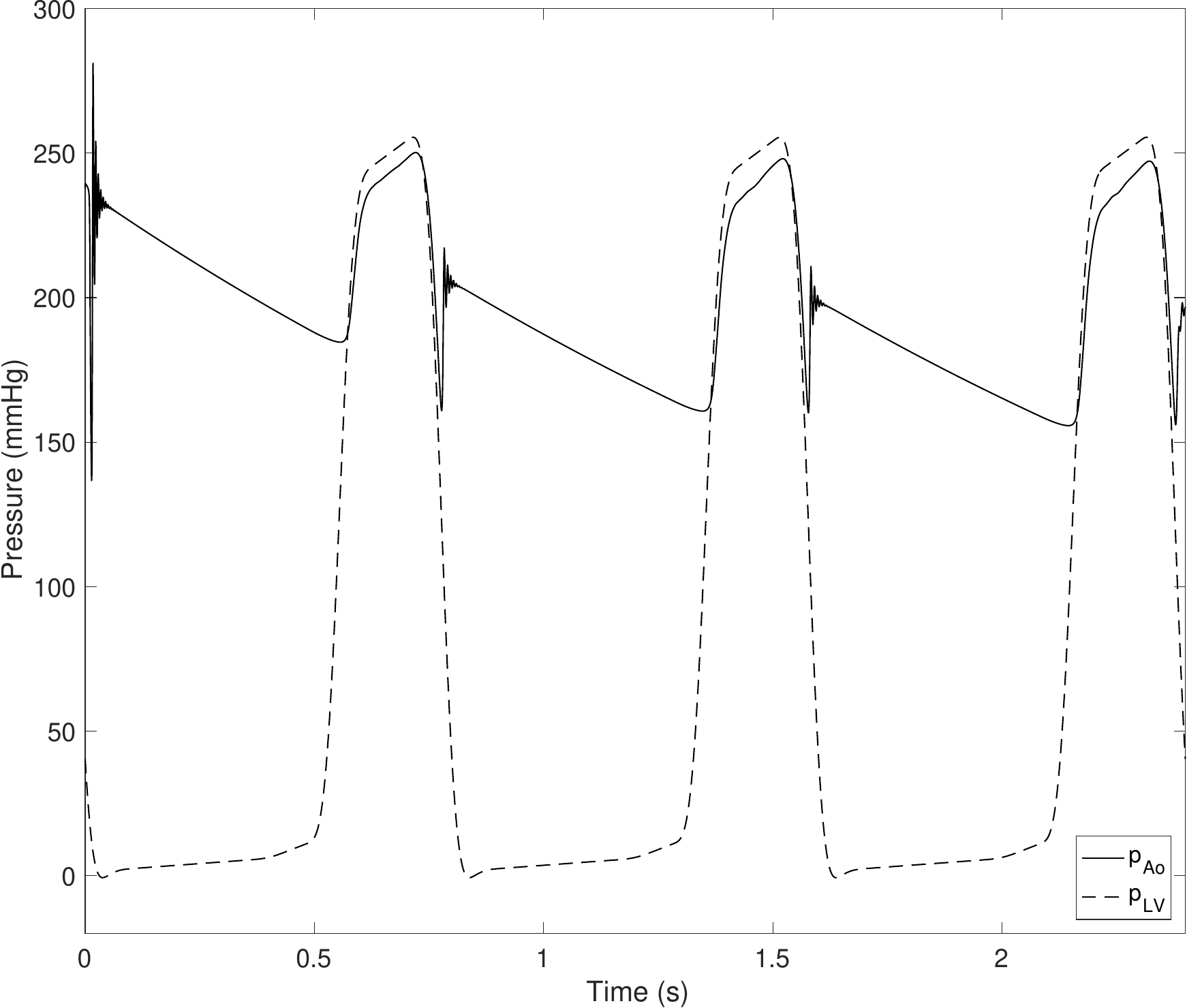} 
\includegraphics[width=0.5\columnwidth]{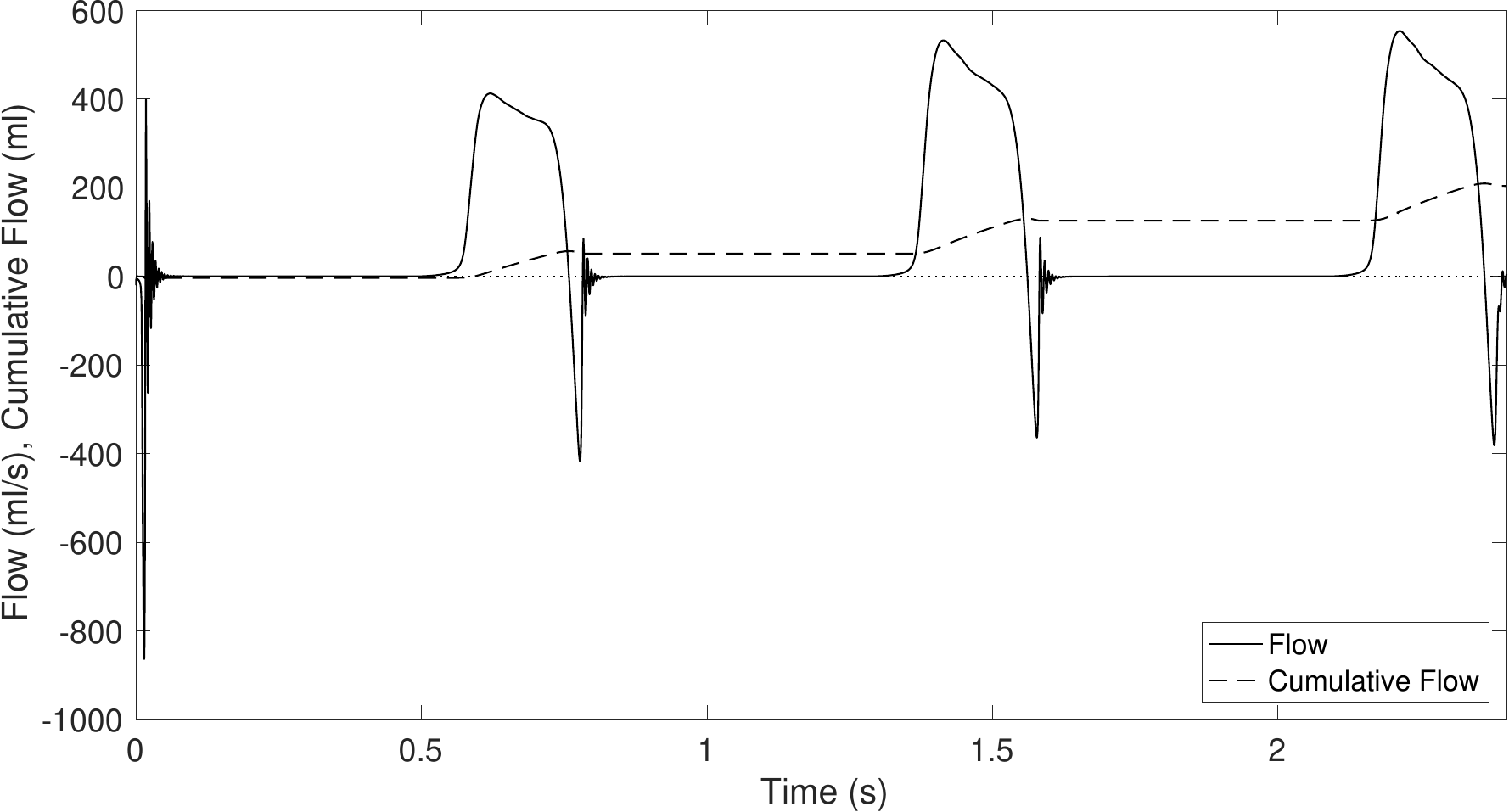} 
\caption{Hypertensive driving pressures and emergent flows.
With much higher loads, the valve remains completely sealed during diastole over three cardiac cycles. 
} 
\label{flow_pressure_high}
\end{figure}

Next, we test the valve under extreme hypertensive, high-pressure conditions. 
Total resistance is doubled, the ratio of proximal to distal resistance is maintained, and capacitance is tuned using equation \eqref{capacitance_formula} with $P_{1} = 160$ mmHg and $P_{2} = 200$ mmHg. 
Note that pulse pressure may rise in conditions such as essential hypertension \cite{carretero2000essential}. 
This gives $R_{p} = 153.62$ ml$^{-1}$ s dynes cm$^{-2}$, $R_{d} = 2364.20$ ml$^{-1}$ s dynes cm$^{-2}$ and $C = 0.0009$ cm$^{5}$ dynes$^{-1}$.
Ventricular pressure remains constant though diastole, then is doubled in systole, as it must rise above the higher systemic systolic pressure that occurs with double resistance. 
Flows and pressures are shown in Figure \ref{flow_pressure_high}. 
Despite extreme hypertensive pressures, the valve seals reliably on all beats. 

\begin{figure}[ht!]
\setlength{\tabcolsep}{1.0pt}
\centering
\begin{tabular}{cccc}
&
hypotensive  & 
physiological  & 
hypertensive \\ 
&
  & 
 pressures & 
 \\ 
\raisebox{210pt}{\includegraphics[width=.04\columnwidth]{colorbar.jpeg}} &
\includegraphics[width=.2\columnwidth]{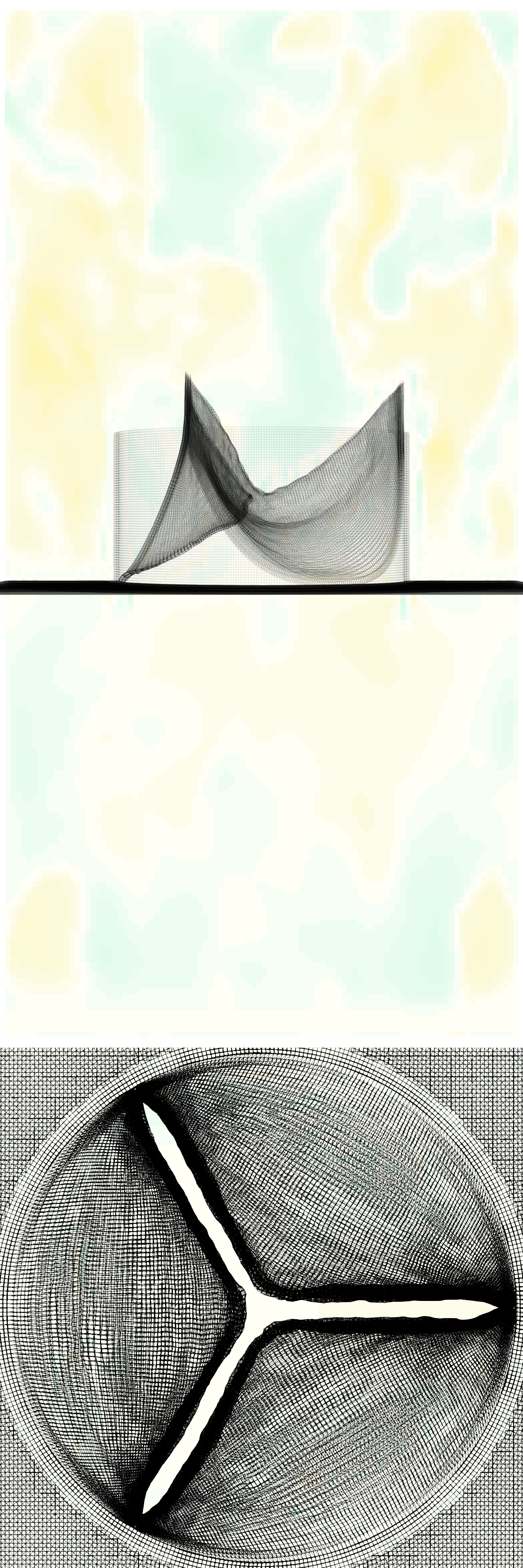}   &
\includegraphics[width=.2\columnwidth]{aortic_2020486_384_f499232_0mm_radial_4mm_circ_basic_rcr_crease_removed_more_stiffness_down_paper1208.jpeg}  &
\includegraphics[width=.2\columnwidth]{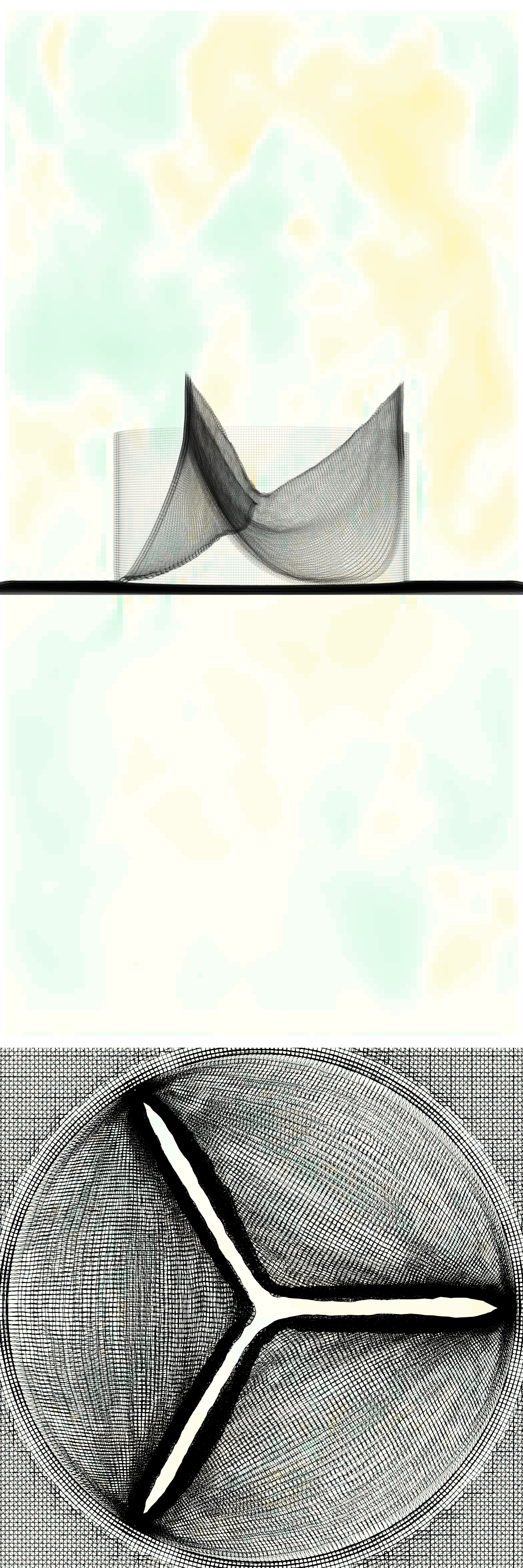}  \\
\end{tabular}
\caption{Slice view of the $z$ component of velocity with low, physiological and high pressure. 
Despite vastly different loading conditions, the valve functions well in each case, with slightly less curvature at low pressure and slightly more at higher pressure, as expected. 
} 
\label{three_pressures}
\end{figure}

Figure \ref{three_pressures} shows the valve in hypotensive, physiological and hypertensive pressures during diastole in the third cardiac cycle. 
In all cases, the loaded configurations are visually similar.
The lowest pressure has the least curvature, which is expected as it is the least loaded, then the model with physiological pressure, then the hypertensive model, which is again expected as it is the most loaded. 
In the hypotensive case, the material model is compliant enough such that with much lower pressure, the valve achieves a similar loaded configuration as in the standard cases. 
In the hypertensive cases, the material, which stiffens more under higher load, deforms only slightly more than under standard pressures. 
This suggests that the nonlinearity of the material enables the valve to function effectively over a range of pressures.

\subsection{Variation in loaded strain}
\label{varied_strain}

In this section, we examine variations in the prescribed loaded strain. 
In experimental studies on the kinematics of the aortic valve, reported values of strain or stretch may vary due to the definition of the reference configuration. 
The reference configuration may be taken to be the open, in vivo state, attached to the aortic root but in a dissected specimen (as in \cite{yap2009dynamic}), excised and floating in liquid, crumpled, or stretched to remove initial buckling. 
Pre-strain alters the deformation between the reference and loaded state, and can dramatically change the reported values of local tangent moduli \cite{rausch2013effect}. 
Since the strains reported in reported in \cite{yap2009dynamic} are measured from the in situ state, there may still be pre-strain present; 
perhaps these models would behave differently if pre-strain was added or potentially excess strain was removed.

For a given loaded state as the solution to equations \eqref{equilbrium_eqn_discrete}, we then ask whether the behavior in FSI simulations is sensitive to the values of strain prescribed. 
We first prescribe smaller strains while maintaining the ratio of circumferential to radial strain, corresponding to longer lengths for all links in the model. The values are 
\begin{align}
E_{c} = 0.1, \quad E_{r} = 0.36 . 
\end{align}
We then prescribe a larger strain, corresponding to a shorter resting lengths for all links in the model, with values 
\begin{align}
E_{c} = 0.2, \quad E_{r} = 0.72 . 
\end{align}
This change preserves exactly the predicted loaded geometry and the force exerted in the loaded geometry, but alters the reference geometry or the model.

\begin{figure}[ht]
\setlength{\tabcolsep}{1.0pt}
\centering
\begin{tabular}{cccc}
&    lower predicted            &  experimentally   &    higher predicted  \\ 
&    strain, longer               &  predicted strain   &   strain, shorter  \\ 
&    reference lengths        &                             &    reference lengths \\ 
\raisebox{210pt}{\includegraphics[width=.04\columnwidth]{colorbar.jpeg}} &
\includegraphics[width=.2\columnwidth]{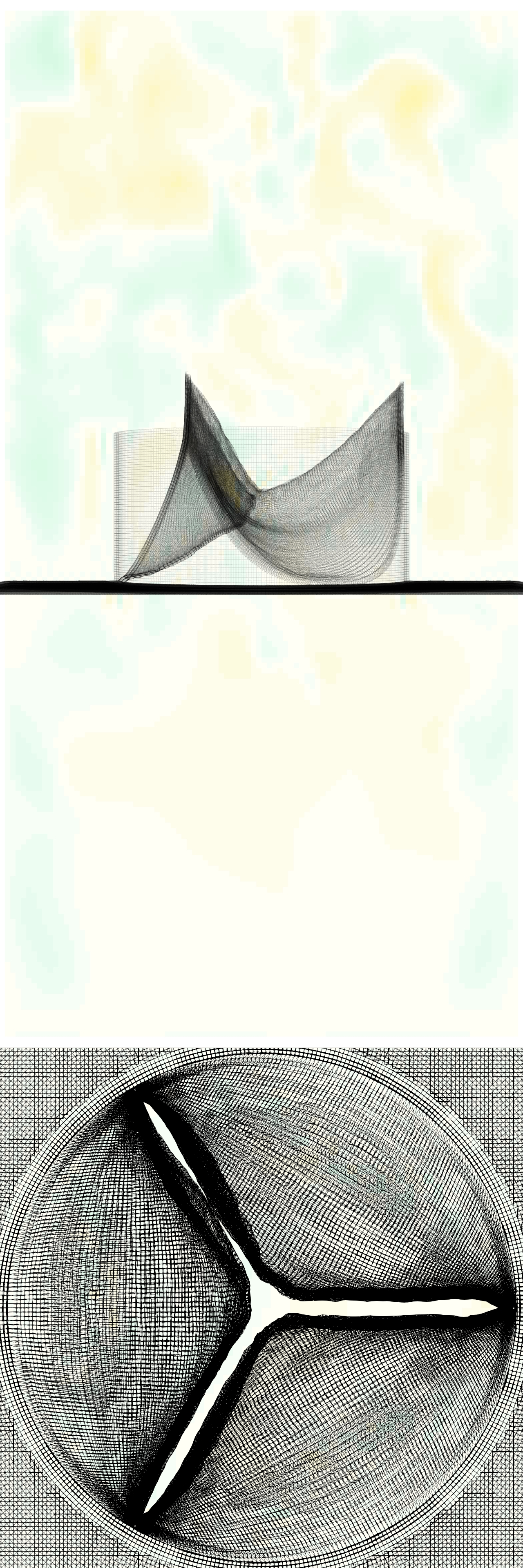}   &
\includegraphics[width=.2\columnwidth]{aortic_2020486_384_f499232_0mm_radial_4mm_circ_basic_rcr_crease_removed_more_stiffness_down_paper1208.jpeg}  &
\includegraphics[width=.2\columnwidth]{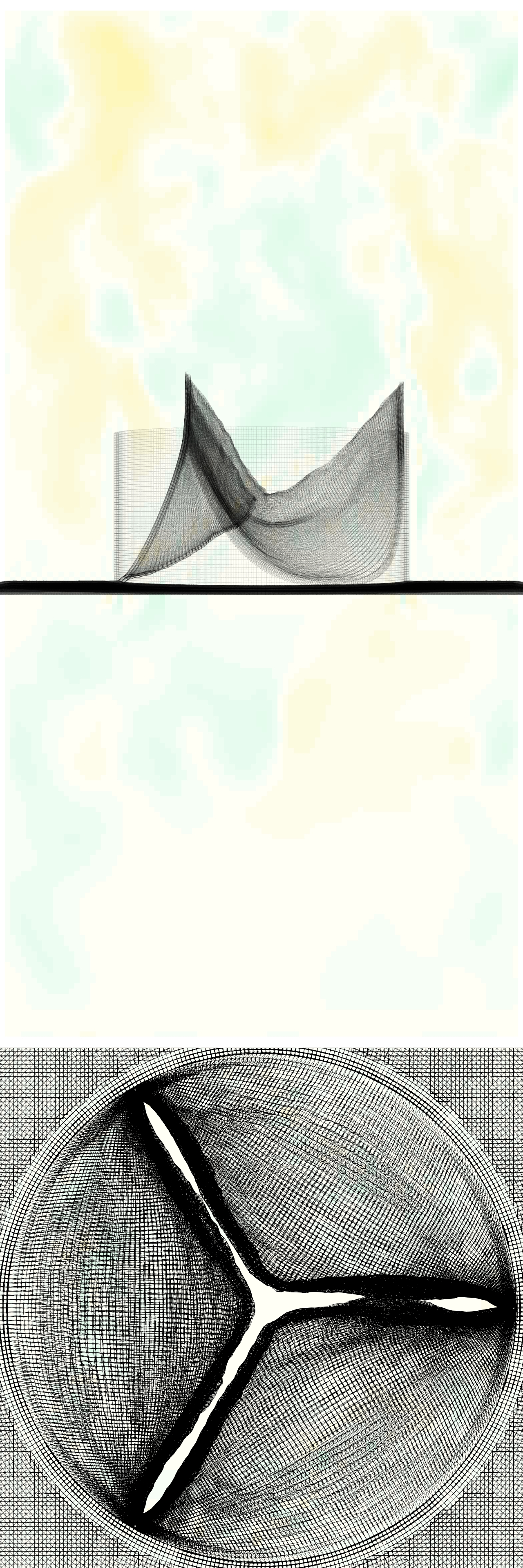}  
\end{tabular}
\caption{Slice view of the $z$ component of velocity with low, physiological and high predicted strain in construction of the model. 
Each has a distinct reference configuration, but identical predicted loaded configurations. 
Despite different reference configurations, the performance during closure of these three models is nearly identical.  
} 
\label{three_strains}
\end{figure}

This change in strain nearly preserves the tangent modulus, the slope of tension with respect to strain, as well. 
Consider two values of the fully-loaded strain $E_{1}$ and $E_{2}$ at which equal tension is achieved. 
By construction, $T_{1}(E_{1}) = T_{2}(E_{2})$ or 
\begin{align}
\kappa_{1} (e^{\lambda E_{1}} - 1) = \kappa_{2} (e^{\lambda E_{2}} - 1), 
\end{align}
so
\begin{align}
\frac{ \left. T_{1}'(E)  \right |_{E = E_{1}} }{   \left. T_{2}'(E)  \right |_{E = E_{2}}  }
= \frac{ \kappa_{1} \lambda e^{\lambda E_{1}}   }{ \kappa_{2} \lambda e^{\lambda E_{2}} } 
= \frac{ (e^{\lambda E_{2}} - 1) e^{\lambda E_{1}}   }{ (e^{\lambda E_{1}} - 1)  e^{\lambda E_{2}} } 
= \frac{ 1 - e^{-\lambda E_{2}}     }{ 1 - e^{-\lambda E_{1}}   } 
\approx 
1 . 
\end{align}
Both the numerator and denominator exponentially approach one away from $E = 0$. 
Since the exponential rate $b = 57.5$, and we consider $E$ to be on the order of $.1$, this ratio is very close to one. 
This implies that changing the loaded strain very nearly maintains the tangent modulus of the predicted loaded configuration.

Flows during closure in the third cycle are shown in Figure \ref{three_strains}. 
The three models have nearly identical performance, including highly similar closed configurations. 
This suggests that the predicted loaded configuration is indeed predictive of valve performance, regardless of alterations to the reference configuration.

\subsection{Variation in constitutive law}
\label{results_constitutive}

\begin{figure}[t!]
\centering
\includegraphics[width=.6\columnwidth]{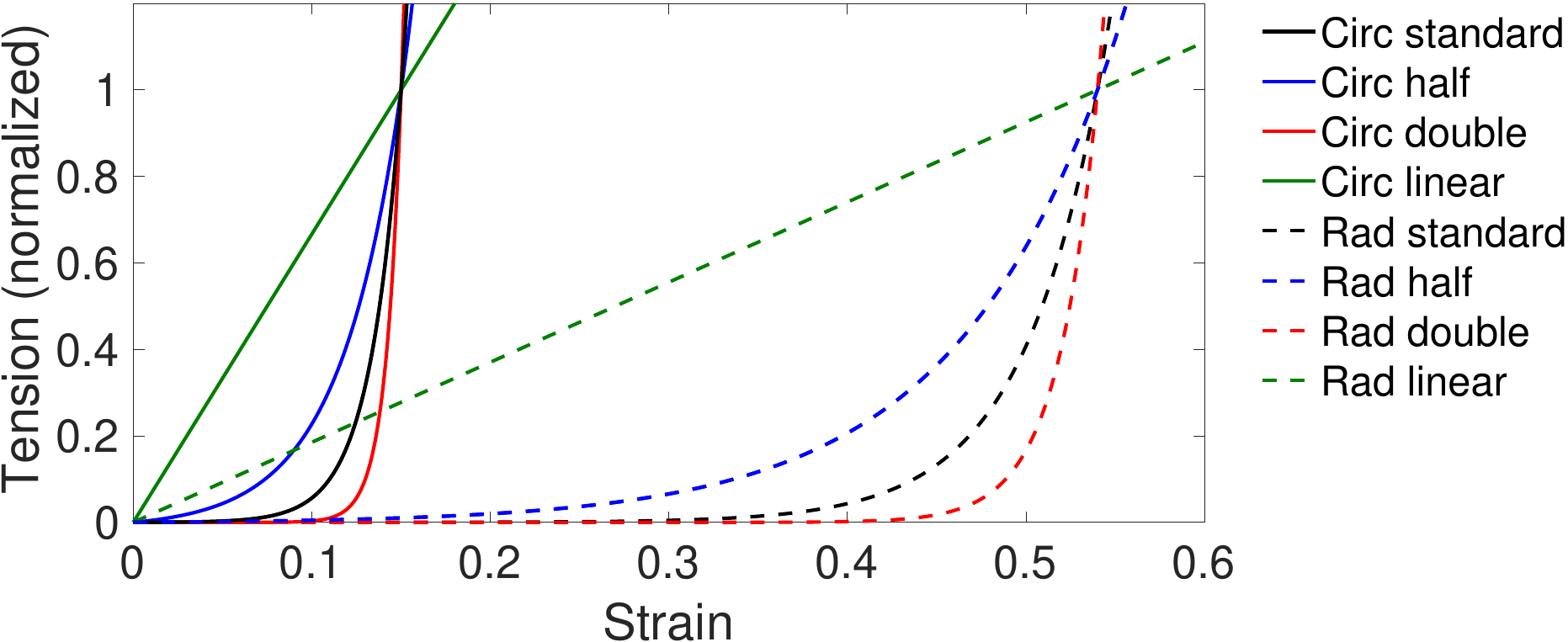} 
\caption{ 
Comparison of constitutive curves, showing laws with experimentally measured exponential rate, a linear law, half the exponential rate and double the exponential rate in the circumferential and radial directions.
The circumferential and radial curves are normalized at the predicted strains of $E_{c} = 0.15$ and $E_{r} = 0.54$, respectively. 
} 
\label{constitutive_comparison.pdf}
\end{figure}

In this section, we vary the constitutive law in the valve. 
We test with a linear law and laws with half and double the experimentally measured exponential rates. 
A comparison of these curves is shown in Figure \ref{constitutive_comparison.pdf}.
First, we apply a linear law to show where the model maintains its performance and where it fails, relative to a model with a more realistic constitutive law. 
Results with this constitutive law are meant to serve as a ``negative example'', and illustrate why nonlinear behavior is required for robust valve function.
We replace equation \eqref{exponential_law} with the linear law
\begin{align}
\tau(E) = \kappa E
\end{align}
The value of tensions for strains lower than $E_{c}$ and $E_{r}$ is greater than the standard model, equal at the strains $E_{c}$ and $E_{r}$, and lower for strains larger than $E_{c}$ and $E_{r}$. 
This creates a lower mean tangent modulus at the predicted strains $E_{c}$ and $E_{r}$. 
The mean circumferential tangent modulus is $1.66 \cdot 10^{7}$ dynes/cm$^{2}$, the mean radial tangent modulus is $ 4.75 \cdot 10^{5}$ dynes/cm$^{2}$, and the ratio of the means is approximately 35.
These moduli are approximately one order of magnitude lower than the tangent moduli of the standard model. 

\begin{figure}[th!]
\setlength{\tabcolsep}{1.0pt}
\centering
\begin{tabular}{cccc}
&
hypotensive  & 
physiological  & 
hypertensive \\ 
&
  & 
 pressures & 
 \\ 
\raisebox{210pt}{\includegraphics[width=.04\columnwidth]{colorbar.jpeg}} & 
\includegraphics[width=.2\columnwidth]{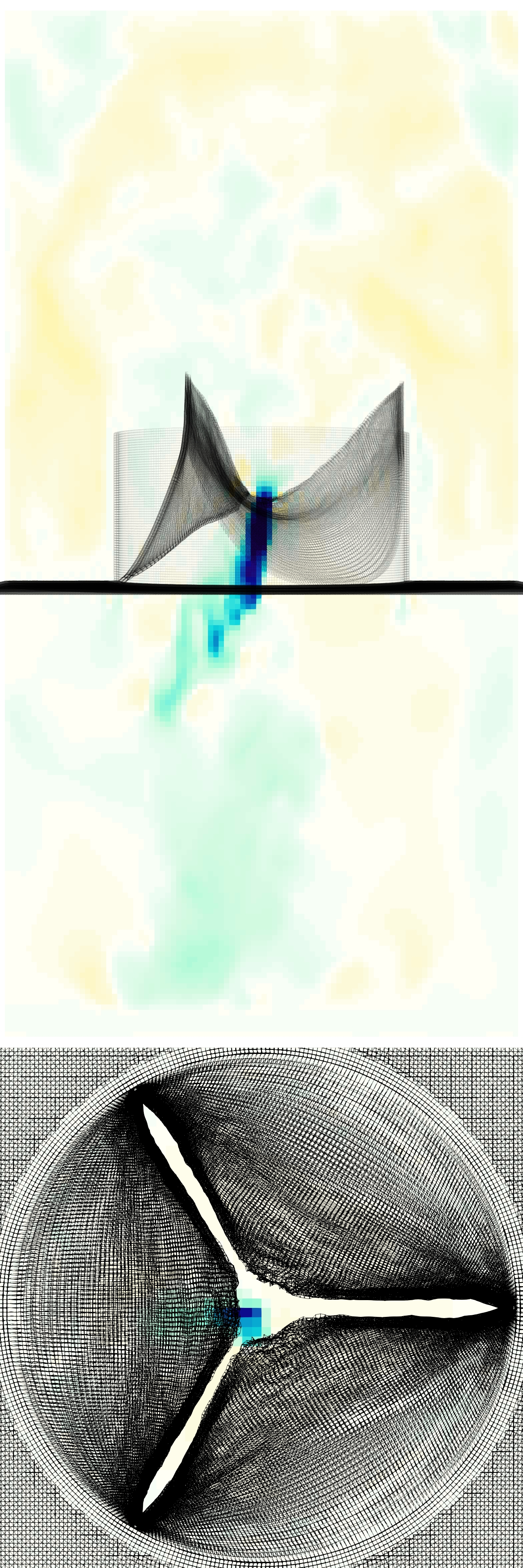}   &
\includegraphics[width=.2\columnwidth]{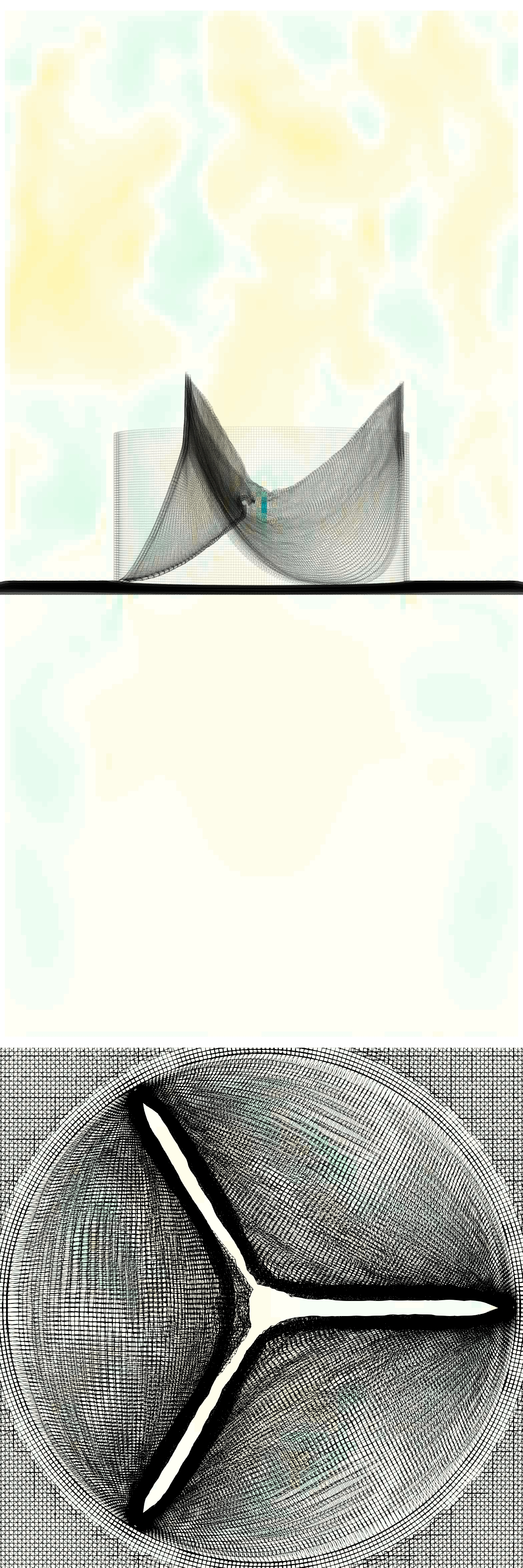}  &
\includegraphics[width=.2\columnwidth]{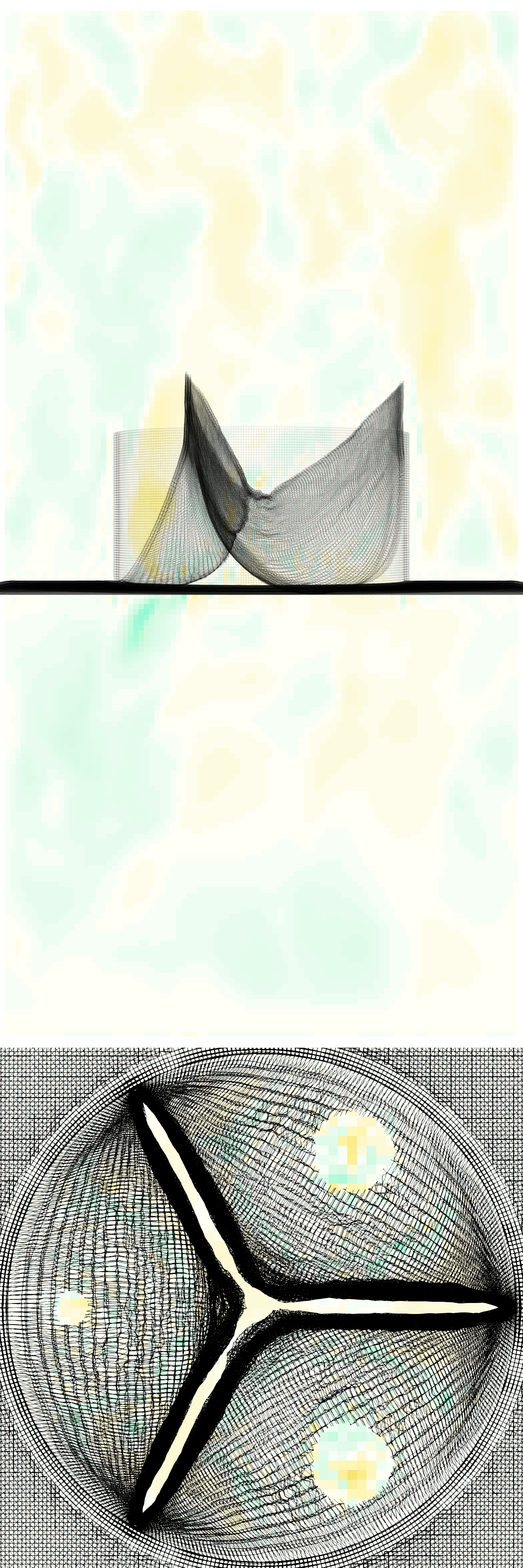}  \\ 
\end{tabular}
\caption{ Results with linear constitutive law under hypotensive pressure (left), physiological pressure (center) and hypertensive pressure (right).
This model leaks under low pressure, seals at physiological pressure, and prolapses at high pressure. 
} 
\label{alternative_constitutive_law_linear}
\end{figure}

Results under low, standard and high pressures for the model with linear constitutive law are shown in Figure \ref{alternative_constitutive_law_linear}. 
At standard pressure, the loaded model is fully sealed and appears qualitatively similar to the standard model. 
At low pressure, the leaflets fail to deform enough to create a good coaptation, and regurgitation results. 
At high pressure, the valve seals, but deforms so much that the belly of the leaflet prolapses below the lowest point on the annulus. 
Aortic valve prolapse may lead to regurgitation or require surgery \cite{10.1016/j.ejcts.2004.05.027}.

Next, we reduce the exponential rates by a factor of one half, representing a less nonlinear material than the standard model. 
The exponential rates are taken to be half of the basic model, as derived from \cite{may2009hyperelastic}.
The circumferential rate takes value $\lambda_{c} = 57.46 / 2 = 28.73$ and the radial rate takes value $\lambda_{r} = 22.40 / 2 = 11.20$. 
The emergent mean circumferential tangent modulus is $7.24 \cdot 10^{7}$ dynes/cm$^{2}$, the mean radial tangent modulus is $ 2.88 \cdot 10^{6}$ dynes/cm$^{2}$, and the ratio of the means is approximately 25.
These values are approximately half of the tangent modulus computed for the standard configuration. 
This is because this change in exponential rate alters the tangent modulus, the slope of tension with respect to strain, as follows. 
Consider two exponential rates $\lambda_{1}$ and $\lambda_{2}$ at which equal tension is achieved at strain $E_{*}$. 
By construction, $T_{1}(E_{*}) = T_{2}(E_{*})$ or 
\begin{align}
\kappa_{1} (e^{\lambda_{1} E_{*}} - 1) = \kappa_{2} (e^{\lambda_{2} E_{*}} - 1), 
\end{align}
so
\begin{align}
\frac{ T_{1}'(E_{*})  }{   T_{2}'(E_{*})   }
= \frac{ \kappa_{1} \lambda_{1} e^{\lambda_{1} E_{*}}   }{ \kappa_{2} \lambda_{2} e^{\lambda_{2} E_{*}} } 
= \frac{ \lambda_{1} (e^{\lambda_{2} E_{*}} - 1) e^{\lambda_{1} E_{*}}   }{ \lambda_{2} (e^{\lambda_{1} E_{*}} - 1)  e^{\lambda_{2} E_{*}} } 
= \frac{ \lambda_{1} (1 - e^{-\lambda_{2} E_{*}})     }{ \lambda_{2} ( 1 - e^{-\lambda_{1} E_{*}} )  } 
\approx \frac{\lambda_{1}}{\lambda_{2}} . 
\end{align}
Thus by taking half the exponential rate, the local tangent moduli are approximately half those of the standard model.

\begin{figure}[p!]
\setlength{\tabcolsep}{1.0pt}
\centering
\begin{tabular}{cccc}
&
hypotensive  & 
physiological  & 
hypertensive \\ 
&
  & 
 pressures & 
 \\ 
\raisebox{210pt}{\includegraphics[width=.04\columnwidth]{colorbar.jpeg}}  & 
\includegraphics[width=.2\columnwidth]{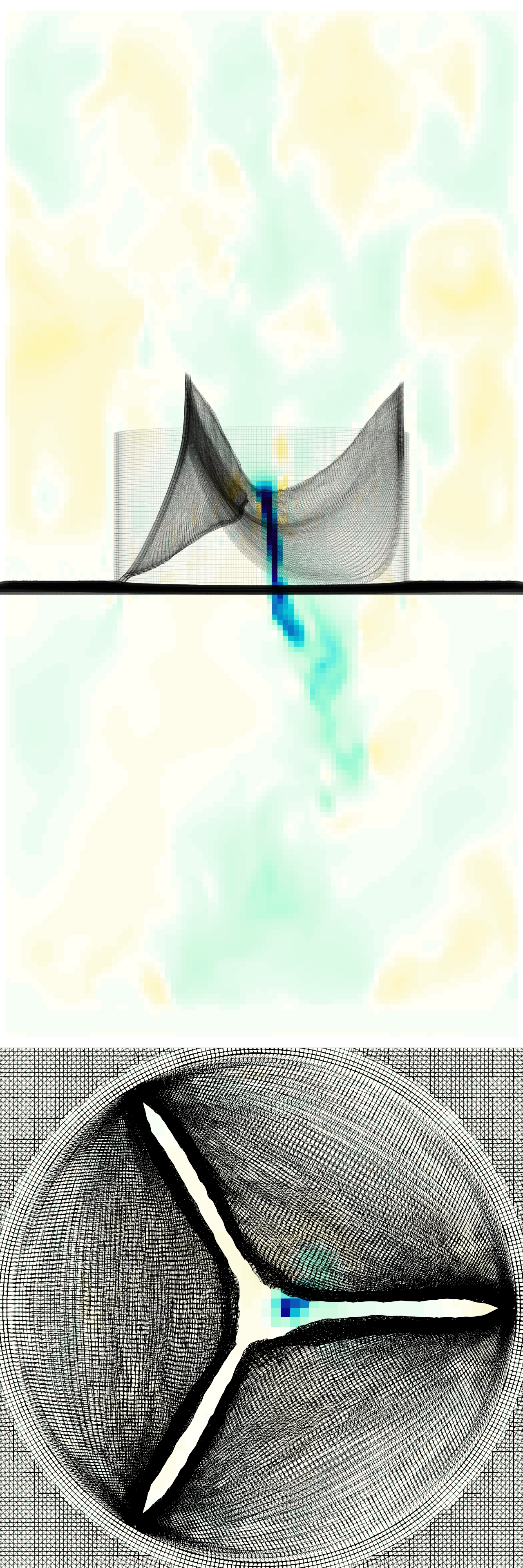} &
\includegraphics[width=.2\columnwidth]{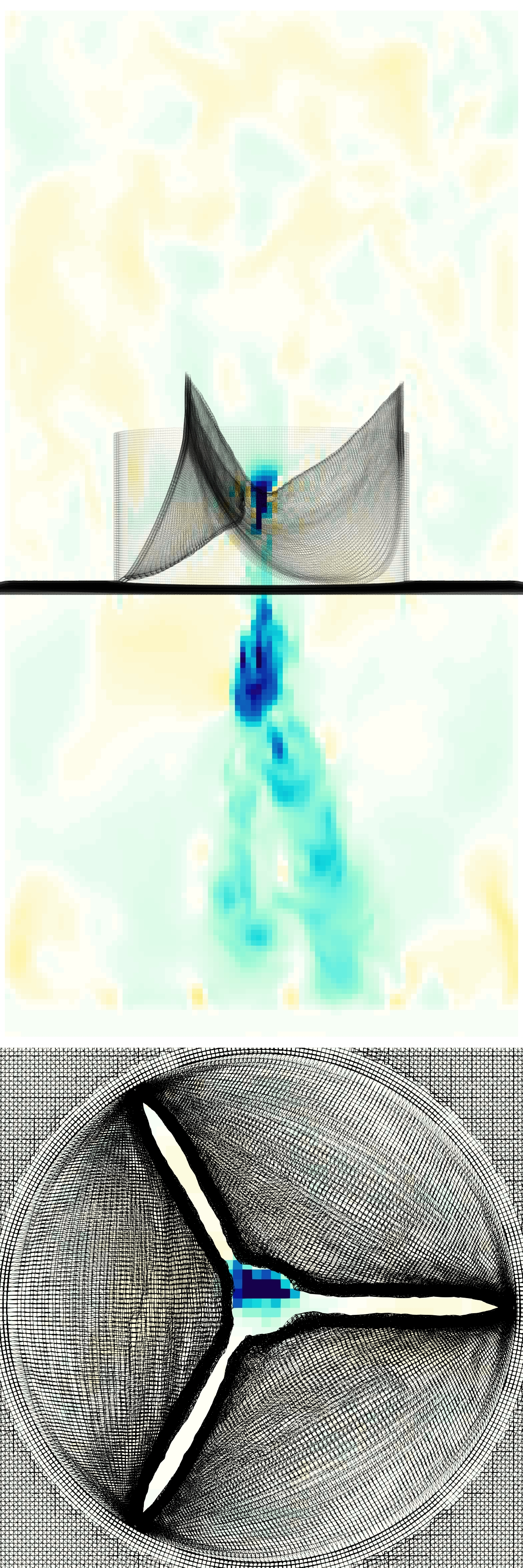}  &
\includegraphics[width=.2\columnwidth]{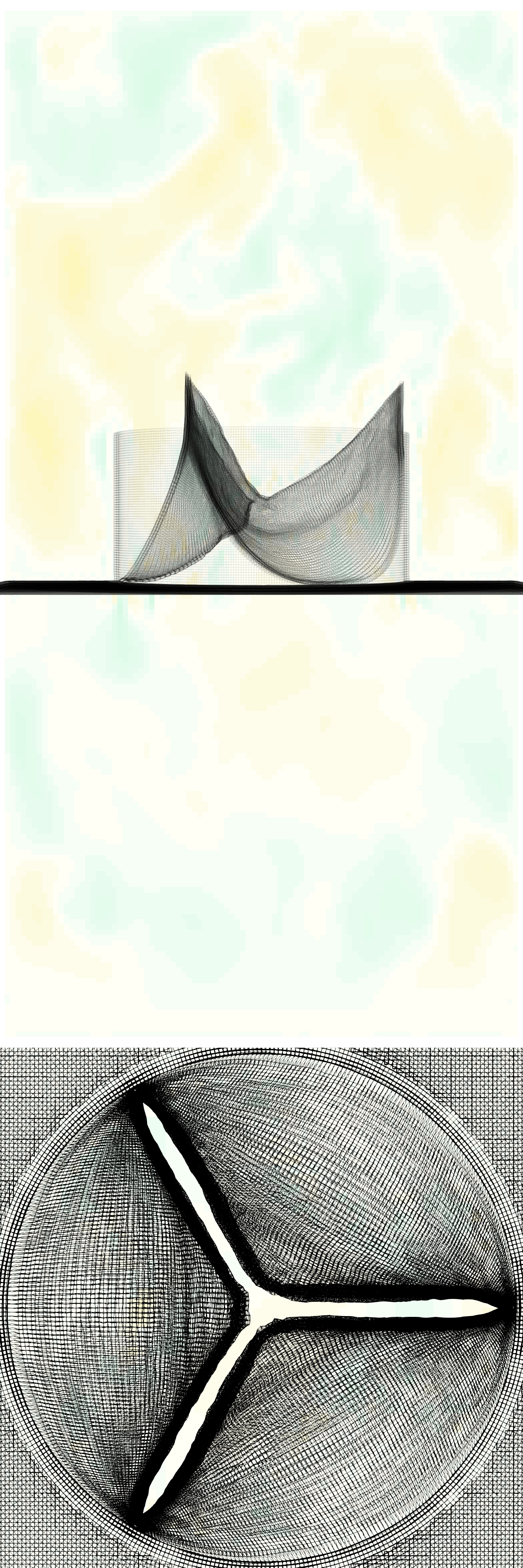}  \\ 
\raisebox{210pt}{\includegraphics[width=.04\columnwidth]{colorbar.jpeg}} & 
\includegraphics[width=.2\columnwidth]{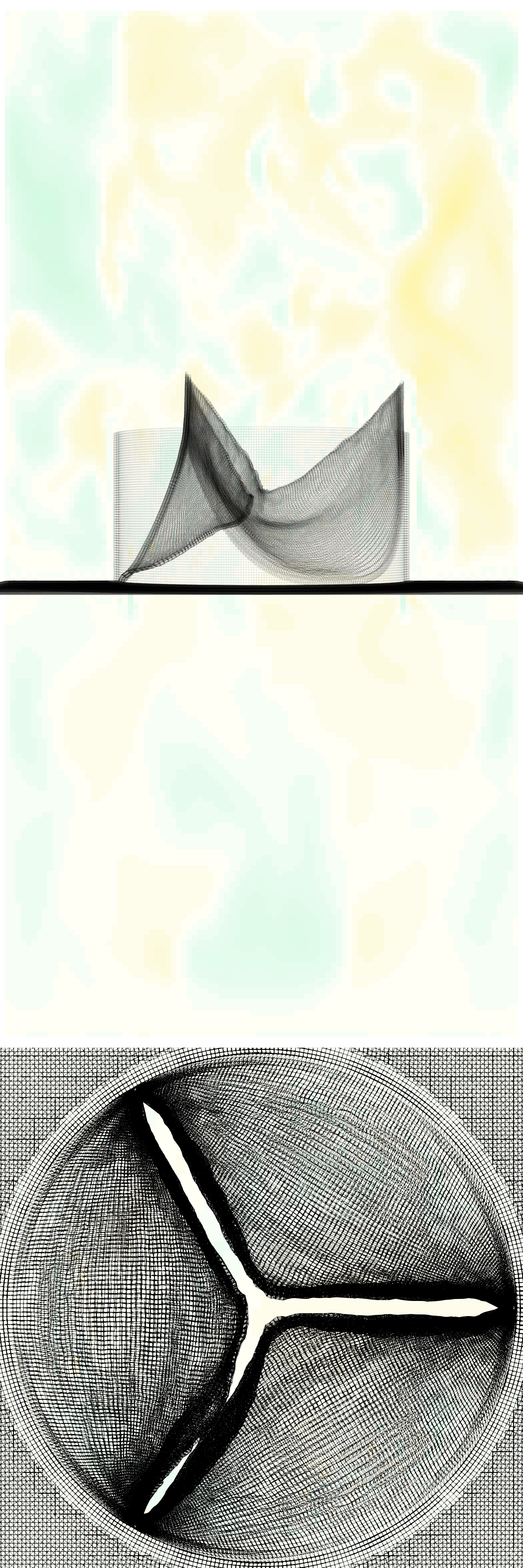} &
\includegraphics[width=.2\columnwidth]{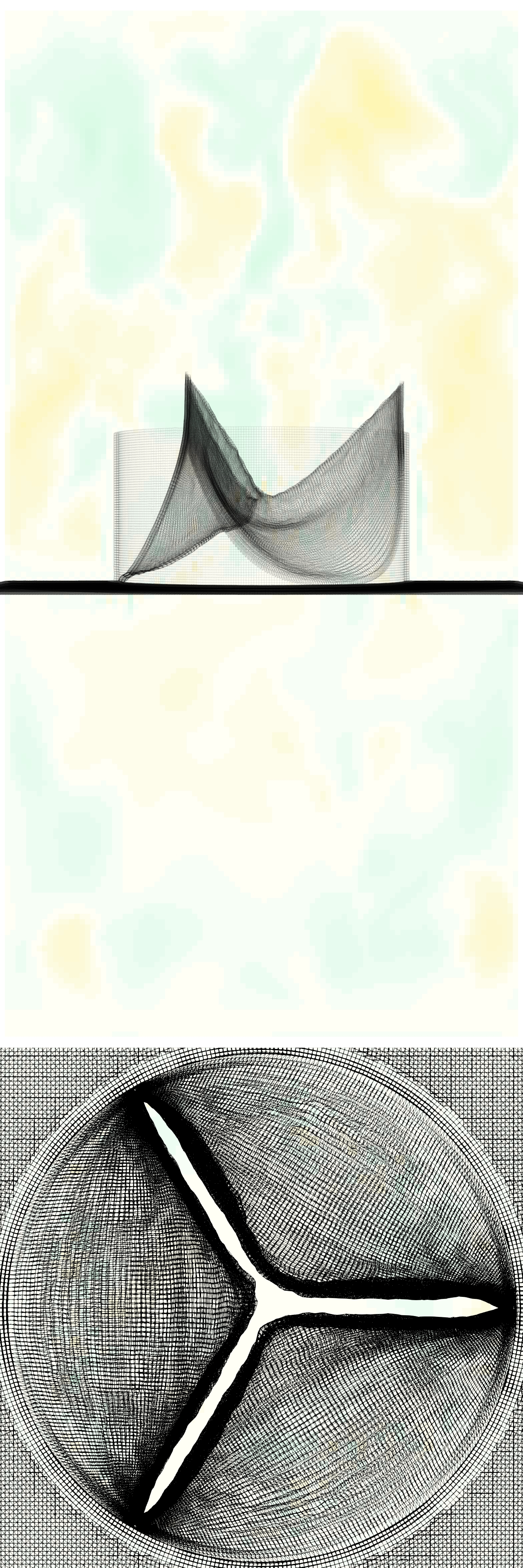} &
\includegraphics[width=.2\columnwidth]{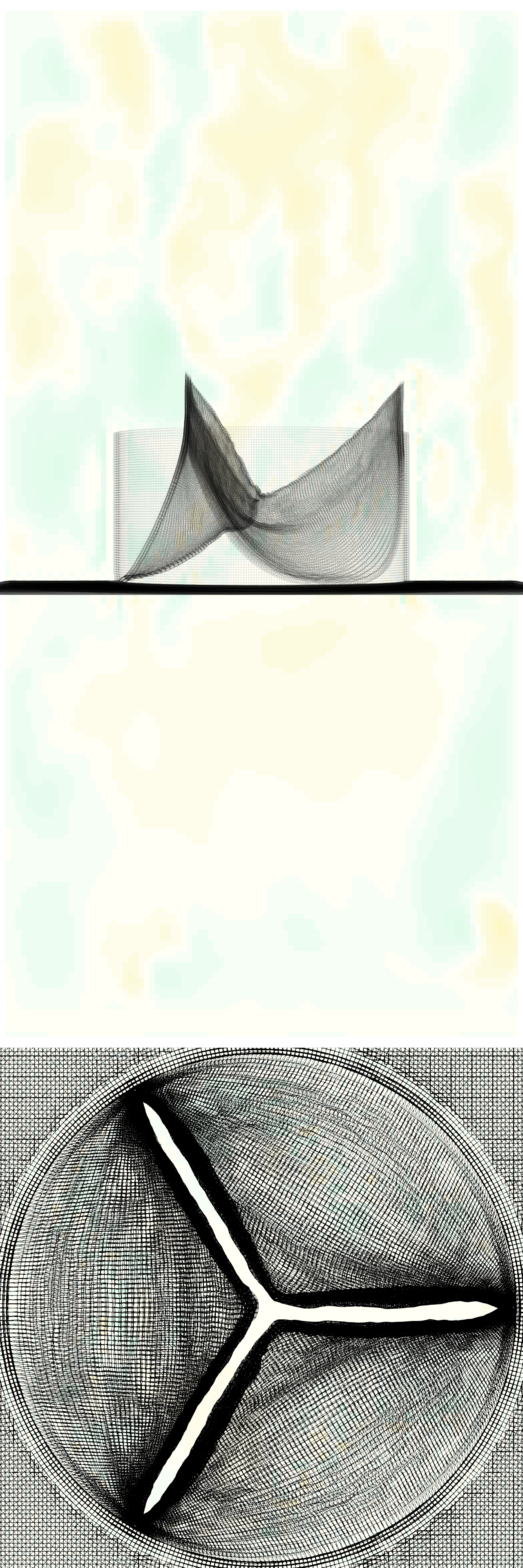}  \\ 
\end{tabular}
\caption{ Results with alternative constitutive laws under hypotensive pressure (left), physiological pressure (center) and hypertensive pressure (right).
The top row shows a model with half the experimentally measured exponential rates in the constitutive law, shows leak at low and standard pressure, but a good seal at high pressure. 
The bottom row shows a model with half the experimentally measured exponential rates in the constitutive law, and this model seals well at all tested pressures. 
} 
\label{alternative_constitutive_law}
\end{figure}

Figure \ref{alternative_constitutive_law} shows results with half exponential rates and approximately half the tangent modulus at the predicted strain values.
At low pressure, the valve develops regurgitation. 
Similarly to the linear model, this model is much stiffer at lower strains than the standard model, and the leaflets do not not coapt well under this smaller load. 
At standard pressures, the model also leaks with a central jet. 
It is not obvious why this model works more poorly than the linear model under standard pressure. 
At high pressure, the valve seals well. 
Despite being less stiff at higher strains, the exponential law is able to generate sufficient tension to support this load.

Last, we double the exponential rates, with values $\lambda_{c} = 57.46 \cdot 2 = 114.91$ and $\lambda_{r} = 22.40 \cdot 2 = 44.79$.
The tangent modulus at the predicted strain is approximately doubled, taking mean values $2.86 \cdot 10^{8}$ and $ 1.15 \cdot 10^{7}$ dynes/cm$^{2}$ circumferentially and radially, respectively. 
Figure \ref{alternative_constitutive_law} shows results at low, standard and high pressures. 
The valve functions well in all cases, with notably little visible difference in the closed configurations in all cases. 
This model is more compliant at strains below the predicted strain, and stiffer still at higher strains. 
This suggests that the model functions best when it is compliant enough to easily obtain the closed configuration, then stiff enough due to nonlinearity around that configuration that it does not strain much more.  

\begin{figure}[ht]
\setlength{\tabcolsep}{1.0pt}
\centering
\begin{tabular}{ccccc}
&
2 mm shorter &  
2 mm extra  &
2 mm shorter  &
2 mm extra  \\ 
&
 free edge  &  
 free edge &
 height &
 height \\ 
\raisebox{210pt}{\includegraphics[width=.04\columnwidth]{colorbar.jpeg}} &
\includegraphics[width=.2\columnwidth]{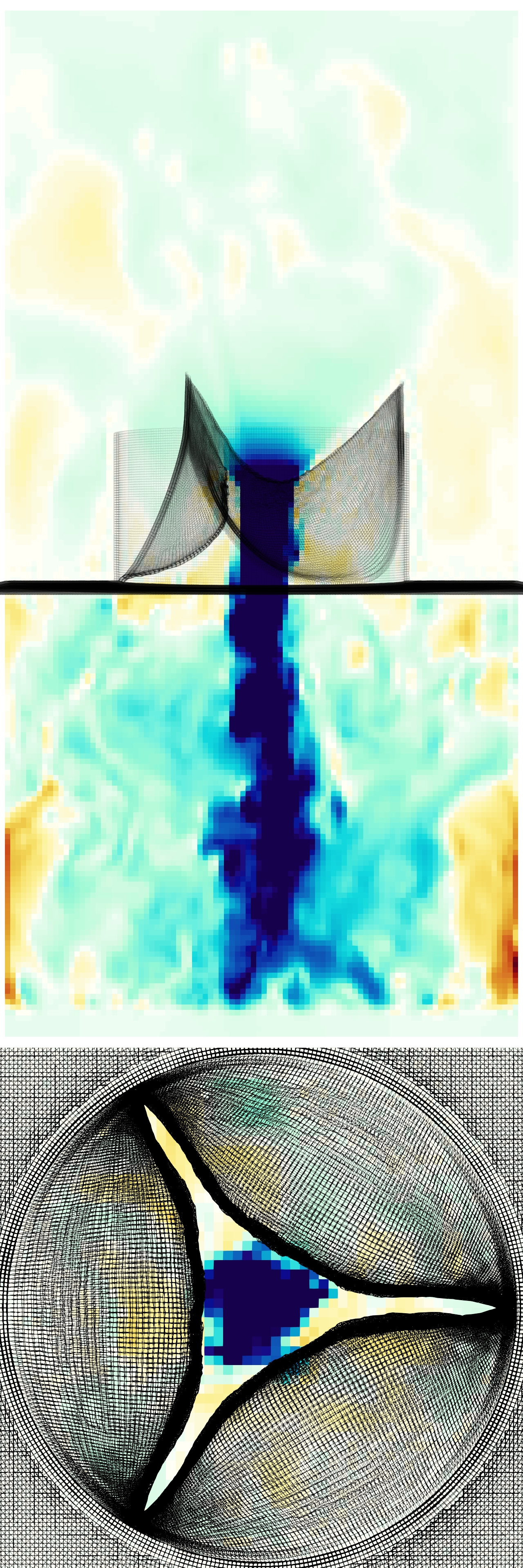}   &
\includegraphics[width=.2\columnwidth]{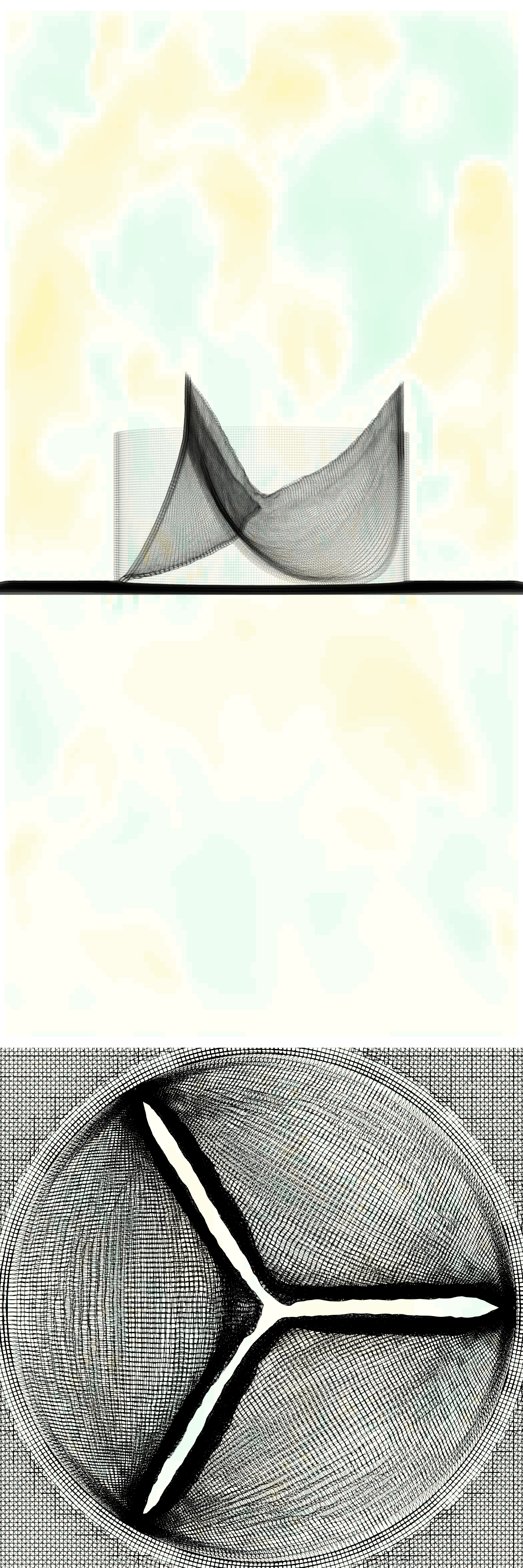} & 
\includegraphics[width=.2\columnwidth]{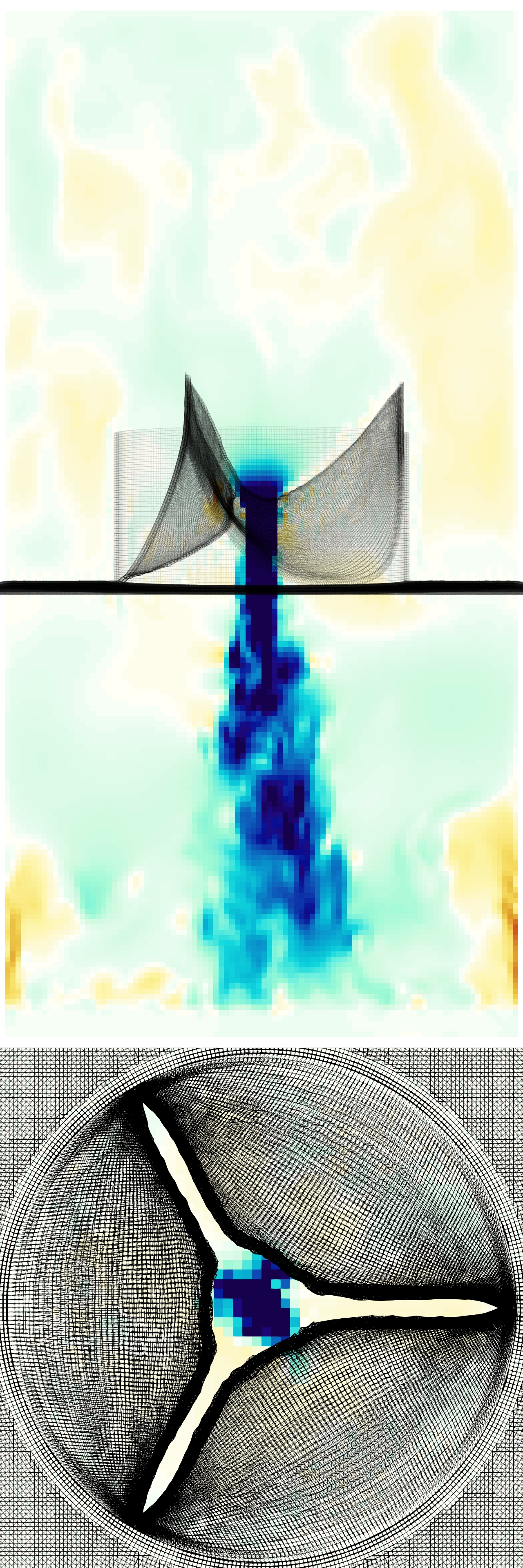} &
\includegraphics[width=.2\columnwidth]{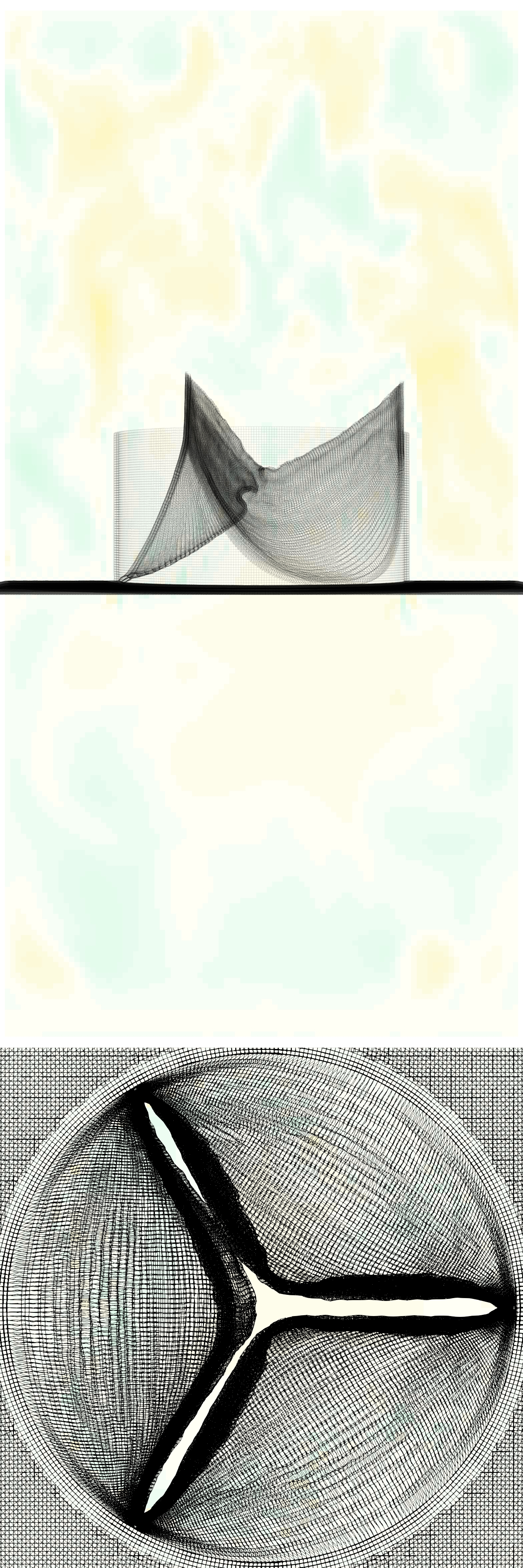}  
\end{tabular}
\caption{Slice view of the $z$ component of velocity showing poor results with altered gross morphology. 
The left panel shows a model with 2mm of predicted loaded length removed at the free edge. 
It leaks catastrophically. 
The center left shows a model with 2mm of predicted loaded length added to the free edge. 
It seals acceptably, but fails to form a good coaptation surface of positive length. 
The center right panel shows a model with 2mm of predicted loaded leaflet height removed.
It shows significant regurgitation. 
The right panel shows a model with 2mm of extra height. While the valve seals acceptably, an unappealing and unphysiological buckle has appeared near the free edge. 
} 
\label{three_failures}
\end{figure}

\subsection{Negative results with adjusted gross morphology}
\label{negative_morphology}

In this section, we test four slightly different geometries under physiological pressures. 
These are all less effective than the standard model geometry. 
This serves to illustrate how slight changes in gross morphology can degrade valve performance. 
Each model is constructed as described in Sections \ref{problem_formulation} and \ref{constitutive}, and the constitutive law is emergent from this process and varies slightly from model to model. 

In the first example, we remove 0.2 cm of length from the free edge of the model in the predicted loaded configuration. 
Note that the removed reference length is less than this value. 
This valve leaks dramatically, forming a large, central jet of backflow. 
In the second example, we add 0.2 cm of length from the free edge of the model, again in the predicted loaded configuration. 
This seals acceptably, but the extra leaflet causes the center coaptation region to not form, meaning that there is a lack of vertically aligned leaflet tissue near the free edge. 
In the third example, we remove 0.2 cm of height from leaflets in the predicted loaded configuration.
As with removing length from the free edge, this leaks dramatically.
In the last example, the leaflet is 0.2 cm taller than the standard example in the predicted loaded configuration. 
This model closes effectively, but a buckle appears at the free edge due to the extra material. 
The valve fails to form a flat coaptation zone, instead coapting only on where the buckling leaflet touches the other leaflets. 
A normal, healthy aortic valve is expected to have a coaptation height of approximately $0.34 r = 0.43$ cm for a 1.25 cm radius valve, in which the leaflets are flush against each other \cite{swanson1974dimensions}. 
We hypothesize that such a buckle may  allow aortic pressure to push on the buckled tissue, and so separate the leaflets then initiate regurgitation. 
Slice views of velocity on these three examples are shown in Figure \ref{three_failures}. 

These results suggest that the valve performance is highly sensitive to gross morphology. 
Too little length on the free edge or in leaflet height can cause the models to leak. 
Any extra height can create buckles and cause a coaptation zone of appropriate length to fail to form.

\subsection{Forward flow comparison}

\begin{figure*}[ht!]
\setlength{\tabcolsep}{0.5pt}
\centering
\begin{tabular}{ccccccccccc}
&
standard & 
linear & 
half rate & 
double rate & 
2 mm  &  
2 mm  &
2 mm  &
2 mm  \\ 
&
 & 
 & 
& 
 & 
shorter   &  
extra  &
shorter &
extra  \\ 
& 
 &
 & 
 & 
 & 
 free edge  &  
 free edge &
 height &
 height \\ 
\raisebox{110pt}{\includegraphics[width=.04\textwidth]{colorbar.jpeg}} &
\includegraphics[width=.117\textwidth]{aortic_2020486_384_f499232_0mm_radial_4mm_circ_basic_rcr_crease_removed_more_stiffness_down_paper1343.jpeg} &
\includegraphics[width=.117\textwidth]{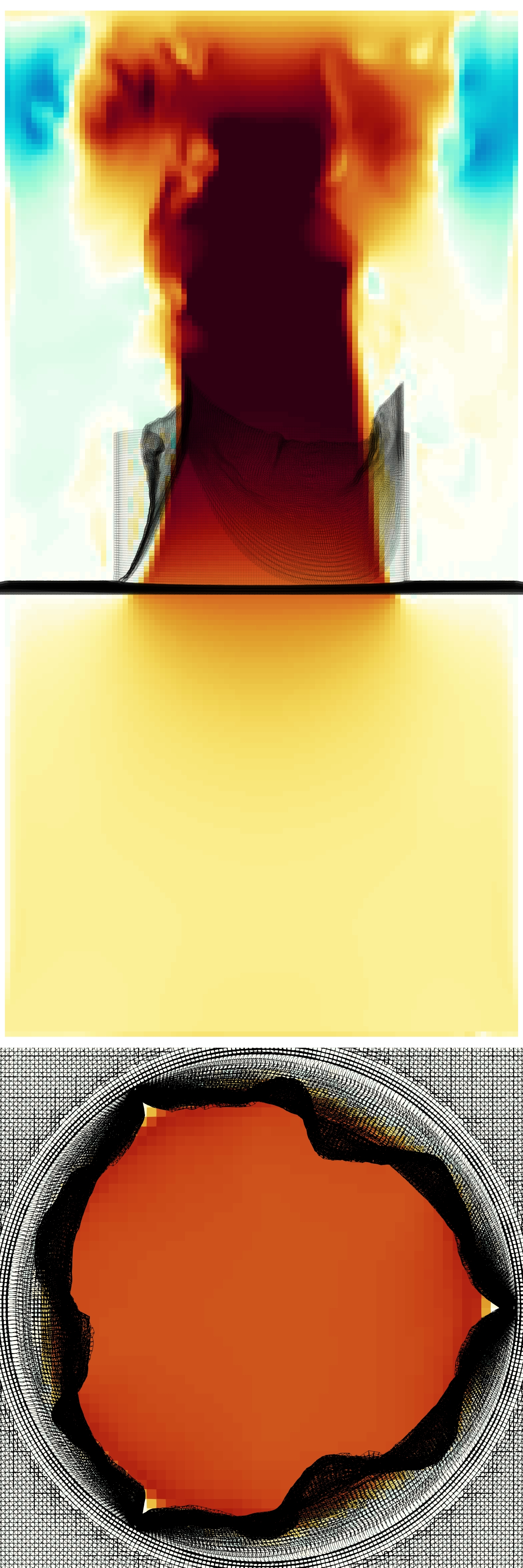} &
\includegraphics[width=.117\textwidth]{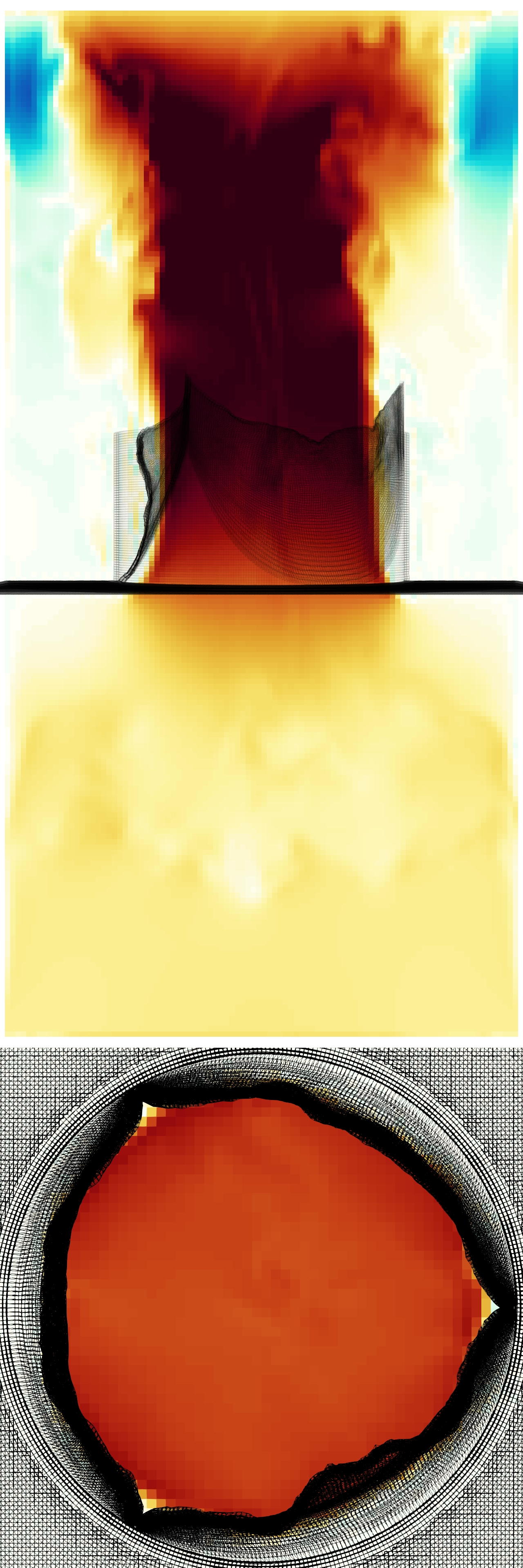} &
\includegraphics[width=.117\textwidth]{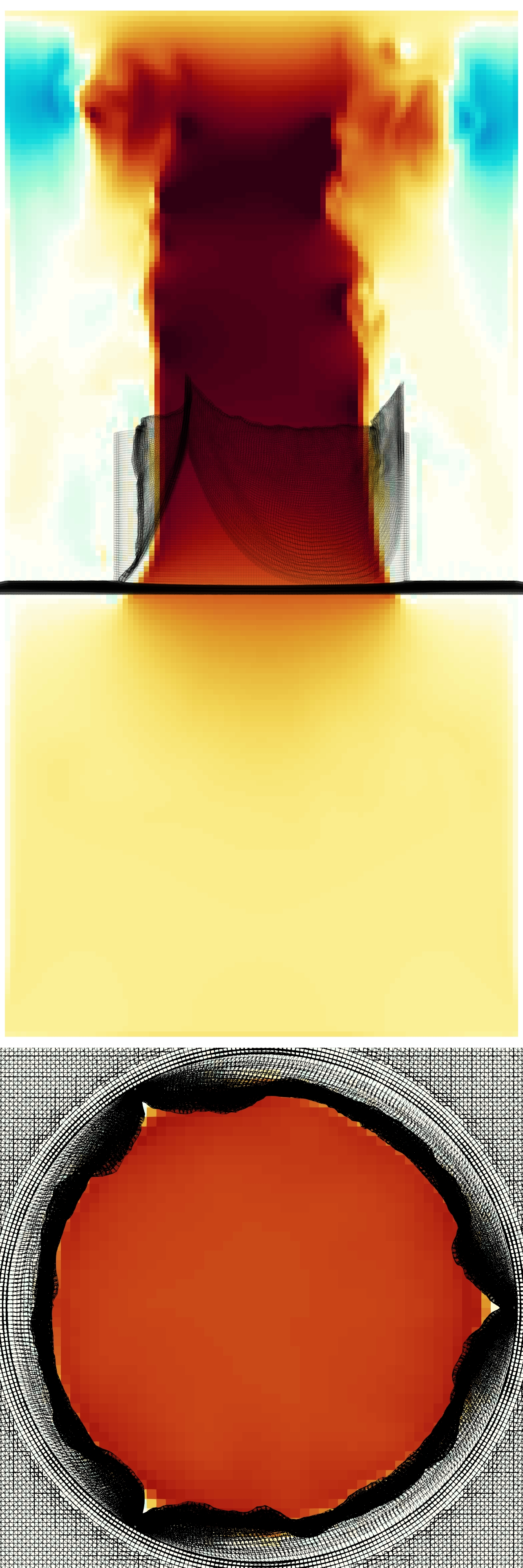} &
\includegraphics[width=.117\textwidth]{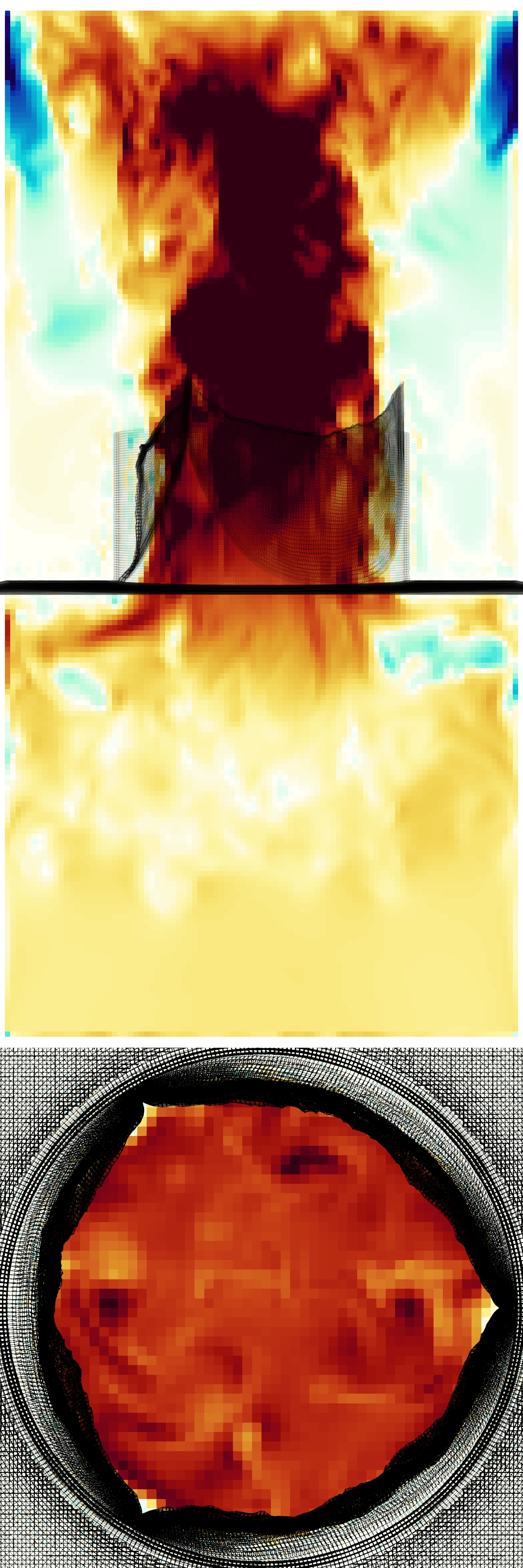}   &
\includegraphics[width=.117\textwidth]{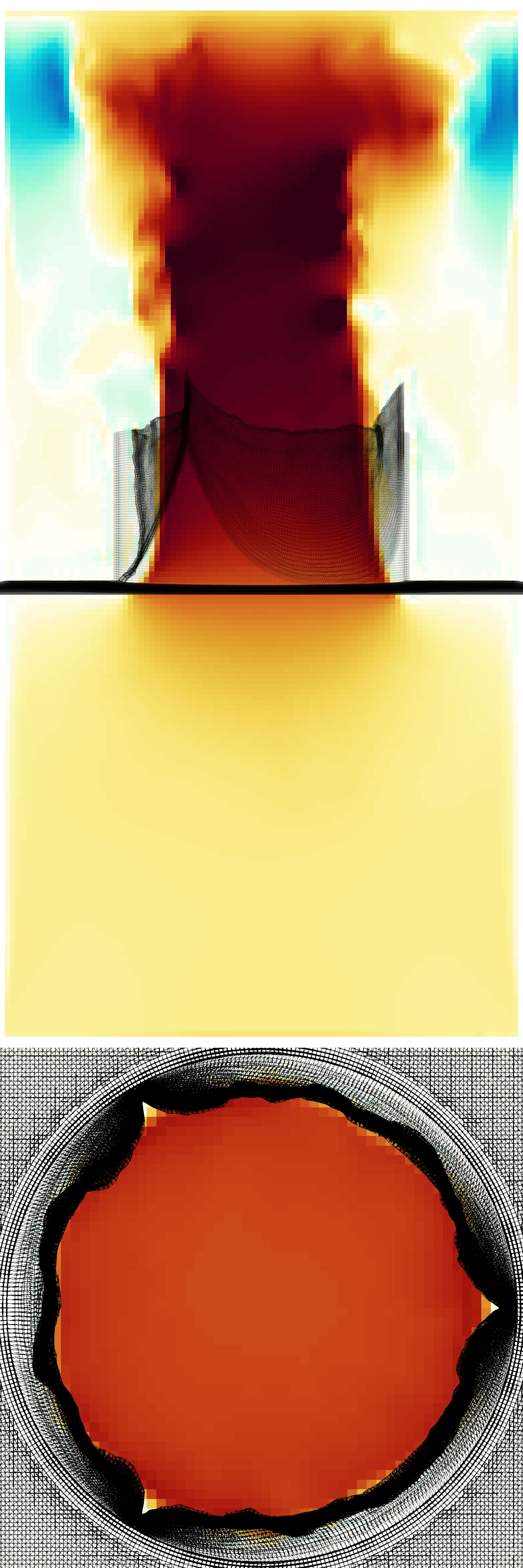} & 
\includegraphics[width=.117\textwidth]{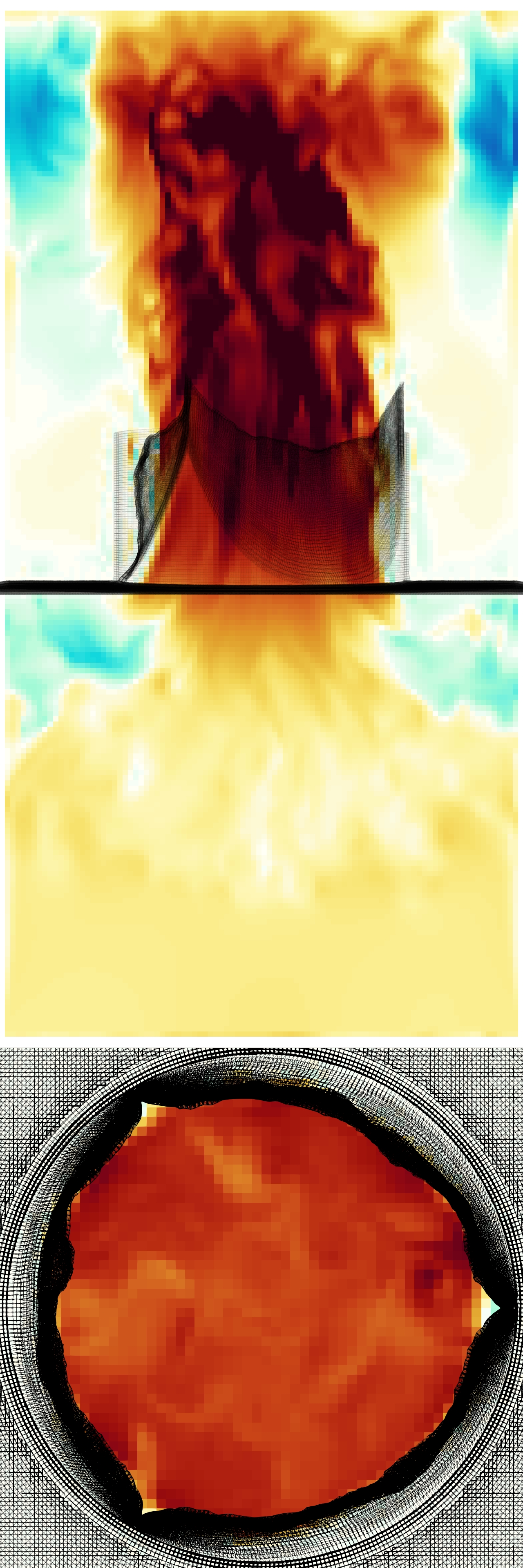} &
\includegraphics[width=.117\textwidth]{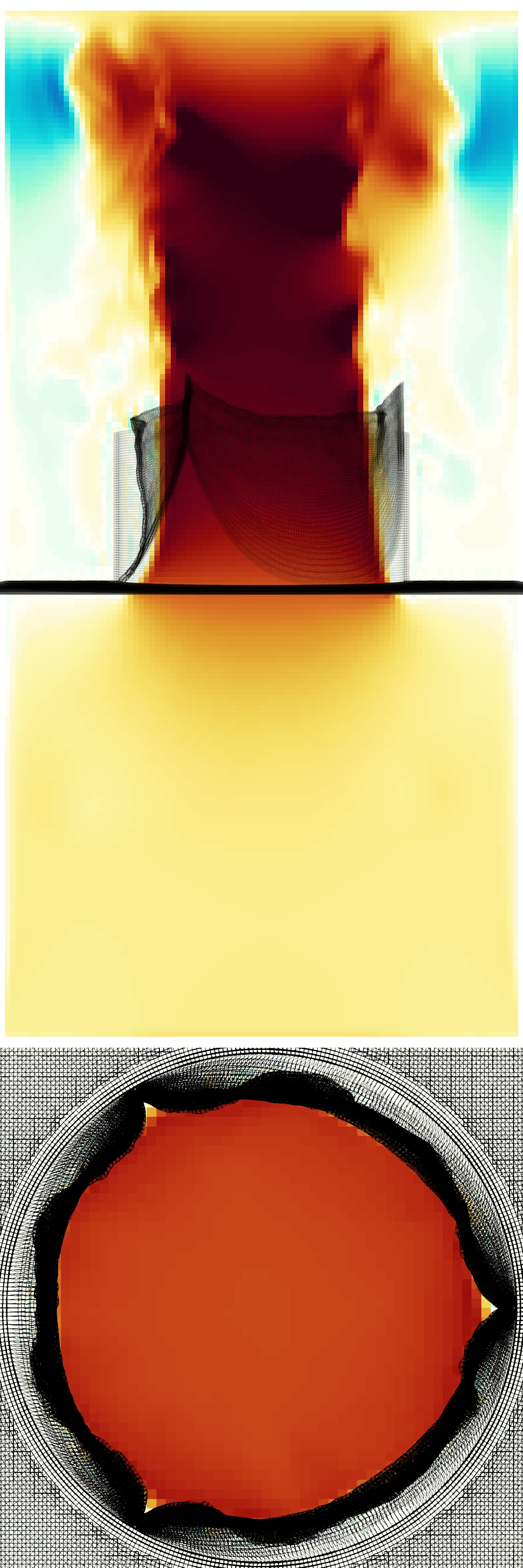} 
\end{tabular}
\caption{Slice view of the $z$ component of velocity with various constitutive laws at peak flow.
Shown are, from left to right, the standard model as discussed in Section \ref{results_fsi}, an unphysiological linear law, 
a law with half the experimentally measured exponential rates, 
a law with double the experimentally measured exponential rates, 
a valve with 2 mm shorter free edge, 
a valve with 2 mm extra free edge,
a valve with 2 mm shorter height 
and 
a valve with 2 mm extra height. 
}
\label{forward_flow_comparisons}
\end{figure*}

A comparison of the velocity field at peak forward flow across a variety of constitutive laws and gross morphologies is shown in Figure \ref{forward_flow_comparisons}.
All of the geometries and constitutive laws that do not regurgitate are qualitatively similar, whereas the models that leak (half exponential rate, shorter free edge and shorter leaflet height) show flow structure that remains from regurgitation during the previous cycle.
(Differing predicted loaded strains are omitted and similar to the standard model.)
The linear law shows a slightly smaller orifice area and a narrower jet. 
The model with half exponential rate, which leaked slightly, shows additional flow structure up- and downstream of the valve, whereas the model with double exponential rate is similar to the standard model. 
The models with reduced free edge length and reduced leaflet height leaked on the previous cycle, and show finer scale flow structures upstream of the valve and in the jet.
The models with extra free edge length and extra height show a qualitatively similar jet to that of the basic model. 

\section{Discussion}
\label{Discussion}

The results of FSI simulations in Section \ref{results_fsi} suggest that the model is robust and effective under physiological, hypo- and hypertensive pressures.
Achieving such function in a variety of conditions is essential for future studies of patient-specific cardiac flow. 

In Section \ref{constitutive_results}, we showed that our model construction produces the fully-loaded mean tangent moduli of $1.43 \cdot 10^{8}$ and $ 5.75 \cdot 10^{6}$ dynes/cm$^{2}$, circumferentially and radially, respectively, for a ratio of approximately 25.
These values are comparable to experimental results. 
Recall that Pham et al. found a circumferential tangent modulus of $9.9 \cdot 10^{7}$ dynes/cm$^{2}$ and a radial tangent modulus of $2.3 \cdot 10^{7}$ dynes/cm$^{2}$ \cite{pham2017quantification}. 
Our mean circumferential, fiber-direction tangent modulus is 44\% stiffer than their experimental results. 
Our radial stiffness is 25\% as stiff as their experimental results, though their experiments were on older human tissue that may become less compliant with age. 
Sauren et al. report an stiffness ratio of approximately 20:1, which is very close to our findings \cite{SAUREN1983327}. 
While a pressure load, strains and exponential rates were prescribed, stiffnesses and anisotropy ratios were not. 
The stiffnesses and ratio of anisotropy that emerges from this process -- solving the equations of equilibrium and tuning the constitutive law to match -- reproduce the material properties and anisotropy ratios that are found in natural aortic valve tissue. 

Our method produces heterogeneous material properties, we hypothesize that heterogeneity assists in good aortic valve function. 
There are ``hot spots'' of higher circumferential tangent modulus near the commissures, and the radial tangent modulus generally decreases moving from the annulus to the free edge. 
Experiments revealed heterogeneity in material properties and thickness \cite{sacks1998aortic,kas1985structure}, as well as variation in the distinct histological layers of the leaflet \cite{10.1115/1.2768111}. 
An experimental study noted the tendinous bunches near the commissures, and decreasing radial tangent modulus toward the free edge \cite{billiar2000biaxial}. 
There is some appearance of radially-oriented fibers near the annulus, that thin and dissipate before the belly of the leaflet \cite{SAUREN198097}. 
Radial collagen fiber bundles were observed on the ventricular side of the leaflet \cite{kas1985structure,swanson1974dimensions}, but other potential causes of material nonlinearity in the radial direction have been proposed \cite{billiar2000biaxial}.  
This suggests an experimental study: measure heterogeneous material properties, produce a map of fiber orientations and local nonlinear force response, then study their functional significance. 
Such a study could validate our predictions of heterogeneity and would advance understanding of the mechanics of the aortic valve.

In Section \ref{varied_strain}, we showed that with a consistent predicted loaded configuration, varying the prescribed strain and thus the reference configuration had little effect. 
This suggests that the loaded configuration, its tension and its tangent modulus are the primary determinants of good valve closure. 
Other studies have used reference lengths based on an excised valve, an excised aorta with the valve still attached or the valve at some point in the cardiac cycle in a living animal.
Some studies may include prestrain to account for these differences. 
None of these differences, however, appear as important having the appropriate tension at a geometry that allows a good seal to form.

In Section \ref{results_constitutive}, we simulated with a linear law, and changed the exponential rates in the constitutive law. 
The linear and half exponential rate functioned poorly, regurgitating at low pressures because they are too stiff at lower strains to achieve good coaptation.
The linear law prolapsed under high pressure.  
These results suggest that the leaflets need sufficient nonlinearity to function well over a large range of pressures. 
This suggests that the aortic valve should operate in two regimes -- very compliant from open to closed and barely loaded, loaded and very stiff once the loaded configuration is achieved to maintain a nearly constant position. 

In Section \ref{negative_morphology}, we showed that under changes in geometry, however, leaks occurred with models that are superficially similar.
With too little length at the free edge, strong central jets of regurgitation formed. 
With too much height on the leaflet, the free edge buckled unphysiologically in the coaptation region. 
This suggests an extremely precise range of lengths at the free edge, in the leaflet belly and in height is required for robust valve function.

These simulations suggest appropriate material properties and gross morphology for prosthetic aortic valve leaflets.
Our conclusions can be summarized as follows: 

\begin{enumerate}

\item Valve performance during closure is primarily determined by the loaded geometry, the force it exerts in the loaded geometry, and the tangent modulus in the loaded geometry. 

\item The reference configuration and pre-strain are not central to the valve closure when the loaded configuration remains similar. 

\item To function over a range of pressures, the leaflets must  behave in a sufficiently nonlinear manner. Materials that are linear or not nonlinear enough may function at physiological pressure but fail at hypotensive, hypertensive pressures or both.

\item The range of free edge lengths and leaflet heights that function well is narrow. Regurgitation occurs in leaflets that are too small and poor coaptation occurs in leaflets that are too large. 

\end{enumerate}

\section{Limitations}
\label{Limitations}

While many in vitro studies use a box-shaped valve tester, downstream hemodynamics are affected by the geometry, and testing in a model aortic root or beating heart would be a step forward. 
Despite testing an extensive variety of constitutive laws, we use a highly-specific, fiber-based constitutive law, in contrast to a more standard, three-dimensional hyperelastic formulation with a volumetric penalty term. 
Valve performance over many cardiac cycles may degrade if the springs connecting layers of the model are not stiff enough.
Bending rigidity is not included, nor is shear tension, though both are expected to be small. 
We leave a direct comparison of such models for future work. 
Anatomical details such as the nodules of Arantius, which may introduce some bending rigidity, were omitted.
Finally, tuning of the models is highly manual; automating the tuning process would be a significant step forward.

\section{Conclusions}
\label{Conclusions}

We began with a near first-principles statement, that the aortic valve leaflets must support a pressure, and derived a corresponding partial differential equation, the solution of which specified the predicted loaded configuration of the valve. 
Parameters were tuned to make the solutions match experimental observations about the gross morphology of loaded aortic valve leaflets and a constitutive law was created. 
This then creates a model suitable for simulation with the IB method. 
When simulated under physiological pressures, the leaflets coapt well, and the valve seals over multiple cardiac cycles, and allows physiological flow rates through. 
When simulated under pressures much lower or much higher than normal, the valve performs similarly. 
Real aortic valves are robust to a variety of loading pressures, and the model appears to mimic this property. 
We then conducted a number of experiments on reference configurations, constitutive laws and gross morphology. 

In conclusion, this method is robust and effective for FSI simulations involving the aortic valve. 
Having such a model is essential for further studies of healthy and pathological patient-specific cardiac flow. 
We hope that this model will be useful for such studies, and that the conclusions  will provide insights for designers of prosthetic aortic valves.

\section{Acknowledgements}

ADK was supported in part by a grant from the National Heart, Lung and Blood Institute \linebreak (1T32HL098049), Training Program in Mechanisms and Innovation in Vascular Disease at Stanford. 
ADK and ALM were supported in part by the National Science Foundation SSI grant \#1663671. 
Computing for this project was performed on the Stanford University's Sherlock cluster with assistance from the Stanford Research Computing Center. 
The authors would like to thank Michael Ma for  providing the image in Figure \ref{anatomy_diagrams}, left panel.

\section*{Appendix} 

\appendix 

\section{Pressure differences across periodic domains}
\label{appendix_pressure}

Here, we show that prescribing a uniform body force on a periodic domain is equivalent to prescribing the value of pressure at the top and bottom of the domain via a change of variables. 
Let $H$ denote the domain height and $P(t)$ denote the desired pressure difference at time $t$. 
Define a body force 
\begin{align}
\mathbf d = \left(0, 0, \frac{P(t) }{H} \right). 
\end{align}
Consider a solution of the Navier Stokes momentum equation on a domain periodic in $z$ subject to this body force with modified pressure $p_{mod}$.
\begin{align}
\rho ( \mathbf u_{t}  + \mathbf u \cdot \nabla \mathbf u ) &= - \nabla p_{mod} + \mu \Delta \mathbf u + \mathbf d . 
\label{periodic_ns}
\end{align}
By periodicity, the pressure at the bottom of the domain $z_{min}$ and top of the domain $z_{max}$ are equal, or 
\begin{align} 
p_{mod}(x,y,z_{min}) = p_{mod}(x,y,z_{max}) .  
\end{align}
Now consider the pressure $p$ defined as 
\begin{align}
p = p_{mod} - \frac{P(t)}{H} z . 
\end{align}
The identical velocity field $\mathbf u$ as in equation \eqref{periodic_ns} and the pressure field $p$ then solve the Navier Stokes momentum equation on a non-periodic domain  
\begin{align}
\rho ( \mathbf u_{t}  + \mathbf u \cdot \nabla \mathbf u ) &= - \nabla p + \mu \Delta \mathbf u , 
\label{nonperiodic_ns}
\end{align}
and further 
\begin{align}
p(x,y,z_{min}) = p(x,y,z_{max}) + P(t) . 
\end{align}
Thus, the pressure $p$ has the desired pressure difference across the domain.

\section{Convergence}
\label{convergence}

Here, we discuss convergence and periodicity of the simulations. 
We show results at two resolutions. 
The fine fluid resolution is $\Delta x_{fine} = 0.046875$ cm, and the fine structure resolution is targeted to half the fluid resolution, or $\Delta s_{fine} = 0.023$ cm. 
The coarse fluid resolution is twice that of the fine resolution, or $\Delta x_{coarse} = 0.09375$ cm, and the coarse structure resolution is targeted to $\Delta s_{coarse} = 0.047$ cm. 
Resolution double the fine resolution is prohibitively expensive to run. 
Simulations are run for four cardiac cycles, rather than three as in the remainder of the paper.

\begin{figure}[t!]
\centering
\includegraphics[width=0.5\columnwidth]{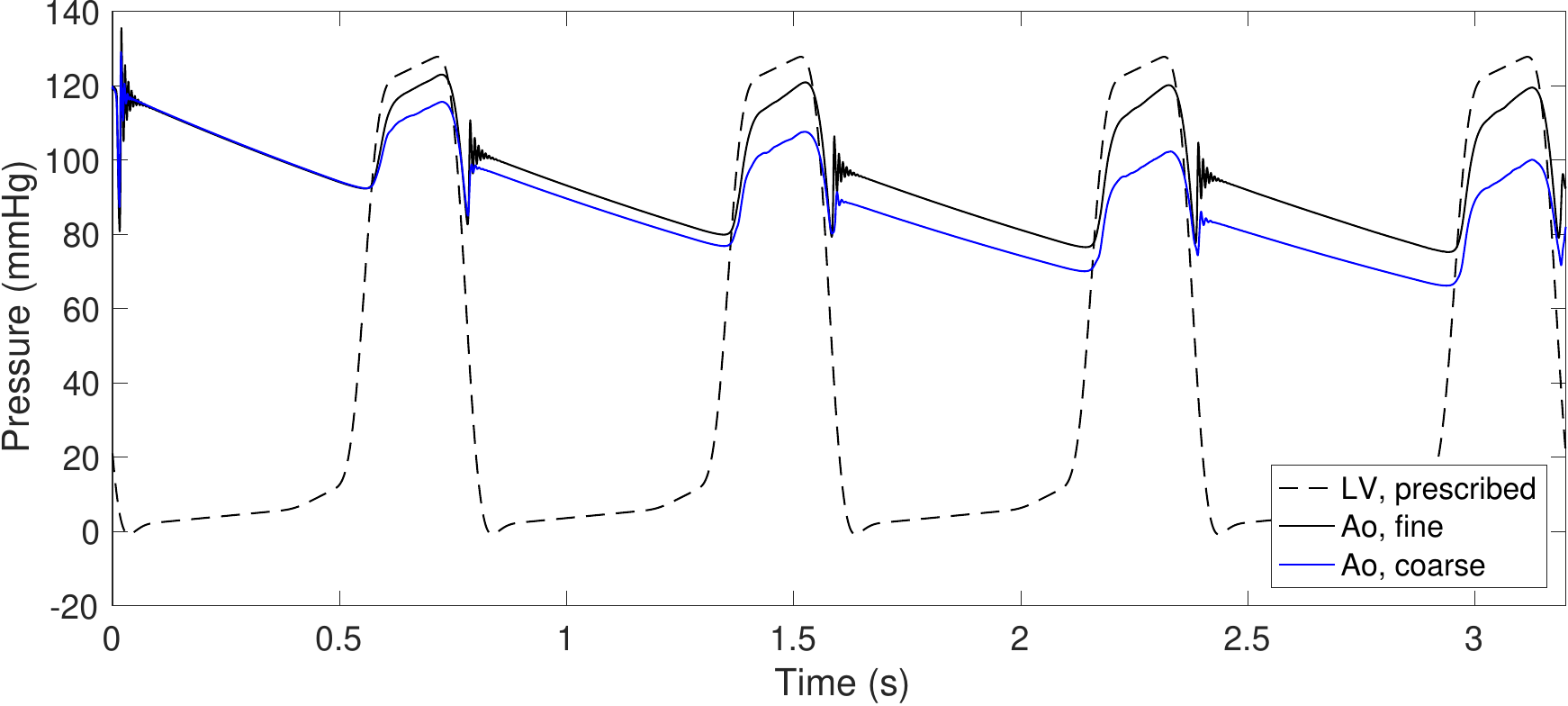} 
\includegraphics[width=0.5\columnwidth]{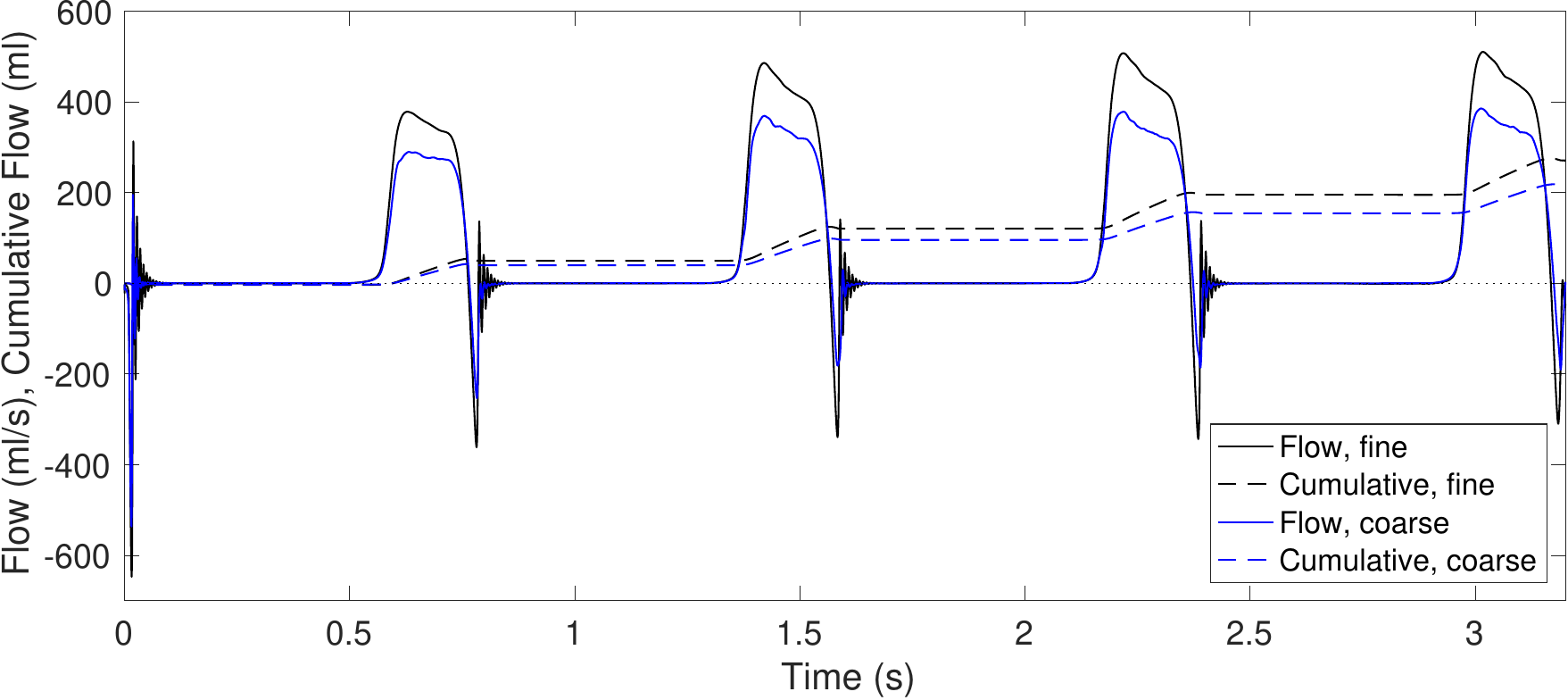} 
\caption{Pressure and flow rates at fine resolution (as used throughout the paper) and coarse resolution. } 
\label{flow_pressure_convergence}
\end{figure}

\begin{figure*}[t!]
\setlength{\tabcolsep}{0.5pt}
\centering
\begin{tabular}{ccccc}
& 
fine &
coarse &
fine &
coarse\\
& 
diastole &
diastole &
systole &
systole \\
\raisebox{210pt}{\includegraphics[width=.04\columnwidth]{colorbar.jpeg}} 
&
\includegraphics[width=.2\columnwidth]{aortic_2020486_384_f499232_0mm_radial_4mm_circ_basic_rcr_crease_removed_more_stiffness_down_paper1208.jpeg} & 
\includegraphics[width=.2\columnwidth]{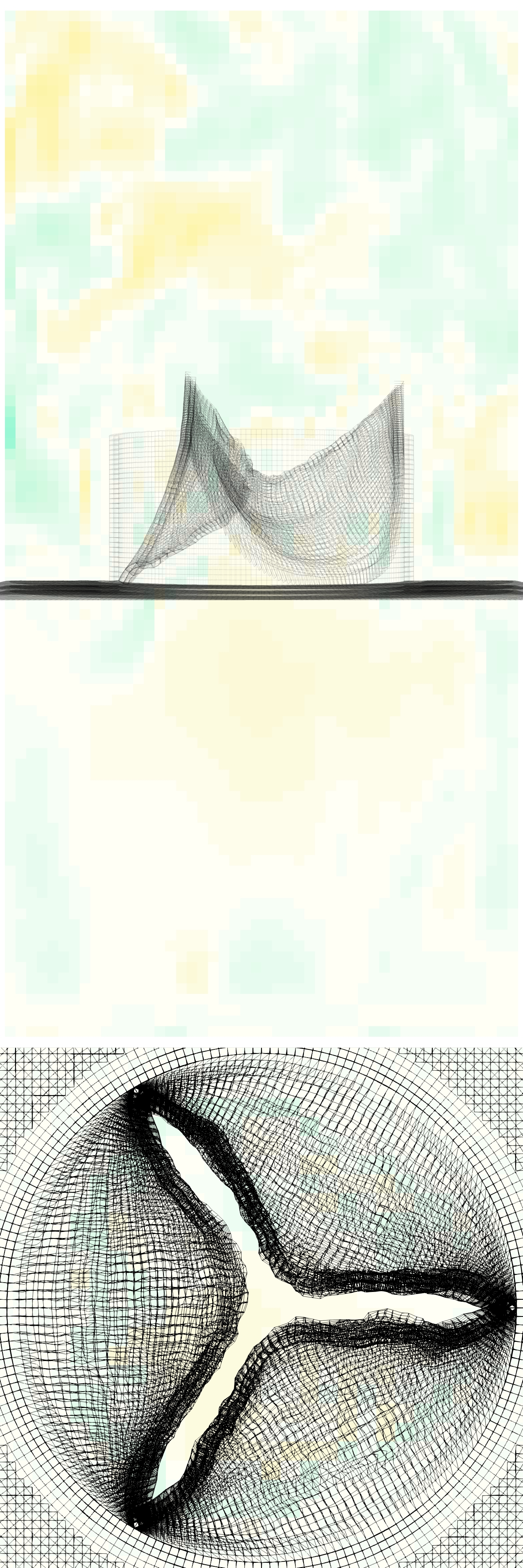} &
\includegraphics[width=.2\columnwidth]{aortic_2020486_384_f499232_0mm_radial_4mm_circ_basic_rcr_crease_removed_more_stiffness_down_paper1343.jpeg} &
\includegraphics[width=.2\columnwidth]{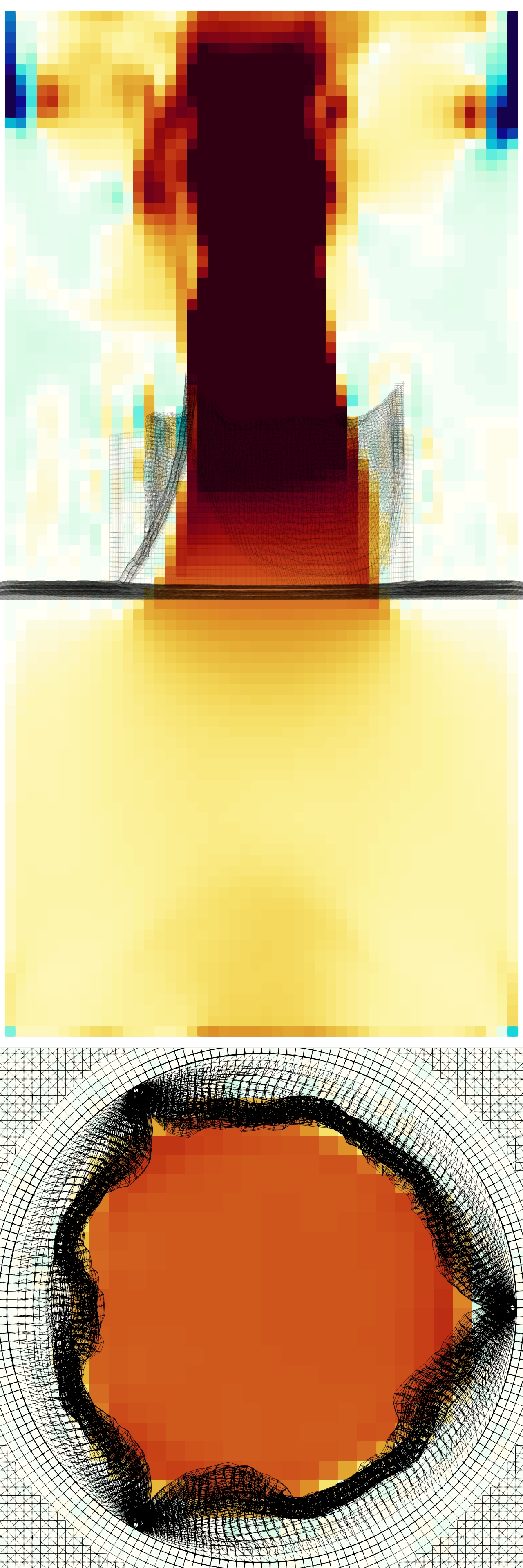} 
\end{tabular}
\caption{Slice view of the $z$ component of velocity during closure and forward flow with fine and coarse resolution. 
} 
\label{flow_convergence}
\end{figure*}

Precise convergence in IB simulations is challenging to achieve for the following reasons, though we achieve some qualitative and quantitative agreement across cycles and resolutions. 
Since the IB method uses diffuse interfaces, the structure interacts with fluid points in the support of the discrete delta function, up to $2.5 \Delta x$ away. 
This causes a decrease in the effective orifice area of the valve, even with the same configuration of the structure itself, and thus a slightly lower flow rate. 
Thus, the fluid/structure domain has a resolution-dependent resistance to forward flow. 
The aortic pressure is determined by the lumped parameter network, which in turn depends on the flow rate through the three-dimensional model. 
Precise, simultaneous periodicity of the ODE and three dimensional flow dynamics is thus challenging to achieve. 
Additionally, the velocity field has physical instabilities since Reynolds numbers are much larger than one, so precise correspondence of flow fields from cycle to cycle is not expected.

The driving pressure and flow rates are shown in Figure \ref{flow_pressure_convergence}. 
After the first cardiac cycle, the fine resolution has aortic minimum and maximum pressures of 75-79 and 119-121 mmHg, respectively. 
The total flow per cycle ranges from 70.7 to 75.5 ml. 
Due to lack of dramatic changes in flow or pressure from the second through fourth cycles, all other simulations are stopped after three cycles and results are presented during the third cycle. 
The coarse resolution, due to increased resistance and lower flow rates driving the lumped parameter network, shows a decreasing trend in aortic pressure and a corresponding increase in flow. 
Thus, at fine resolution the results are closer to periodic in time than at coarse resolution.

The velocity field at fine and coarse resolution during two points in the cardiac cycle is shown in Figure \ref{flow_convergence}. 
The valve and flow fields during closure appear similar at both resolutions. 
The leaflets are closer together in the fine version, and about twice as far apart given the thicker discrete delta function in the coarse version. 
During forward flow, both flows show a strong jet. 
Vortices and flow structure appear in the fine resolution jet. 
The coarse resolution valve opens slightly less, and the effective thickening from the IB method is greater, so the coarse resolution jet is narrowed and has a more uniform appearance. 
Despite substantially more visible structure in the fine resolution version, both resolutions produce qualitatively similar flow fields.

{\small
\bibliographystyle{acm}
\bibliography{aortic_refs.bib}
}

\end{document}